\documentclass[11pt,twoside,a4paper,pdf]{article}
\usepackage{graphicx,color}
\usepackage{times}
\usepackage{booktabs}
\usepackage{cite}
\usepackage{a4wide}
\usepackage{xspace}
\usepackage{amsmath, amssymb}
\usepackage[small,bf]{caption} 
\usepackage[colorlinks=true, pdftex, pdftitle={Efficient Matrix-Element Matching with Sector Showers}, pdfauthor={J.J. Lopez-Villarejo and Peter Zeiler Skands}, pdfsubject={Monte Carlo generators}, pdfkeywords={Monte Carlo, generators, QCD, Parton Showers, HEP,VINCIA}]{hyperref}
\usepackage{subfig}

 \newcommand{\figscale}{1.0}
 \newenvironment{eqn}{\begin{equation}}{\end{equation}}
 \newenvironment{eqnar}{\begin{eqnarray}}{\end{eqnarray}}
 \newcommand{\capt}[1]{\caption{#1}}

\newcommand{\eqRef}[1]{eq.~\eqref{#1}\xspace}

\newcommand{\eqsRef}[1]{eqs.~\eqref{#1}\xspace}

\newcommand{\secRef}[1]{section~\ref{#1}\xspace}
\newcommand{\SecRef}[1]{Section~\ref{#1}\xspace}

\newcommand{\tabRef}[1]{tab.~\ref{#1}\xspace}

\newcommand{\figRef}[1]{fig.~\ref{#1}\xspace}
\newcommand{\FigRef}[1]{Fig.~\ref{#1}\xspace}

\newcommand{\figsRef}[1]{figs.~\ref{#1}\xspace}

\newcommand{\IntA}{{{\cal A}}}
\def\dash{\hbox{-\kern-.02em}}

\newcommand{\abar}{\ensuremath{\bar{a}}}
\newcommand{\atrial}{\ensuremath{a_{\mathrm{trial}}}}
\newcommand{\abartrial}{\ensuremath{\bar{a}_{\mathrm{trial}}}}
\newcommand{\atrialemit}{\ensuremath{a_{\mathrm{trial\dash{}emit}}}}

\newcommand{\mrm}[1]{\mathrm{#1}}
\newcommand{\tsc}[1]{\textsc{#1}}

\newcommand{\pT}{\ensuremath{p_\perp}\xspace}
\newcommand{\TeV}{\,\mbox{Te\kern-0.2exV}}
\newcommand{\GeV}{\,\mbox{Ge\kern-0.2exV}}
\newcommand{\MeV}{\,\mbox{Me\kern-0.2exV}}
\newcommand{\keV}{\,\mbox{ke\kern-0.2exV}}
\newcommand{\eV}{\,\mbox{e\kern-0.2exV}}
\newcommand{\qbar}{\bar{q}}
\renewcommand{\d}[1]{\ensuremath{\mrm{d}#1}}

\newcommand{\Ar}{\tsc{Ariadne}\xspace}

\newcommand{\Mg}{\tsc{MadGraph}\xspace}
\newcommand{\Py}{\tsc{Pythia}\xspace}
\newcommand{\Sh}{\tsc{Sherpa}\xspace}
\newcommand{\Vc}{\tsc{Vincia}\xspace}

\newlength{\tabcolsepsave}
\newenvironment{loopsnlegs}[1][t]{
\setlength{\tabcolsepsave}{\tabcolsep}
\setlength{\tabcolsep}{0pt}
\begin{tabular}{cc}\parbox[c]{1.1em}{\rotatebox{90}{\small $\ell$ (loops)}}&%
\begin{tabular}[#1]}{
\end{tabular}\\[-1mm]
 & \small $k$ (legs)
\end{tabular}%
\setlength{\tabcolsep}{\tabcolsepsave}
}
\newcommand{\cbox}[2]{%
\begin{minipage}[c]{1.4cm}%
\center%
{%
\parbox[c]{1.4cm}{\includegraphics*[width=1.4cm]{#1.pdf}}}%
\end{minipage}%
\hspace*{-1.4cm}%
\begin{minipage}[c]{1.4cm}
\center
#2%
\end{minipage}}
\newcommand{\pqcd}[2][]{\ensuremath{\sigma^{(#1)}_{#2}}}

\newcommand{\gbox}[1]{\cbox{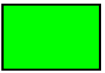}{#1}}
\newcommand{\wbox}[1]{\cbox{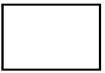}{#1}}

\newcommand{\gybox}[1]{\cbox{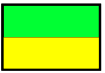}{#1}}

\newcommand{\ybox}[1]{\cbox{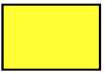}{#1}}

\begin{document}

\thispagestyle{empty}
\begin{minipage}{\textwidth}\vspace*{-5mm}
\flushright
CERN-TH-2011-202\\
MCNET-11-22
\end{minipage}
\vskip8mm
\begin{center}
\Large{\bf Efficient Matrix-Element Matching with Sector Showers}\\[5mm]
\end{center}
\vskip5mm
\begin{center}
{\large J.J.~Lopez-Villarejo and P.~Skands}\\[2mm]\small
Theoretical Physics, CERN CH-1211,
Geneva 23, Switzerland
\end{center}
\vskip5mm
\begin{center}
\parbox{0.83\textwidth}{\small
\textbf{Abstract} --- 
A Markovian shower algorithm based on ``sector antennae'' is
presented and its main properties illustrated. Tree-level full-color
matrix elements can be automatically incorporated in the algorithm and are
re-interpreted as process-dependent $2\to n$ antenna functions. In hard
parts of phase-space, these functions generate tree-level
matrix-element corrections to the shower. In soft parts, they should
improve the logarithmic accuracy of it.  The number of
matrix-element evaluations required per order of matching is 1, 
with an unweighting efficiency that remains very high for arbitrary
numbers of legs. Total  rates can be augmented to NLO precision in a
straightforward way. As a
proof of concept, we present an implementation in  the publicly
available \Vc plug-in to the \Py 8 event generator, for hadronic
$Z^0$ decays including tree-level matrix elements through ${\cal
  O}(\alpha_s^4)$.\vspace*{3mm}
}
\end{center}
\vskip8mm
\section{Introduction}
Perturbative calculations of high-energy processes typically start
from the calculation of one or more Matrix 
Elements (MEs) for specific signal and background processes. By virtue
of the factorization theorem, such  ``hard'' or ``short-distance'' 
partonic processes can be factored off from lower-scale physics and
computed in a systematic way. 
At Leading Order (LO), the 
procedure is standard textbook material and
it has also by now been 
highly automated, by the advent of general-purpose tools like \textsc{CalcHep} \cite{Pukhov:2004ca},
\textsc{CompHep} \cite{Boos:2004kh}, \textsc{MadGraph} \cite{Alwall:2007st}, 
and others   
\cite{Kanaki:2000ey,Krauss:2001iv,Moretti:2001zz,Bahr:2008pv,Gleisberg:2008fv}.  

To the simple LO picture, several corrections must be added in order
to obtain more realistic and accurate descriptions that can be compared
with experimental observables. On the perturbative side, one may
access these corrections either by  
computing more coefficients in the fixed-order
expansion explicitly --- as in higher-order calculations --- or by
approximating them via  
infinite-order resummations --- as in parton showers. 

\begin{figure}[t]
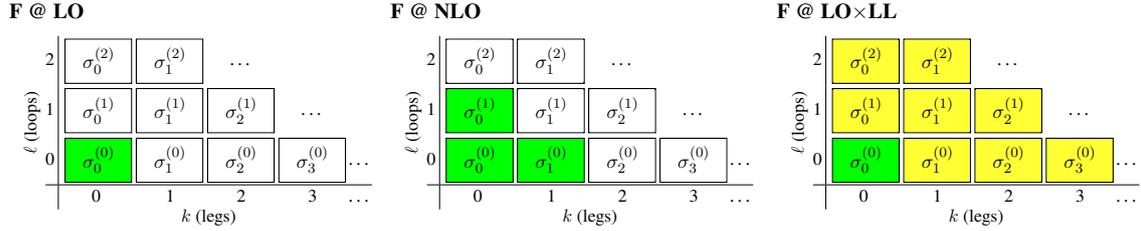

\begin{center}
\scalebox{0.67}{
\begin{tabular}{l}
\large\bf F @ LO\\[2mm]
\begin{loopsnlegs}[c]{p{0.25cm}|ccccc}
 \small 2&~\wbox{\pqcd[2]{0}} & \wbox{\pqcd[2]{1}} & \ldots \\[2mm]
 \small 1&~\wbox{\pqcd[1]{0}} & \wbox{\pqcd[1]{1}}  
   & \wbox{\pqcd[1]{2}} & \ldots \\[2mm]
 \small 0&~\gbox{\pqcd[0]{0}} & \wbox{\pqcd[0]{1}} 
   & \wbox{\pqcd[0]{2}} 
   & \wbox{\pqcd[0]{3}} & \ldots \\
\hline
& \small 0 & \small 1 & \small 2 & \small 3 & \ldots
 \end{loopsnlegs}\end{tabular}\hspace*{-0.15cm}
\begin{tabular}{l}
\large\bf F @ NLO\\[2mm]
\begin{loopsnlegs}[c]{p{0.25cm}|ccccc}
 \small 2&~\wbox{\pqcd[2]{0}} & \wbox{\pqcd[2]{1}} & \ldots &
\\[2mm]
 \small 1&~\gbox{\pqcd[1]{0}} & \wbox{\pqcd[1]{1}}  
   & \wbox{\pqcd[1]{2}} & \ldots \\[2mm]
 \small 0&~\gbox{\pqcd[0]{0}} & \gbox{\pqcd[0]{1}} 
   & \wbox{\pqcd[0]{2}} &\wbox{\pqcd[0]{3}} & \ldots \\
\hline
& \small 0 & \small 1 & \small 2 & \small 3 & \ldots
 \end{loopsnlegs}
\end{tabular}\hspace*{-0.15cm}
\begin{tabular}{l}
\large\bf F @ LO$\times$LL\\[2mm]
\begin{loopsnlegs}[c]{p{0.25cm}|ccccc}
 \small 2&~\ybox{\pqcd[2]{0}} & \ybox{\pqcd[2]{1}} & \ldots & 
\\[2mm]
 \small 1&~\ybox{\pqcd[1]{0}} & \ybox{\pqcd[1]{1}}  
   & \ybox{\pqcd[1]{2}} & \ldots \\[2mm]
 \small 0&~\gbox{\pqcd[0]{0}} & \ybox{\pqcd[0]{1}} 
   & \ybox{\pqcd[0]{2}} &\ybox{\pqcd[0]{3}} & \ldots \\
\hline
& \small 0 & \small 1 & \small 2 & \small 3 & \ldots
 \end{loopsnlegs}
\end{tabular}}
\caption{Illustration of the part of the perturbative series covered
  by {\sl (left)} a calculation at LO, {\sl (middle)} a calculation at
  NLO, and {\sl (right)} a calculation at LO combined with an LL
  resummation. Darker shaded (green) boxes indicate exact coefficients
  while lighter shaded (yellow) boxes indicate LL approximations to
  them. 
\label{fig:loopsnlegs1}}
\end{center}
\end{figure}

To help
illustrate the complementarity of these two approaches, for an
arbitrary final state, ``$F$'', we shall use 
 diagrams such as the ones shown in
\figRef{fig:loopsnlegs1}, from  \cite{Skands:2011pf}, in which 
the horizontal and vertical axes indicate the numbers of additional
legs $(k)$ and
loops $(\ell)$ beyond LO, respectively.
The notation $\sigma^{(\ell)}_k$ is used to represent the sum
of contributions to the cross section for fixed $k$ and $\ell$. 
These  are separately divergent for
$k+\ell\ge 1$, and only the sum over all coefficients with fixed
$k+\ell=n$ is finite. (See \cite{Skands:2011pf} for a more pedagogical
introduction to this type of diagrams.)

In the left-hand pane of \figRef{fig:loopsnlegs1}, 
the part of the series covered by the LO matrix element has been
shaded. In a Next-to-Leading-Order (NLO) calculation, two further
coefficients are computed 
exactly, as illustrated in the middle pane, but all other coefficients
are neglected. Conversely, in a Leading-Logarithmic (LL) resummation, 
infinite numbers of both legs and loops can be included, but
only the leading singular parts of each coefficient will be correctly
accounted for, which we illustrate by giving the boxes
corresponding to coefficients with $k+\ell\ge 1$ a lighter (yellow) 
shading in the right-hand pane of \figRef{fig:loopsnlegs1}. 

\begin{figure}[t]
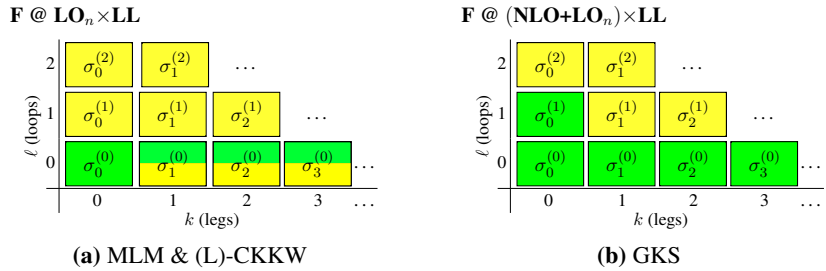

\centering
\subfloat[MLM \& (L)-CKKW\label{fig:ckkw}]{
\scalebox{0.67}{
\begin{tabular}{l}
{\large\bf F @ LO$_n\times$LL}\\[2mm]
\begin{loopsnlegs}[c]{p{0.25cm}|ccccc}
 \small 2&~\ybox{\pqcd[2]{0}}&~\ybox{\pqcd[2]{1}} &  \ldots & 
\\[2mm]
 \small 1&~\ybox{\pqcd[1]{0}} & \ybox{\pqcd[1]{1}}  &\ybox{\pqcd[1]{2}}  
   & \ldots \\[2mm]
 \small 0&~\gbox{\pqcd[0]{0}} & \gybox{\pqcd[0]{1}} 
   & \gybox{\pqcd[0]{2}} &  \gybox{\pqcd[0]{3}} & \ldots \\
\hline
& \small 0 & \small 1 & \small 2 & \small 3 & \ldots
 \end{loopsnlegs}
\end{tabular}}} \hspace*{5mm}
\subfloat[GKS\label{fig:gks}]{
\scalebox{0.67}{
\begin{tabular}{l}
{\large\bf F @ $($NLO+LO$_{n})\times$LL}\\[2mm]
\begin{loopsnlegs}[c]{p{0.25cm}|ccccc}
 \small 2&~\ybox{\pqcd[2]{0}} & \ybox{\pqcd[2]{1}} & \ldots & 
\\[2mm]
 \small 1&~\gbox{\pqcd[1]{0}} & \ybox{\pqcd[1]{1}}  
   & \ybox{\pqcd[1]{2}} & \ldots \\[2mm]
 \small 0&~\gbox{\pqcd[0]{0}} & \gbox{\pqcd[0]{1}} 
   & \gbox{\pqcd[0]{2}} &\gbox{\pqcd[0]{3}} & \ldots \\
\hline
& \small 0 & \small 1 & \small 2 & \small 3 & \ldots
 \end{loopsnlegs}
\end{tabular}}}
\caption{Illustration of the part of the perturbative series covered
  by {\sl (left)} MLM \& (L)-CKKW matching, and {\sl
    (right)}  GKS matching.}
\end{figure}

Several approaches for ``matching'' the two kinds of approaches are
already in widespread use. Here, we shall focus on the 
matching of parton showers to LO matrix elements with large numbers of
additional legs. For this problem, there are essentially two dominant
approaches, called MLM~(see \cite{Alwall:2007fs} for a description) 
and
(L)-CKKW~\cite{Catani:2001cc,Krauss:2002up,Lonnblad:2001iq,Mrenna:2003if,Lavesson:2007uu}, 
see 
\cite{Buckley:2011ms} for a recent pedagogical review. 
Two main limiting factors in these approaches are that the computational speed
 falls off steeply with the number of matched partons, and that 
both approaches only apply matching above a certain ``matching
scale'', below which the pure shower is used for all
multiplicities. This is illustrated in \figRef{fig:ckkw}, in which the
matched multiplicities are shown with half light (yellow) and half
dark (green) shading.

An alternative multileg matching strategy
was recently proposed by GKS~\cite{Giele:2011cb}. This algorithm uses
an antenna-based parton shower \cite{Giele:2007di,Ridder:2011dm} 
as its underlying phase-space
generator\footnote{The
idea of using a shower for phase-space generation also underlies the
SARGE~\cite{Draggiotis:2000gm} and GENEVA~\cite{Bauer:2008qj}
algorithms.}. 
Matching to matrix elements is performed by applying 
multiplicative corrections to the branching probability at each step
of the algorithm, an approach that was first pioneered in \cite{Bengtsson:1986hr}
for matching to a single
additional parton in PYTHIA~\cite{Sjostrand:2006za}. The
generalization to multiple legs proposed in \cite{Giele:2011cb} relies
explicitly on unitarity to remove the need for a matching scale also 
beyond the first matched leg. In principle, it can therefore be
applied over all of phase-space up to the highest matched
multiplicity, though for reasons of algorithmic speed, a low matching
scale may still be imposed beyond the first few additional legs. The
part of the perturbative series that can be covered by this approach
is illustrated in \figRef{fig:gks}. We note that the one-loop correction
to the lowest multiplicity can also be included (as indicated on the
figure), since the GKS formalism essentially reduces to the
POWHEG one~\cite{Nason:2006hfa,Frixione:2007vw} at this order.

In terms of algorithmic speed, the number of individual shower paths
that populate each $n$-parton phase-space point is a deciding
factor within the GKS formalism, since at least one 
$(n-1)$-parton matrix element has to be evaluated for each
path. Showers based on partons and/or partitioned dipoles 
(such as Altarelli-Parisi
\cite{Altarelli:1977zs,Marchesini:1983bm,Bengtsson:1986et,Sjostrand:2004ef} 
or Catani-Seymour 
\cite{Catani:1996vz,Nagy:2007ty,Dinsdale:2007mf,Schumann:2007mg} ones)  
produce one term per color charge in the event, i.e., 
 one per quark and two per gluon. 
Showers based on antennae
\cite{Gustafson:1987rq,Lonnblad:1992tz,Winter:2007ye,Giele:2007di}  
are slightly more economic, producing
only one term per color-connected \emph{pair} of color charges. This difference is
still comparatively insignificant, however, compared to the
proliferation of terms caused by the fact that only strongly
ordered paths can contribute (using whatever definition of
  ordering the particular shower algorithm's authors prefer):
to see which paths actually
contribute to each $n$-parton configuration, one must check the
ordering condition all the way back to the Born configuration,
including the effects of any previous matching steps. Even for the 
antenna-based showers, this makes the number of paths contributing to
the $m^\mrm{th}$ branching grow like $m!$, which quickly becomes
intractable. The main improvement proposed in \cite{Giele:2011cb} 
was to replace the strong-ordering condition by a smoothly damped and 
strictly Markovian equivalent, which eliminates the factorial. This
brings the number of terms produced at each order down to a linear
dependence on $m$, resulting in substantial speed gains for high
parton multiplicities. 

There is, however, an alternative formulation of the antenna language
\cite{Kosower:1997zr,Kosower:2003bh}, 
for which only one term contributes to each phase-space point. We
refer to this as ``sector'' antennae, to distinguish them from the
``global'' antennae used in
\cite{GehrmannDeRidder:2005cm,Giele:2011cb}. The two kinds differ in how the 
collinear singularities of gluons are partitioned among neighboring
antennae. In the global approach, the gluon-collinear singularity
is partitioned such that two neigbouring antennae each contain
``half'' of it; their sum reproduces the full singularity. In
the sector case, \emph{both} of the neighboring antennae contain the
full collinear singularity, but only one of them (typically the most
singular one) is allowed to
contribute to each $(n+1)$-parton phase-space point. This divides up
the $(n+1)$-parton phase-space into a number of ``sectors'' inside
each of which only a single 
antenna contributes. 

In this paper, we present a complete shower formalism based on such sector
antennae, including an adaptation and implementation of 
GKS matching. We note that another formalism dealing with sector 
showers has also been proposed
\cite{Larkoski:2009ah,Larkoski:2011fd}. While those works 
also treat polarization and mass effects, which are neglected here,
they do not contain an explicit implementation or matching strategy, which
are included here and have been made publicly available in the VINCIA
code~\cite{vincia}, a plug-in to the PYTHIA~8 event
generator~\cite{Sjostrand:2007gs}, starting from
VINCIA version 1.0.26.

In \secRef{sec:conventions}, we briefly summarize some convenient
notation choices we shall use in the remainder of the paper. In
\secRef{sec:antennae}, we present the sector antenna functions that
have been implemented in VINCIA, including a discussion of their
ambiguous non-singular terms and the choice of sector decomposition
criterion. Some comparisons to
higher-multiplicity tree-level matrix elements are also given, 
to investigate how the quality of the approximation evolves with
parton multiplicity. In \secRef{sec:shower}, we adapt the VINCIA
shower formalism to sector antennae, including trial branchings and GKS
matching. \SecRef{sec:results} contains some basic validation
comparisons, to show that the
implementation gives sensible results. We also present a speed
comparison between various different matching strategies, quantifying
the improvement obtained for the GKS matching in the sector
case. Finally, in \secRef{sec:conclusion}, we round off with
conclusions and an outlook.

\section{Conventions \label{sec:conventions}}

Dipole-antenna showers
\cite{Gustafson:1987rq,Lonnblad:1992tz,Winter:2007ye,Giele:2007di} 
are based on nested $2\to3$ splitting
processes, with an on-shell, Lorentz-invariant phase-space factorization
taking place at each step 
\cite{Kosower:1997zr,Kosower:2003bh}.
Following \cite{GehrmannDeRidder:2005cm,Giele:2011cb}, 
we label the participants  in a $2\to3$
dipole-antenna branching 
by $I K \to i j k$. By energy-momentum conservation we have
$s_{ijk}=s_{IK}=(p_I+p_K)^2\equiv s$.  
We denote the dimensionless (scaled) 
post-branching invariants by
\begin{eqnarray}
y_{ij} = \frac{s_{ij}}{s} = \frac{2 p_i\cdot p_j}{s}&\hspace*{3mm} &
y_{jk} = \frac{s_{jk}}{s} = \frac{2p_j\cdot p_k}{s}~,
\end{eqnarray}
and we define the \pT of a $2\to3$ splitting process in the same way as
in \textsc{Ariadne} \cite{Lonnblad:1992tz}, 
\begin{equation}
\pT^{2} \ = \ \frac{s_{ij}s_{jk}}{s} \ = \ y_{ij} \, y_{jk} \, s~.
\end{equation}
\begin{figure}[t]
\centering
\subfloat{\includegraphics*[scale=0.56]{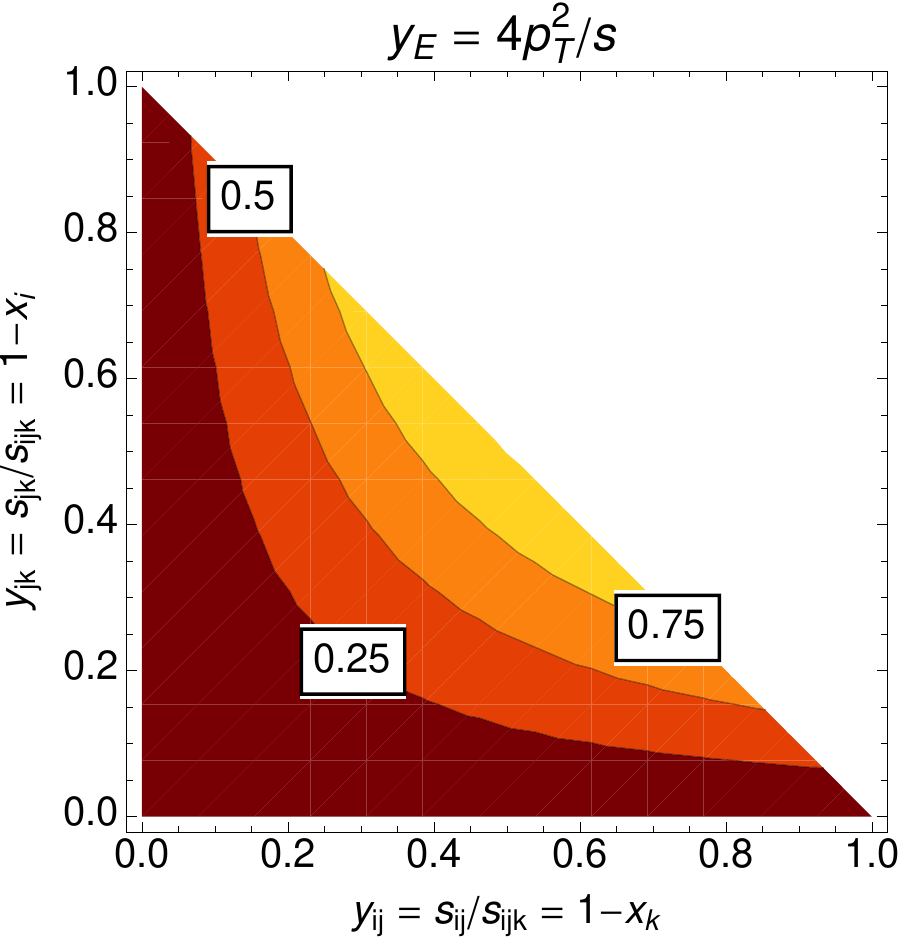}}\hspace*{8mm}
\subfloat{\includegraphics*[scale=0.56]{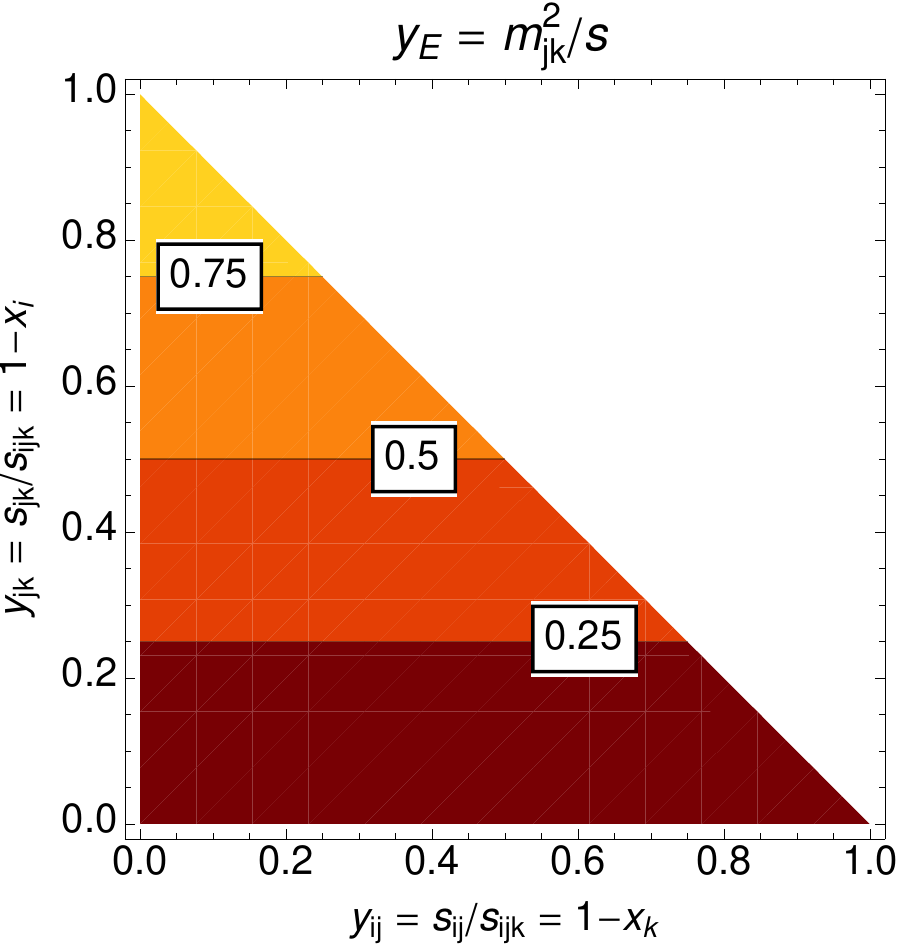}}
\caption{Contours of constant value of the variables $4p_\perp^2/s$
  and $m_{jk}^2/s$ over the $IK\to ijk$ phase-space triangle, with
  $y_{ij}=s_{ij}/s$ and $y_{jk}=s_{jk}/s$ on the $x$ and $y$
  axes, respectively. \label{fig:variables}}
\end{figure}
Contours of constant value of $4p_\perp^2/s$ (normalized so that its
maximal value is unity) are illustrated in the left-hand pane of
\figRef{fig:variables}. For comparison, contours of constant
$m_{jk}^2$, which we shall use for processes involving $g\to q\bar{q}$
splittings below, are illustrated in the right-hand pane of the
figure.  

For the $2\to 3$ antenna functions, we shall use the following general
notation, which is intended to be analogous to that used for parton
distribution functions (PDFs), for the emission of a parton of type
\emph{j} from a parent dipole of type \emph{IK}, 
\begin{equation} 
a^{\mrm{type}(\mrm{order})}_{j/IK}(p_i,p_j,p_k)
\end{equation}
where ``type(order)'' specifies the type (sector or global) and loop
order of the function, and the arguments $p_i$, $p_j$, and $p_k$ represent
the final-state momenta of the corresponding color-ordered post-branching
particles. As is the case for PDFs, one or more of the super-
and subscripts can be omitted when they are obvious from context. This 
results in a fully general but nevertheless quite compact notation, 
which we summarize in \tabRef{tab:notationAnt}, including a comparison
to the notation used for global antennae in
\cite{GehrmannDeRidder:2005cm}. 
\begin{table}[t]
\begin{center}
\[
\begin{array}{ccc|cc|c}\hline\hline
\mbox{Branching} & & \mbox{Compact} & \mbox{Global} &  \mbox{GGG \cite{GehrmannDeRidder:2005cm}}
 & \mbox{Sector}\\
\mbox{type} & & \mbox{form} &  \mbox{form}&  \mbox{notation} & \mbox{form}\\[2mm]
q\bar{q} \to qg\bar{q} & : & 
a_{g/q\bar{q}} &  a_{g/q\bar{q}}^{\mrm{gl}(0)} & a_3^0 &a_{g/q\bar{q}}^{\mrm{sct}(0)}  \\[2mm] 
qg \to qgg & : & 
a_{g/qg} &  a_{g/qg}^{\mrm{gl}(0)} & d_3^0 &  a_{g/qg}^{\mrm{sct}(0)} \\[2mm]
qg \to q\bar{q}'q' & : & 
a_{\bar{q}'/qg} &  a_{\bar{q}'/qg}^{\mrm{gl}(0)} & \frac12E_3^0 &
a_{\bar{q}'/qg}^{\mrm{sct}(0)} \\[2mm]
gg \to ggg & : & 
a_{g/gg} &  a_{g/gg}^{\mrm{gl}(0)} & f_3^0  &  a_{g/gg}^{\mrm{sct}(0)}\\[2mm]
gg \to g\bar{q}q & : & 
a_{\bar{q}/gg} &  a_{\bar{q}/gg}^{\mrm{gl}(0)} & \frac12G_3^0 &
a_{\bar{q}/gg}^{\mrm{sct}(0)} \\
\hline\hline
\end{array}
\]
\caption{Notation for antenna functions, including comparisons to 
  the notation used  in 
  \cite{GehrmannDeRidder:2005cm}. \label{tab:notationAnt}} 
\end{center}
\end{table}
Part of the motivation for introducing this notation, apart from the
desire to be able to distinguish clearly between sector and global
functions when necessary, is that it generalizes easily to include more branching
types, such as ones involving photons in QED, 
and to higher-order antenna functions, such as $2\to 4$ ones. 
We also note that, e.g., $a_{g/qg}=a_{g/g\qbar}$, 
by charge conjugation, with the appropriate permutation of 
invariants ($s_{qg} \leftrightarrow s_{g\qbar}$). 

The antenna functions have dimension GeV$^{-2}$. It is often convenient to work with a 
color- and coupling-stripped variant, which we label $\bar{a}$,
defined by 
\begin{equation} 
a_{j/IK} = 
g_s^2 {\cal C}_{j/IK}\  \bar{a}_{j/IK}
\end{equation}
where $g_s^2=4\pi\alpha_s$ and ${\cal C}_{j/IK}$ is the color factor
assigned to the $IK \to 
ijk$ branching, defined in the normalization convention of 
\cite{Giele:2011cb},  
such that, in the leading-color limit, ${\cal C}_{g} \to N_C$ and ${\cal
  C}_q \to 1$.

When comparing to collinear (Altarelli-Parisi) splitting functions
\cite{Altarelli:1977zs}, we define 
$z$ as the momentum fraction of the radiated parton in the collinear
limit:
\begin{eqnarray}
z &\equiv& \frac{ E_j}{ E_j+E_k}=\frac{ E_j}{ E_K},~ \mrm{for} ~p_j \| p_k \\
& \equiv & \frac{ E_j}{ E_i+E_j}=\frac{ E_j}{ E_I},~ \mrm{for} ~p_i \| p_j ~,
\end{eqnarray}  
The AP splitting functions for $g\to gg$ and $g \to q\qbar$ are then
\begin{eqnarray}
P_{gg\rightarrow G}(z) & =& 2\left[\frac{z}{(1-z)}+\frac{(1-z)}{z}+z(1-z)\right] ~, \\
P_{q\bar{q} \rightarrow G}(z) & =& \left[z^{2}+(1-z)^{2}\right]~,
\end{eqnarray}
which we shall use when analyzing the collinear singular limits
of the corresponding global and sector antenna functions below. 

\section{The Sector Antennae \label{sec:antennae}}

\begin{figure}[t]
\centering
\includegraphics*[scale=0.46]{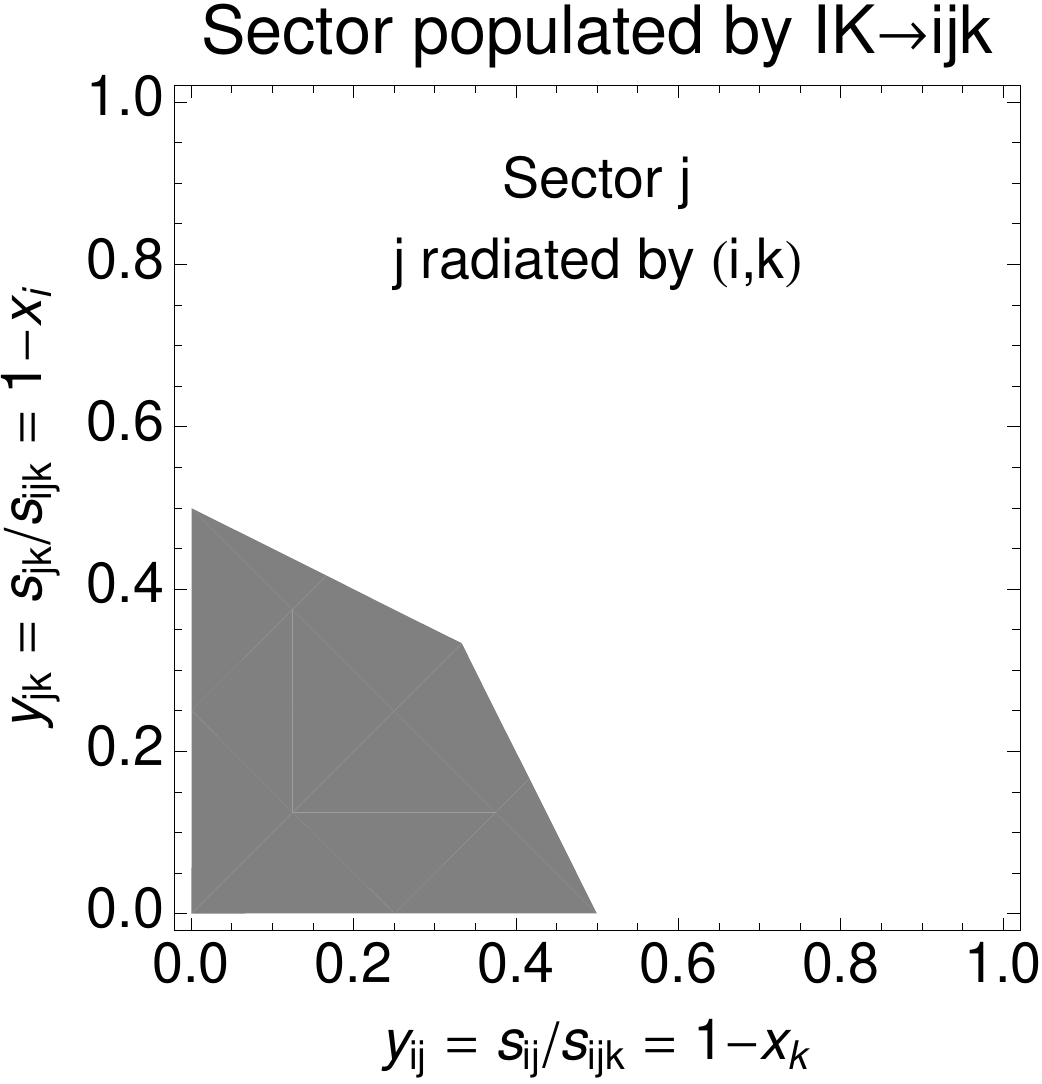} \ 
\includegraphics*[scale=0.46]{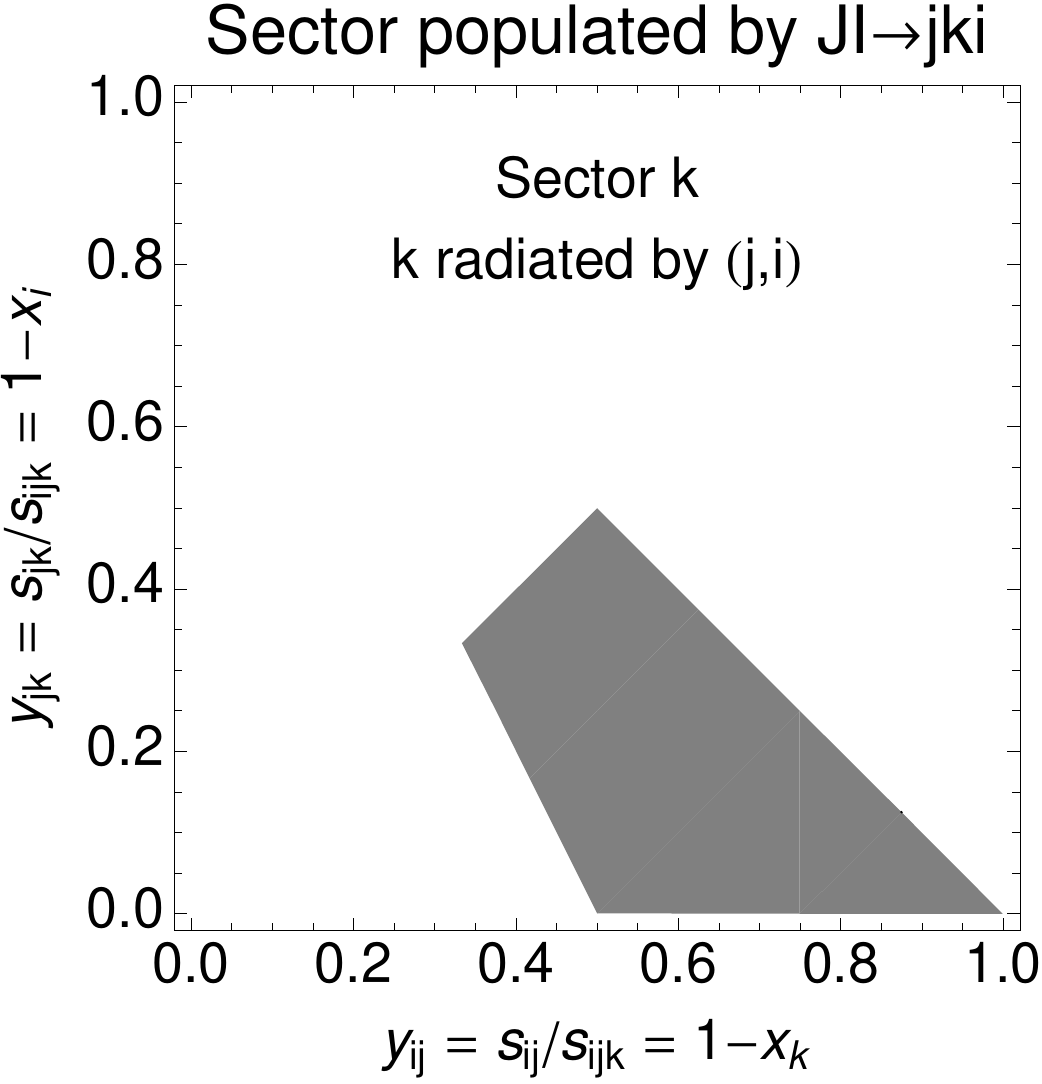} \ 
\includegraphics*[scale=0.46]{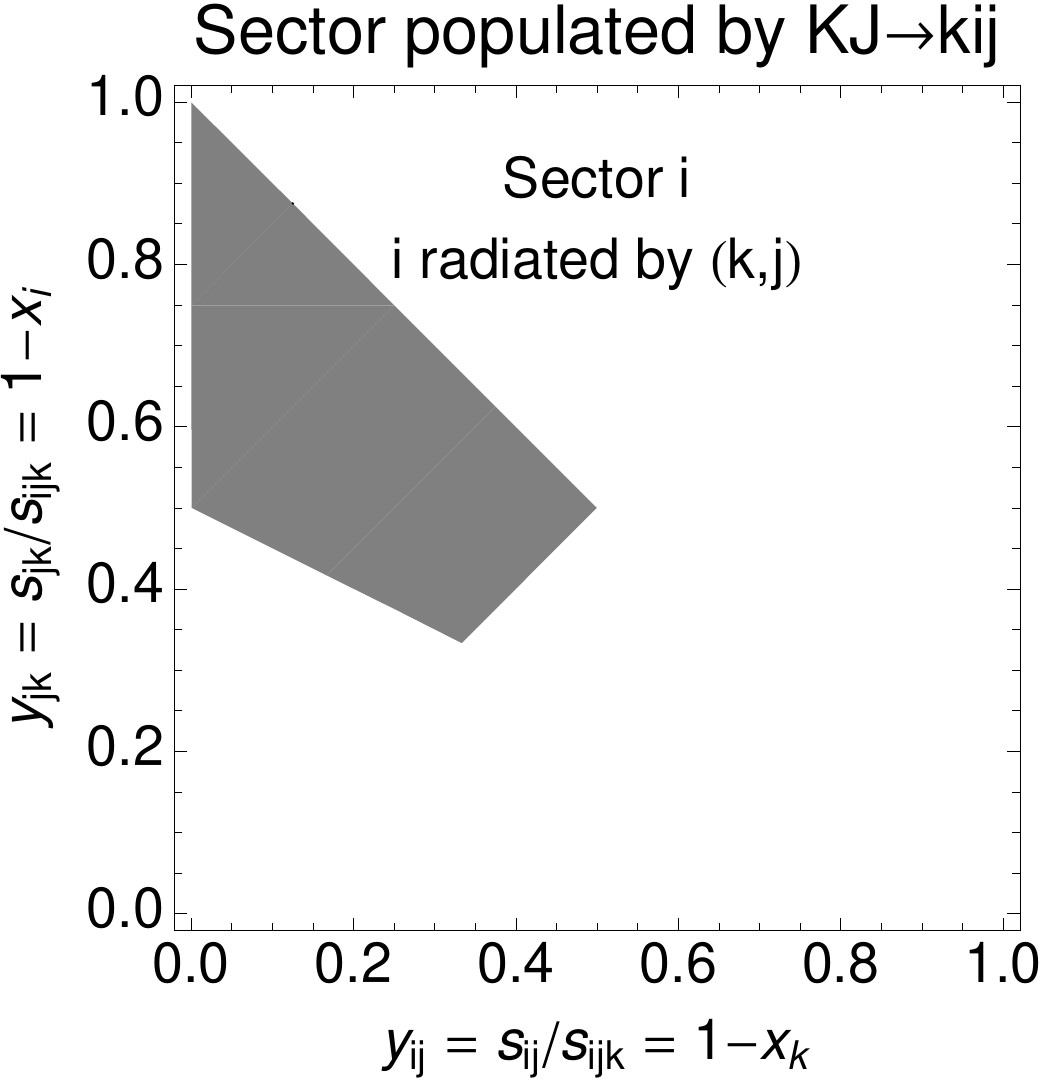}
\caption{Illustration of the three phase-space sectors in a color-singlet $g_ig_jg_k$
  configuration, using transverse momentum to discriminate between
  sectors \cite{Giele:2011cb}. \label{fig:sectors}} 
\end{figure}

In the sector approach, only one antenna contributes to any given
phase-space point, as opposed to several overlapping ones in the global
antenna case. The three different phase-space sectors that occur for 
$gg\to ggg$  (with cyclic color connections, as in $H\to gg$) are
illustrated in \figRef{fig:sectors}~\cite{Giele:2011cb}.
In a global shower, the $IK\to ijk$ antenna, shown in the left-hand
pane, would be allowed to fill the entire branching phase-space,
which is defined by the triangle $y_{ij}+y_{jk} \le 1$. In the sector
case, however, it only fills the part of phase-space in which the
transverse momentum of 
$j$ with respect to $i$ and $k$ is smaller than that of either of the
two other possible combinations (assuming  transverse
momentum is what is used to separate the sectors, a point we return to
below). The remaining part of phase-space is not empty --- it is
filled by the two complementary permutations of the $i$, $j$, and $k$
partons, as shown in the middle and right-hand panes of the figure. 
The coefficients of the 
singular terms of the antenna functions must necessarily reflect this
reorganization. The double pole, located at the origin of the plots in
\figRef{fig:sectors}, is contained entirely within the $IK
\to ijk$ antenna, and can therefore be carried over 
from the global case without modification. The
single-pole terms, however, change to account for collinear radiation now being
produced by a single antenna rather than two overlapping ones. 

In  \secRef{sec:poles}, we discuss how the singularity structure of the individual
antennae is modified and derive a complete set of sector antenna
functions. In \secRef{sec:tests}, we compare these functions to
fixed-order matrix elements for $Z\to 4$, $5$, and $6$ partons. 
In \secRef{sec:finite}, we discuss 
the ambiguities remaining concerning non-singular (and non-universal)
terms. Finally, in \secRef{sec:decomposition}, we compare 
various options for how to partition phase-space into sectors.

\subsection{Singularity Structure \label{sec:poles}}

In the so-called ``planar'' (leading-color) limit, which is used to represent
color flow in parton-shower event generators, gluons are viewed as
composed of a triplet and an antitriplet color charge, which are part
of two separate color dipoles. For instance, in a $qg\bar{q}$
configuration, there will be one color dipole stretched between the
$qg$ pair and one stretched between the $g\bar{q}$ pair. The full
collinear singularity of the gluon is obtained by summing over the
two. In the global antenna approach, radiation from both pairs is
allowed to contribute over all of phase-space. In the sector
approach, \emph{either} the $qg$ pair \emph{or} the $g\bar{q}$ one
contributes to each $qgg\bar{q}$ phase-space point. In order for the
two approaches to reproduce the same collinear limit, the sector
antennae must include those collinear
terms that would be generated by their neighbors in the global case.   

As our starting point, we take  
the GGG global antennae \cite{GehrmannDeRidder:2005cm}. The
$q\qbar\to qg\qbar$ antenna is the same for global and sector decompositions, since
there are no neighboring antennae in this case. 
In the terminology of our conventions, 
\begin{equation}
a_{g/q\bar{q}}^{\mrm{sct}}=a_{g/q\bar{q}}^{\mrm{gl}}.
\end{equation}
In the $qg\to qgg$ (or $g\qbar \to gg\qbar$) case,
there is the collinear limit on the edge of the parent gluon to be
dealt with. In this limit there is a mapping $\ensuremath{z\to1-z}$
between the antenna and its neighboring antenna.
A single global antenna thus compares to the full $g\to gg$ splitting
function in the collinear limit as follows \cite{GehrmannDeRidder:2005cm},
\begin{equation}
\bar{a}_{g/qg}^{\mrm{gl}}(p_i,p_j,p_k)\stackrel{s_{jk}\to
  0}{\longrightarrow}\frac{1}{s_{jk}}\left(P_{gg\rightarrow
  G}(z)-\frac{2z}{1-z}-z(1-z)  \right) \ + \ {\cal O}(1),
\end{equation}
where the ${\cal O}(1)$ ambiguity due to non-singular terms is
unimportant for the limiting behavior. For a corresponding sector antenna, we
would want to reproduce the full splitting function in this limit,
i.e., just the first term in the equation above.
We ensure this by simply adding back the ``missing'' singular 
pieces to $a_{g/qg}^{\mrm{gl}}$. In the collinear limit, for massless
particles $p_j\|p_k$, we have 
\begin{equation}
 y_{ij} \equiv \frac{s_{ij}}{s} 
\rightarrow \frac{p_i \cdot  p_j}{p_i \cdot (p_j+p_k)} = \frac{ E_j}{
  E_j+E_k} \equiv z~.
\end{equation}
Thus, we obtain for $a_{g/qg}^{\mrm{sct}}$, 
\begin{equation}
\bar{a}_{g/qg}^{\mrm{sct}}(p_i,p_j,p_k)\equiv
\bar{a}_{g/qg}^{\mrm{gl}}(p_i,p_j,p_k)+\frac{1}{s}\frac{1}{y_{jk}}\left(\frac{2y_{ij}}{1-y_{ij}}+y_{ij}(1-y_{ij})\right).
\end{equation}

The $gg\to ggg$ antenna is analogous to the previous
one, simply considering both edges instead of just one. We define the
sector antenna as
\begin{equation}
\bar{a}_{g/gg}^{\mrm{sct}}\equiv \bar{a}_{g/gg}^{\mrm{gl}}+\frac{1}{s}\left[\frac{1}{y_{jk}}\left(\frac{2y_{ij}}{1-y_{ij}}+y_{ij}(1-y_{ij})\right)+\frac{1}{y_{ij}}\left(\frac{2y_{jk}}{1-y_{jk}}+y_{jk}(1-y_{jk})\right)\right].
\end{equation}
With a small amount of algebra, we arrive at the following generic
sector generalization of global gluon emission antennae,
\begin{eqnarray}
\bar{a}_g^{\mrm{sct}}(p_i,p_j,p_k) & \equiv &
\bar{a}_{g}^{\mrm{gl}}(p_i,p_j,p_k)
+ \frac{1}{s_{ijk}}
\Bigg[ \delta_{Ig}\left(\frac{2}{y_{ij}(1-y_{jk})} - \frac{2}{y_{ij}} +
  (1-y_{jk})\frac{y_{jk}}{y_{ij}} \right) \nonumber\\[1mm] &
  & \hspace*{3cm} \ + \
  \delta_{Kg}\left(\frac{2}{y_{jk}(1-y_{ij})} - \frac{2}{y_{jk}}  + (1-y_{ij})\frac{y_{ij}}{y_{jk}}\right) \Bigg]
~,\label{sectordecomposition}
\end{eqnarray}
 with 
$\delta_{Ig} = 1$ ($\delta_{Kg}=1$) if parton $I$ ($K$) 
is a gluon and 0 otherwise. We note that, for the results reported on
later in this paper, we set the finite terms in the global-antenna
parts, $\bar{a}_g^{\mrm{gl}}$, to zero, in the
parametrization of \cite{Ridder:2011dm}.

For the antennae that involve splitting of a gluon into quarks, the
only divergence is the one associated with the collinear limit in
which the quark-antiquark pair become collinear (partons denoted as
$j$ and $k$). Moreover, this limit is represented by the splitting
function $P_{q\qbar\rightarrow G}(z)$. One can then take the following
definition for these sector antennae, which has the correct limit:

\begin{equation}
\bar{a}_{\bar{q}'/qg}^{\mrm{sct}}(p_i,p_j,p_k)=\bar{a}_{\bar{q}'/gg}^{\mrm{sct}}(p_i,p_j,p_k)\equiv \frac{1}{s}\left[\frac{y_{ij}^{2}+y_{ik}^{2}}{y_{jk}}\right].
\end{equation}

The corresponding global antennae are identical to these, modulo a
factor $1/2$ due to the fact that two neighboring antennae add up to the
same limit.

With this notation for the coefficients, our sector antennae
can be expressed as in tab.~\ref{tab:antenna_coefficients} (under ``VS''), 
where we also compare to three global antenna sets, including the
default \Vc\ ones \cite{Ridder:2011dm}, 
the ones used by Gehrman-Gehrman-Glover (GGG)
\cite{GehrmannDeRidder:2005cm}, 
and the set used by \tsc{Ariadne} \cite{Gustafson:1987rq,Lonnblad:1992tz}. 
We also note that the singular coefficients given here agree
with those obtained for polarized sector antennae in \cite[Tab.~1]{Larkoski:2009ah}, 
when the latter are summed over polarizations.

\begin{table}[t]
\begin{center}{
\begin{tabular}{lr|rrrrrr|rr|rrr}
\toprule 
$\times$ & \hspace*{-2mm}$\frac{1}{y_{ij}y_{jk}}$
& $\frac{1}{y_{ij}}$ & $\frac{1}{y_{jk}}$ & $\frac{y_{jk}}{y_{ij}}$ &
$\frac{y_{ij}}{y_{jk}}$
& $\frac{y_{jk}^2}{y_{ij}}$ & $\frac{y_{ij}^2}{y_{jk}}$ & $\frac{1}{y_{jk}(1-y_{ij})}$&
$\frac{1}{y_{ij}(1-y_{jk})}$ & $1$ & $y_{ij}$ & $y_{jk}$\\[2mm]
\hline
\multicolumn{4}{l}{\textsc{VS}  (sector)}\\
$q\bar{q}\to qg\bar{q}$ & 2 & -2 & -2 & 1 & 1 & 0 & 0 &0&0& 0 & 0 & 0\\
$qg\to qgg$ & 2 & -2 & -4 & 1 & 2 &0 & -2 &2&0& 0 & 0 & 0\\
$gg\to ggg$ & 2 & -4 & -4 & 2 & 2 &-2 & -2 &2&2& 0 & 0 & 0\\
$qg\to q\bar{q}'q'$& 0 & 0 & 1 & 0 & -2 & 0 & 2 & 0 & 0 & -2 & 2& 1 \\
$gg\to g\bar{q} q$& 0 & 0 & 1 & 0 & -2 & 0 & 2 & 0 & 0 & -2 & 2 & 1\\
\hline
\multicolumn{5}{l}{GRS (global; default in \textsc{Vincia})}\\
$q\bar{q}\to qg\bar{q}$ & 2 & -2 & -2 & 1 & 1 & 0 & 0 & 0 & 0 & 0 & 0 & 0\\
$qg\to qgg$& 2 & -2 & -2 & 1 & 1 & 0 & -1 & 0 & 0 & $2$ & $-1$ & $0$\\
$gg\to ggg$& 2 & -2 & -2 & 1 & 1 &-1 & -1 & 0 & 0 & $2$ & 0 & 0 \\
$qg\to q\bar{q}'q'$& 0 & 0 & $\frac12$ & 0 & -1 & 0 & 1 & 0 & 0 & $-0.7$ & 1& $\frac12$ \\
$gg\to g\bar{q} q$& 0 & 0 & $\frac12$ & 0 & -1 & 0 & 1 & 0 & 0 & $-0.7$ & 1 & $\frac12$\\
\hline
\multicolumn{4}{l}{\textsc{GGG} (global)}\\
$q\bar{q}\to qg\bar{q}$ & 2 & -2 & -2 & 1 & 1 & 0 & 0 & 0 & 0 & 0 & 0 & 0\\
$qg\to qgg$& 2 & -2 & -2 & 1 & 1 & 0 & -1 & 0 & 0 & $\frac52$ & -1 & -$\frac12$\\
$gg\to ggg$& 2 & -2 & -2 & 1 & 1 &-1 & -1 & 0 & 0 & $\frac83$ & -1 & -1 \\
$qg\to q\bar{q}'q'$& 0 & 0 & $\frac12$ & 0 & -1 & 0 & 1 & 0 & 0 & -$\frac12$ & 1& 0 \\
$gg\to g\bar{q} q$& 0 & 0 & $\frac12$ & 0 & -1 & 0 & 1 & 0 & 0 & -1 & 1 & $\frac12$\\
\hline
\multicolumn{4}{l}{\textsc{Ariadne} (global)}\\
$q\bar{q}\to qg\bar{q}$ & 2 & -2 & -2 & 1 & 1 & 0 & 0 & 0 & 0 & 0 & 0 & 0\\
$qg\to qgg$& 2 & -2 & -3 & 1 & 3 & 0 & -1 & 0 & 0 & 0 & 0 & 0\\
$gg\to ggg$& 2 & -3 & -3 & 3 & 3 &-1 & -1 & 0 & 0 & 0 & 0 & 0\\
$qg\to q\bar{q}'q'$& 0 & 0 & $\frac12$ & 0 & -1 & 0 & 1 & 0 & 0 & -1 & 1  & $\frac12$\\
$gg\to g\bar{q} q$& 0 & 0 & $\frac12$ & 0 & -1 & 0 & 1 & 0 & 0 & -1 & 1  & $\frac12$\\
\bottomrule
\end{tabular}}
\caption{Table of coefficients for sector (VS) and global
  (GRS \cite{Ridder:2011dm}, GGG \cite{GehrmannDeRidder:2005cm}, \Ar
  \cite{Lonnblad:1992tz}) antenna functions.  
\label{tab:antenna_coefficients}}
\end{center}
\end{table}

\subsection{Comparison to tree-level matrix elements \label{sec:tests}}
In order to examine the quality of the approximation furnished 
by a shower based on the antennae derived in the previous subsection, 
independently of the shower code itself, we follow the approach used 
for global antennae in \cite{Skands:2009tb,Giele:2011cb,massive}. That is, we use 
\textsc{Rambo} \cite{Kleiss:1985gy} (an implementation of which
has been included in \textsc{Vincia}) to generate a large number
of evenly distributed 4-, 5-, and 6-parton phase-space points. For
each phase-space point, we use \textsc{MadGraph}~\cite{Murayama:1992gi,Alwall:2007st}
to evaluate the leading-color $Z\to n$ matrix
element squared (suitably modified to be able 
to switch subleading color terms on and off). 
We then compute the corresponding antenna-shower 
approximation, expanded to tree level, in the same phase-space point,
in the following way: using a clustering algorithm that contains the
exact inverse of the default 
\textsc{Vincia} $2\to3$ kinematics map \cite{Giele:2007di}, we
perform $m$ clusterings of the type $(i,j,k)\to(I,K)$ in a way
that exactly reconstructs the intermediate $(n-m)$-parton configurations
that would have been part of the shower history for each $n$-parton
test configuration. Summing over all possible such clusterings (in the
global case), we may compute the nested products of $2\to 3$ antenna
functions that produce the tree-level shower approximation. 
Finally, we form the ratio between this approximation and the LO
matrix element, as a measure of the amount of over- or under-counting
by the shower, with values greater than unity
corresponding to over-counting and vice versa. 

The sector approach is characterized by the existence of only one
possible path from a given final parton configuration back to any
previous step in the 
shower, resulting in an unequivocal clustering sequence, which in turn
produces a single nested product of antennae. To define which sector 
is clustered in each step, one must choose a partitioning variable. Our
default sector decomposition prescription (studied in more
detail in \secRef{sec:decomposition}) is based on the variable  
\begin{equation}
Q_{\mrm{s}_j}^{2}\equiv\left\{ \begin{array}{l}
p_{\perp j}^2 \ =\ \frac{s_{ij}s_{jk}}{s}\mbox{ for }j\mbox{ a gluon\,\,}\\[2mm]
\tilde{s}_{jk} \ =\ s_{jk}\frac{\sqrt{s_{ij}}}{2
  \sqrt{s}}\mbox{ for }(j,\, k)\mbox{ a quark-antiquark pair}\\[2mm]
\tilde{s}_{ij} = s_{ij}\frac{\sqrt{s_{jk}}}{2 \sqrt{s}}\mbox{ for
}(i,\, j)\mbox{ a quark-antiquark pair,}\end{array}\right. \label{eq:QSvariable}
\end{equation}
which is calculated for each set of three color-connected partons in the
configuration (treating same-flavor $\bar{q}q$ combinations as being
color-connected for this purpose). The three-parton cluster with the
smallest value of $Q_\mrm{s}^{2}$ gets clustered. The aforementioned
shower-to-matrix-element ratio, for the reaction $Z\to q_1 g_2 g_3
\qbar_4$, is then
\begin{equation} 
R_{4}^\mrm{sct}=\left\{ \begin{array}{ll}
\displaystyle\frac{\left( a_{g/qg}^{\mrm{sct}}(1,2,3)~
  a_{g/q\qbar}^{\mrm{sct}}(\widehat{12},\widehat{23},4)\right)|M_{2}(E_{\mrm{cm}}^2)|^{2}}{|M_{4}(1,2,3,4)|^{2}}
& \mbox{, if }p_{\perp 2}^{2}<p_{\perp 3}^{2}\\[5mm]
\displaystyle\frac{\left( a_{g/g\qbar}^{\mrm{sct}}(2,3,4) ~ a_{g/q\qbar}^{\mrm{sct}}(1,\widehat{23},\widehat{34})\right)|M_{2}(E_\mrm{cm}^2)|^{2}}{|M_{4}(1,2,3,4)|^{2}} & \mbox{, otherwise}\end{array}\right.
 \label{eq:R4Esector} 
\end{equation}
where hatted variables $\widehat{\i{}\j}$ denote clustered momenta,
$|M_n|^2$ denote the color-ordered $n$-parton matrix elements and
$E_\mrm{cm}=m_Z$ is the total invariant mass of the $n$-parton
system. The numerators of eq.~(\ref{eq:R4Esector}) thus reproduce the
shower approximation expanded to tree level, phase-space point by
phase-space point, for an arbitrary choice of kinematics map,
$(i,j,k)\to (\widehat{\i{}\j},\widehat{\j{}k})$. For compactness, we
do not give the explicit forms of $R_5$ and $R_6$, to which we shall
also compare in the following;  the relevant
generalizations are straightforward. (Note: we do not consider $R_3$, 
since the $q\bar{q}\to qg\bar{q}$ antenna functions can be chosen to
reproduce the $Z\to 3$ LO matrix element exactly.) 

We  compare to three different variants of the global
approach  \cite{Giele:2011cb}: 
unordered, strongly ordered, and smoothly ordered, as follows.

Firstly, we consider an ``unordered'' shower, where all possible
histories/paths are allowed. The $R_4$ ratio above then becomes \cite{Skands:2009tb}
\begin{equation} 
R_{4}^\mrm{gl.unord} =  \frac{|M_2(E_\mrm{cm}^2)|^2 \bigg( a_{g/qg}^{\mrm{gl}}(1,2,3)  a_{g/q\qbar}^{\mrm{gl}}(\widehat{12},\widehat{23},4)  + ~ a_{g/g\qbar}^{\mrm{gl}}(2,3,4)  a_{g/q\qbar}^{\mrm{gl}}(1,\widehat{23},\widehat{34})\bigg)}{|M_4(1,2,3,4)|^2} ~. \label{eq:R4unord} 
\end{equation}

\begin{figure}[t]
\centering
\includegraphics*[scale=0.75]{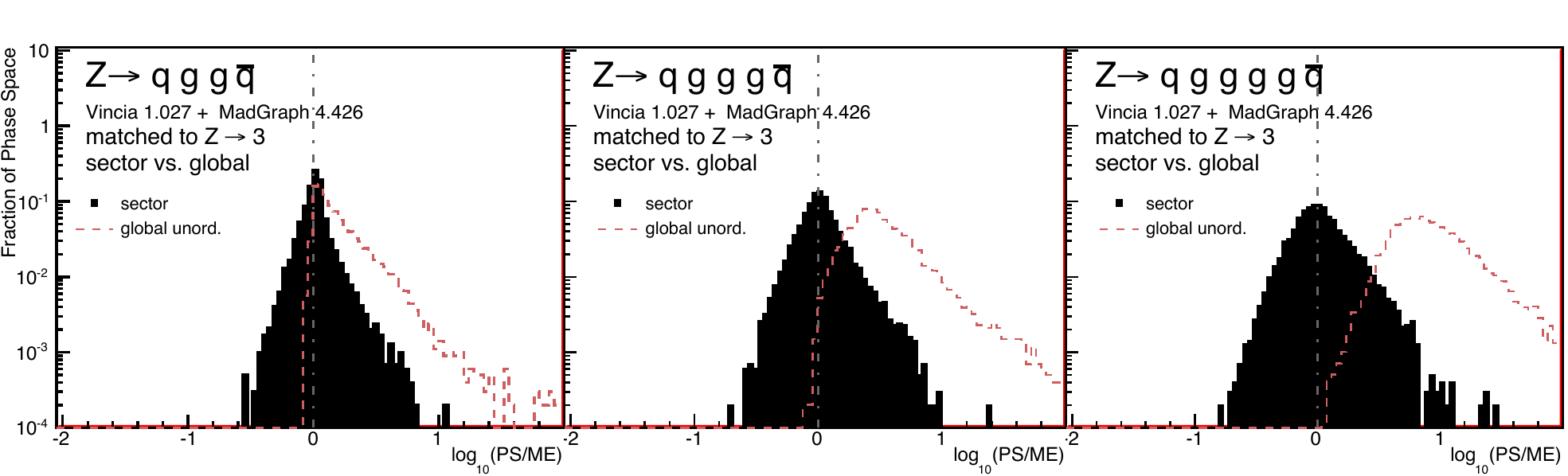}
\caption{Comparison between global unordered and sector shower
  approximations to LO matrix elements  for $Z\to
  q\bar{q}\,+\,$gluons. Distributions of
  $\log_{10}(\mrm{PS}/\mrm{ME})$ in a flat phase-space 
  scan, normalized to unity.}
\label{secvsglo_gluons_pureshower_glUNORDvssector}
\end{figure}
In \figRef{secvsglo_gluons_pureshower_glUNORDvssector}, 
the sector shower approximation, \eqRef{eq:R4Esector}, 
using the ``VS'' sector antennae defined in the previous subsection, is
shown as a filled
solid histogram, for $Z\to q\bar{q}$ + 2 (left), 3 (middle), and 4
(right) gluons. The unordered global approximation,
\eqRef{eq:R4unord}, using the default \Vc~antenna
  functions \cite{Ridder:2011dm}, is shown with dashed lines
The $x$ axis shows the distribution of 
$\log_{10}(R_n)$ obtained in the flat phase-space scan, with the
middle (zero) corresponding to $R_n = \mrm{PS}/\mrm{ME} = 1$. 
One sees that the ``naive'' unordered global approach (dashed
histogram) generates a large tail of overcounting of the matrix
elements, to the right of zero, and that this overcounting grows worse
with parton multiplicity, whereas the sector antennae produce a more evenly
distributed ratio, whose central value is fairly stable as the number
of partons increases, while only its width grows (reflecting the added
uncertainty coming from having several branchings in a row).  

For the global showers, it is essentially the overcounting illustrated
by the dashed histograms in
\figRef{secvsglo_gluons_pureshower_glUNORDvssector} 
that makes it mandatory to impose an ordering condition in the shower
(beyond that of energy-momentum conservation, which is already present
in the nested antenna phase-spaces),
to obtain a reasonable average approximation.  In \Vc, two types of
ordering of the global shower algorithm are possible, called strong and
smooth, see
\cite{Giele:2011cb,lhpimp} for details. With strong ordering in
\pT, for instance, the \pT of each consecutive radiation has to be
strictly smaller than 
that of the previous one. Therefore not all histories or paths are
allowed; there even exist points in phase-space for which \emph{no}
path can possibly contribute, called ``dead zones''. Again, for
the reaction $Z\to q_1 g_2 g_3 \qbar_4$, we have  
\begin{eqnarray} 
R_{4}^\mrm{gl.ord} & = & \frac{|M_2(E_\mrm{cm}^2)|^2}{|M_4(1,2,3,4)|^2}
\bigg(\Theta({p}_{\perp \widehat{23}}-p_{\perp 2})  a_{g/qg}^{\mrm{gl}}(1,2,3)
a_{g/q\qbar}^{\mrm{gl}}(\widehat{12},\widehat{23},4) \nonumber \\ &
& \hspace*{2.5cm} + ~\Theta(p'_{\perp \widehat{23}}-p_{\perp 3})
a_{g/g\qbar}^{\mrm{gl}}(2,3,4)
a_{g/q\qbar}^{\mrm{gl}}(1,\widehat{23},\widehat{34})\bigg)~, \label{eq:R4E} 
\end{eqnarray} 
where the ordering conditions depend on 
\begin{equation} \begin{array}{rclcrcl} p_{\perp2} & = & \pT(1,2,3)
    & ; & p_{\perp \widehat{23}} & = &
    \pT(\widehat{12},\widehat{23},4)\\[2mm] p_{\perp 3} & = & \pT(2,3,4) &
    ; & p'_{\perp\widehat{23}} & = &
    \pT(1,\widehat{23},\widehat{34}) \end{array}~.\label{eq:4pAB} 
\end{equation} 
Smooth ordering 
basically replaces the strong-ordering $\Theta$ 
functions above by a smooth suppression factor that goes to unity in the
strongly ordered soft/collinear limits and to zero for highly
``unordered'' branchings, see \cite{Giele:2011cb,massive,lhpimp} for further
details. In this case, there are no strict dead zones; unordered
branchings are merely suppressed, not forbidden. 

\begin{figure}[t]
\centering
\includegraphics*[scale=0.75]{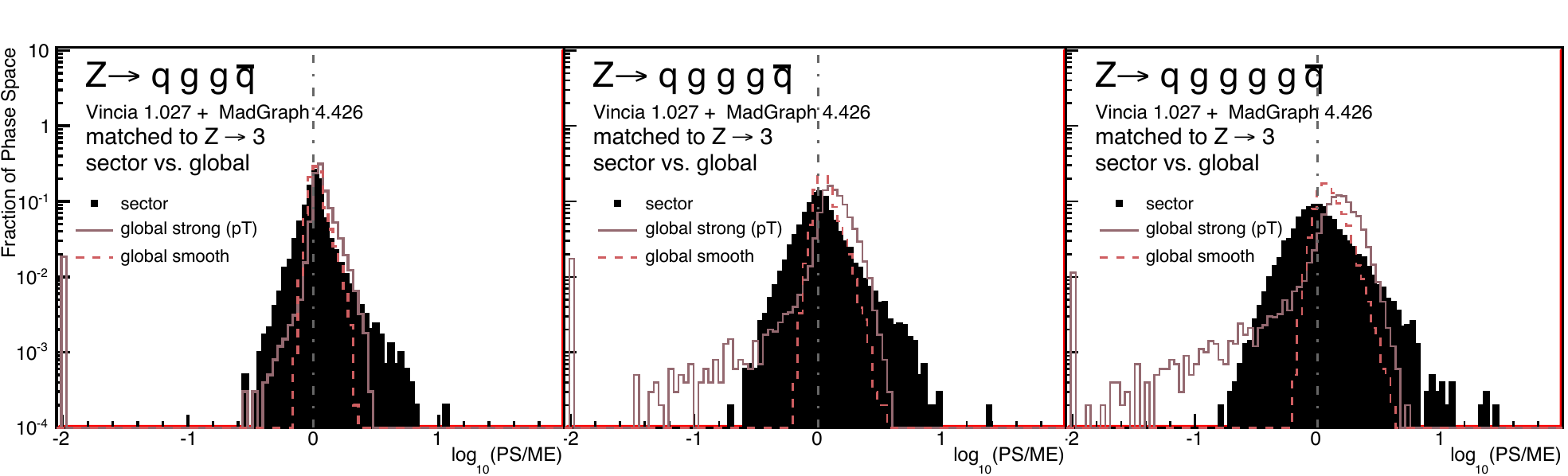}
\caption{Comparison between global strong, global smooth and sector shower approximations to LO matrix elements, for $Z\to
  q\bar{q}\,+\,$gluons. Distributions of
  $\log_{10}(\mrm{PS}/\mrm{ME})$ in a flat phase-space 
  scan.
  Spikes on the far left represent the underflow bin ---
  dead zones in the shower approximations.} 
\label{secvsglo_gluons_pureshower}
\end{figure}
In \figRef{secvsglo_gluons_pureshower}, we show the same sector
approximation as above, while the global approximation has been
replaced by strong (solid lines) and smooth (dashed lines) ordering in \pT,
respectively. Here, we see that the ordered global showers also
generate peaks that extend roughly symmetrically around $\log(R)=0$,
although the strong-ordering condition does produce a tail of large 
undercounting at higher multiplicities. We also see that the
distributions generated by the global showers are somewhat narrower,
indicating a slightly better average agreement, than those of their
sector counterparts. For the strongly ordered case, this comes at the
price of a dead zone, of course, illustrated at the far left-hand edge
of each panes.

\begin{figure}[t]
\centering
\includegraphics*[scale=0.75]{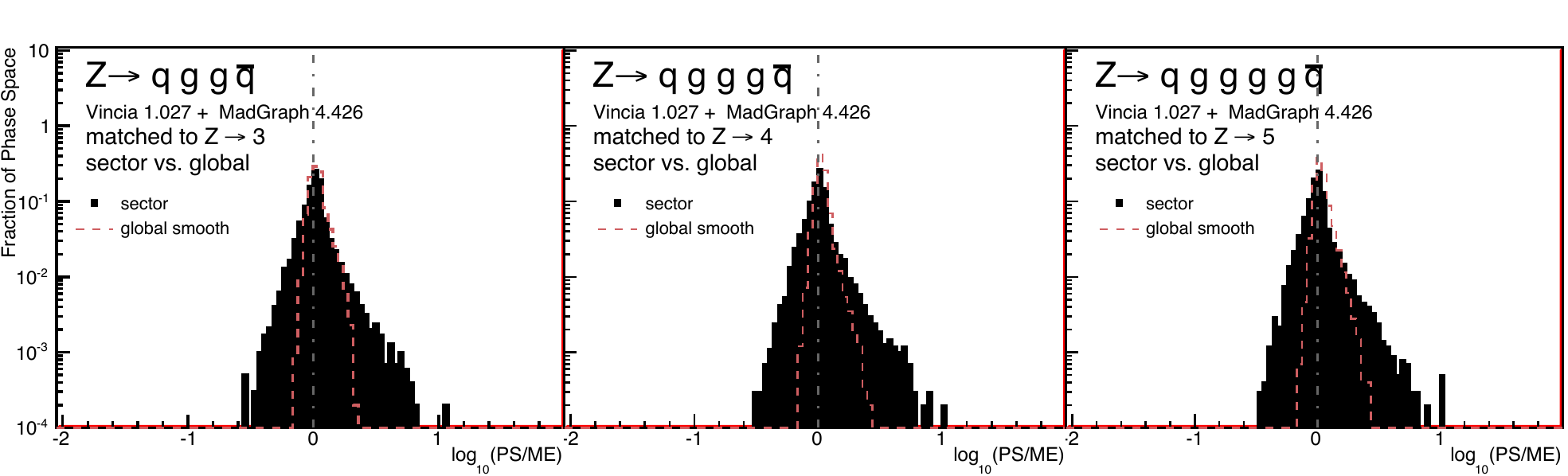}
\caption{Comparison between global smooth and sector shower matched
  approximations to LO matrix elements, for $Z\to
  q\bar{q}\,+\,$gluons. Distributions of
  $\log_{10}(\mrm{PS}/\mrm{ME})$ in a flat phase-space 
  scan.
  }
\label{secvsglo_gluons_matched}
\end{figure}
The trend that the smoothly ordered global shower gives a somewhat
narrower distribution remains when we include matching 
through to the $(n-1)$-parton LO matrix element at each step (see section
\ref{2ton_matching}). This is illustrated in
\figRef{secvsglo_gluons_matched}, in which each pane thus 
only reflects the last  branching step, 
rather than the whole shower history. 
Note that, since GKS matching has only 
been developed for the smoothly ordered shower,  strong
ordering is not shown in this figure. Despite the slightly wider
tails, we nonetheless conclude that the
sector shower furnishes an acceptable overall approximation, without any 
dead or substantially under- or overcounted tails.

Secondly, we look at processes for which the
gluon-splitting antennae $qg \to q\qbar q$ and $g\qbar \to \qbar q
\qbar$ contribute. 
Specifically, we compare to the leading-color matrix elements squared
for 
$Z \to q\qbar q \qbar$ and $Z \to q g \qbar q \qbar$, with the other
color-ordering, $Z \to q\qbar q g \qbar$, identical by charge conjugation. 
Since the leading singular structure of 
gluon-splitting antennae is less pronounced (a single pole, as compared to the double
pole for gluon emission), mismatches at the subleading level
become relatively more important. We therefore expect an overall worse
agreement with the matrix elements than in the gluon-emission case.

\begin{figure}[t]
\centering
\includegraphics*[scale=0.5]{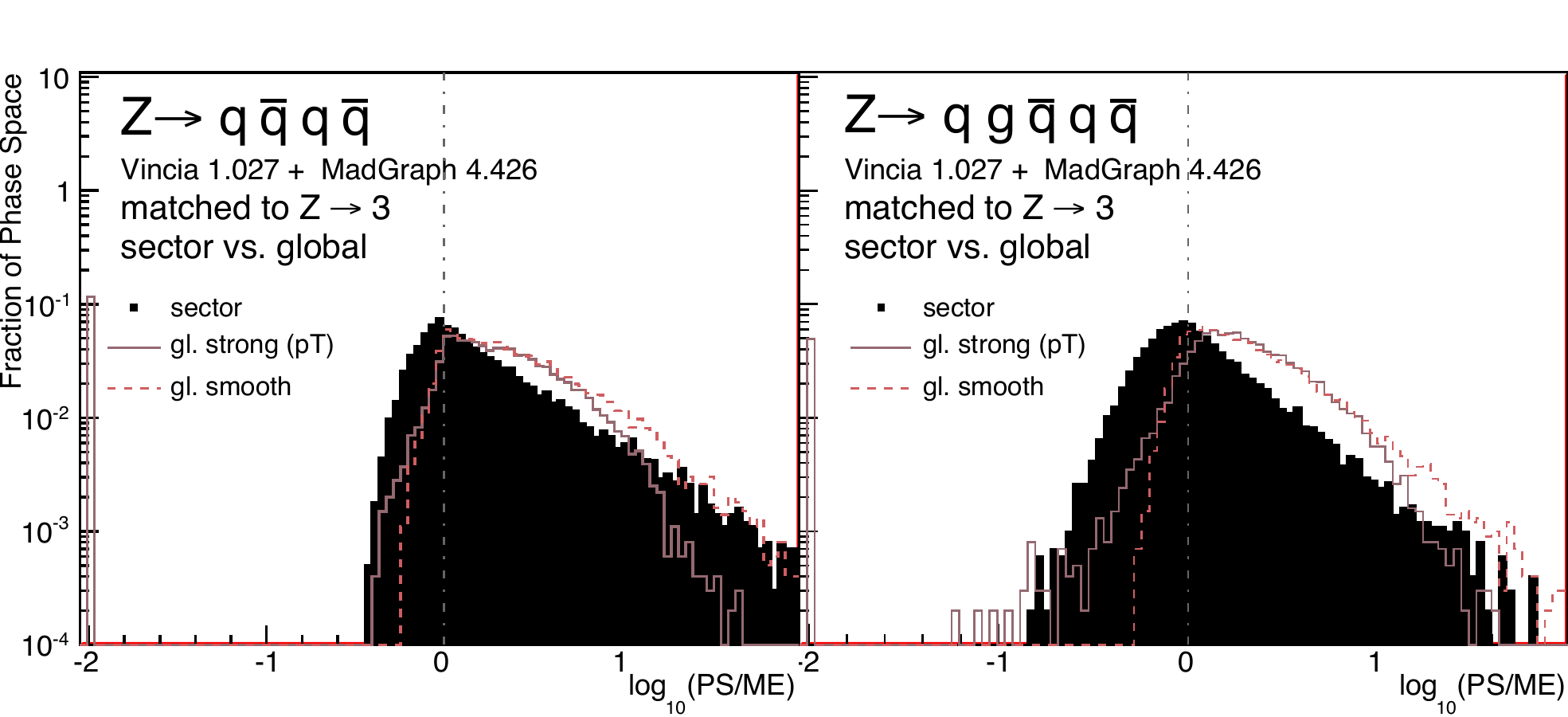}
\caption{Comparison between global strong, global smooth and sector
  shower approximations to LO 
  matrix elements, for 
  processes involving a $g\to q\bar{q}$ splitting, without applying
  the  Ariadne factor to the shower approximation. 
  Spikes on the far left represent the underflow bin. }
\label{secvsglo_quarks_noAri_pureshower}
\end{figure}
A naive application of gluon-splitting antennae, in the same way as in 
\eqRef{eq:R4E}, results in the distributions shown in
\figRef{secvsglo_quarks_noAri_pureshower}. Clearly, a
large overcounting is produced already at the first order of the $g\to
q\bar{q}$ process, shown in the left-hand pane. 
Within the Lund dipole model, this was
identified as due to gluon-screening effects between neighboring
dipole-antennae, which are not taken properly into account when adding
them independently. The perturbative cascade implemented in the
\Ar\ program therefore uses the following factor to modify its gluon
splitting probabilities \cite{Lonnblad:1992tz},
\begin{equation}
P_\mrm{ari} = \frac{2m^2_N}{m_P^2 + m^2_N}~,
\end{equation}
where $m_P^2=m_{IK}^2=s$ is the invariant mass squared of the parent
dipole-antenna and $m^2_N$ is that of the
neigboring one that shares the splitting gluon. 
Thus, e.g., if the preceding branching
was collinear, with $m_N^2\to 0$, this factor
produces a very strong suppression also inside the $m_P^2$
antenna. See also \cite{massive} for more discussion of 
this issue in the global-shower context.

\begin{figure}[t]
\centering
\includegraphics*[scale=0.75]{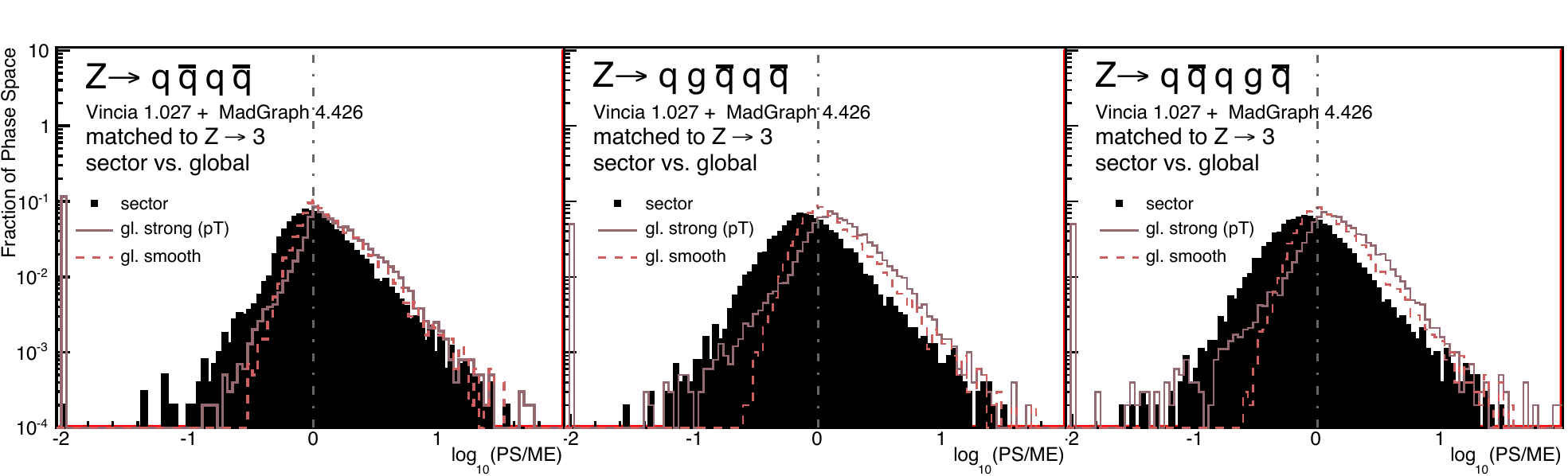}
\caption{Comparison between  global strong, global smooth and sector
  shower approximations to LO matrix
  elements, for 
  processes involving a $g\to q\bar{q}$ splitting,  including 
  the  Ariadne factor in the shower approximation. 
  Spikes on the far left represent the underflow bin. The configurations
$qg\qbar g\qbar$ and $q\qbar q g \qbar$ are related by charge
  conjugation and give the same result up to statistical precision; we
  will only plot one of them in the following. 
}
\label{secvsglo_quarks_Ari_pureshower}
\end{figure}
In \figRef{secvsglo_quarks_Ari_pureshower}, we include the ``Ariadne
factor'', $P_\mrm{Ari}$, on the gluon-splitting antennae. While the
resulting distributions are still significantly broader than their
gluon-emission counterparts in \figRef{secvsglo_gluons_pureshower},
they now show a significantly more symmetric peak around $\log(R)\sim
0$. For completeness, we also show the alternative color-ordering,
$Z\to q\bar{q}qg\bar{q}$ in the right-hand pane, noting that it is
indeed identical to the $Z\to qg\bar{q}q\bar{q}$ one within the
statistical precision. 

\begin{figure}[t]
\centering
\includegraphics*[scale=0.5]{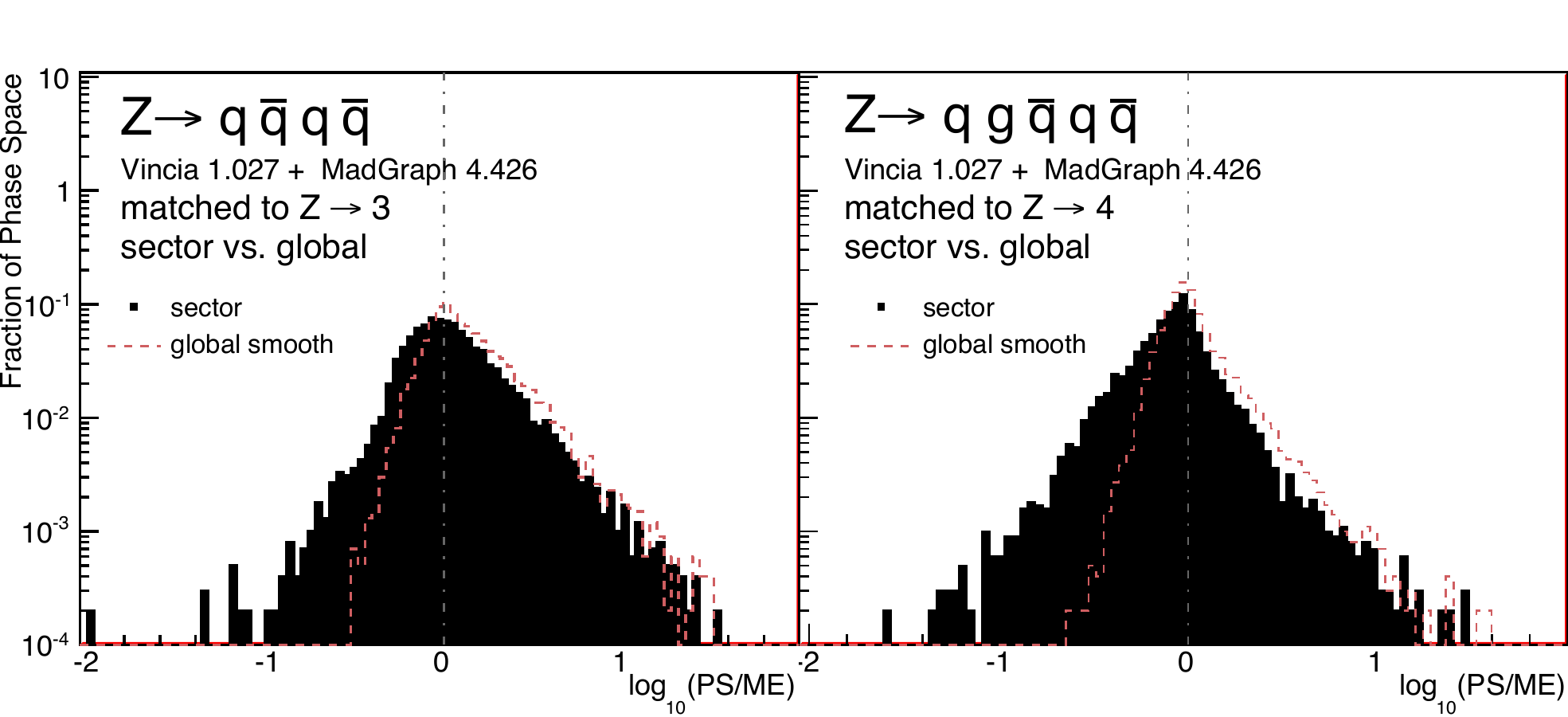}
\caption{Comparison between global smooth and sector
  shower approximations to LO matrix elements, for 
  processes involving a $g\to q\bar{q}$ splitting, including 
  the  Ariadne factor in the shower approximation.
  }
\label{secvsglo_quarks_matched}
\end{figure}
Finally, in \figRef{secvsglo_quarks_matched}, we show how the
distribution at the 5-parton level changes when matching to 4 partons
is included. The peak then becomes significantly sharper than at the
4-parton level, primarily due to the greater relative accuracy of the
gluon-emission antenna that fills one of the phase-space sectors in
that step.

\subsection{Finite Terms \label{sec:finite}}

The arbitrariness of all non-singular (``finite'') 
terms in the antenna functions
was already mentioned in \secRef{sec:poles}: the universal
leading-logarithmic approximation
furnished by the shower is only exact in the soft and
  collinear regions; in the hard region of phase-space, process-dependent
  subleading terms become important. In order to fully
  specify a set of 
  antenna functions, their finite terms must therefore also be defined,
  keeping in mind that even zero is as arbitrary a choice as
  any other, and that the choice depends explicitly on the parametrization 
  used to write the singular parts of the antennae. 

To cite a few examples, the finite parts of the GGG antennae
\cite{GehrmannDeRidder:2005cm} 
are simply the leftovers from the specific matrix elements
that were used to derive those functions in
\cite{GehrmannDeRidder:2004tv,GehrmannDeRidder:2005hi,GehrmannDeRidder:2005aw}. In
the \Ar\ and \Vc\ codes, the current defaults are based on comparisons
to $Z$ decay matrix elements. They should thus work especially well
for that process, chosen since it is the main reference for final-state
showering, but they could in principle 
do less well for other processes. 

For simplicity, we here set all finite coefficients of the
gluon-emission antennae to zero, as summarized in \tabRef{tab:antenna_coefficients}. 
To illustrate the indeterminacy associated with this choice, we
compare this choice (labeled ``central'') with two other sets\footnote{the choices for the finite-terms of the gluon splitting antennae in \tabRef{tab:antenna_coefficients} situate them at the edge of the positivity condition in some regions of phase-space; we do not consider a ``minus'' set for these antennae.} labeled
``minus'' and ``plus'', defined by
\begin{eqnarray}
\mbox{gluon emission~~~:~~~ }\bar{a}_g^\mrm{plus/minus} & = & \bar{a}_g^\mrm{central} \pm \frac{y_{ij}+y_{jk}}{s} ~,\\
\mbox{gluon splitting~~~:~~~ }\bar{a}_{q}^\mrm{plus}~~~~~~~~~~ & = & \bar{a}_q^\mrm{central} + \frac{1}{2s} ~,
\end{eqnarray}
with finite-term variations motivated partly by the finite
terms of the other antenna sets listed in
\tabRef{tab:antenna_coefficients}. 

\begin{figure}[t]
\centering
\includegraphics*[scale=0.75]{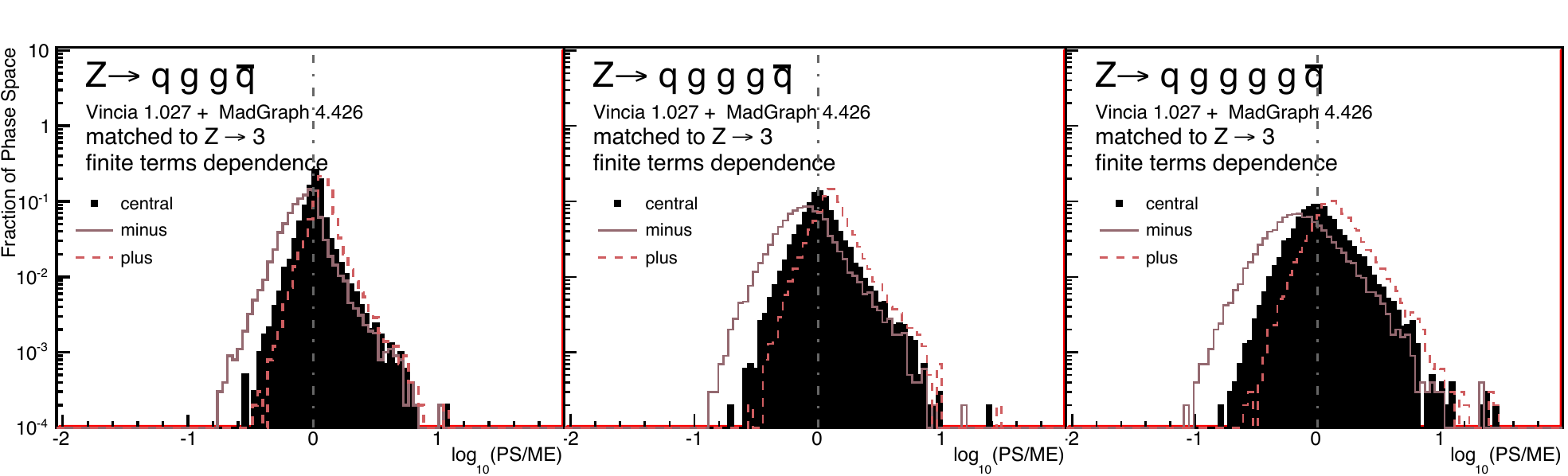}
\caption{Illustration of the impact of finite terms in the sector
  shower approximation, for $Z\to q\bar{q}\,+\,$gluons.
  The
  default sector antennae are shown in the solid filled histogram,
  with ``minus'' (solid lines) and ``plus'' (dashed lines) 
  variations defined in the text. 
  }
\label{finiteterms_gluons_pureshower}
\end{figure}

\begin{figure}[t]
\centering
\includegraphics*[scale=0.5]{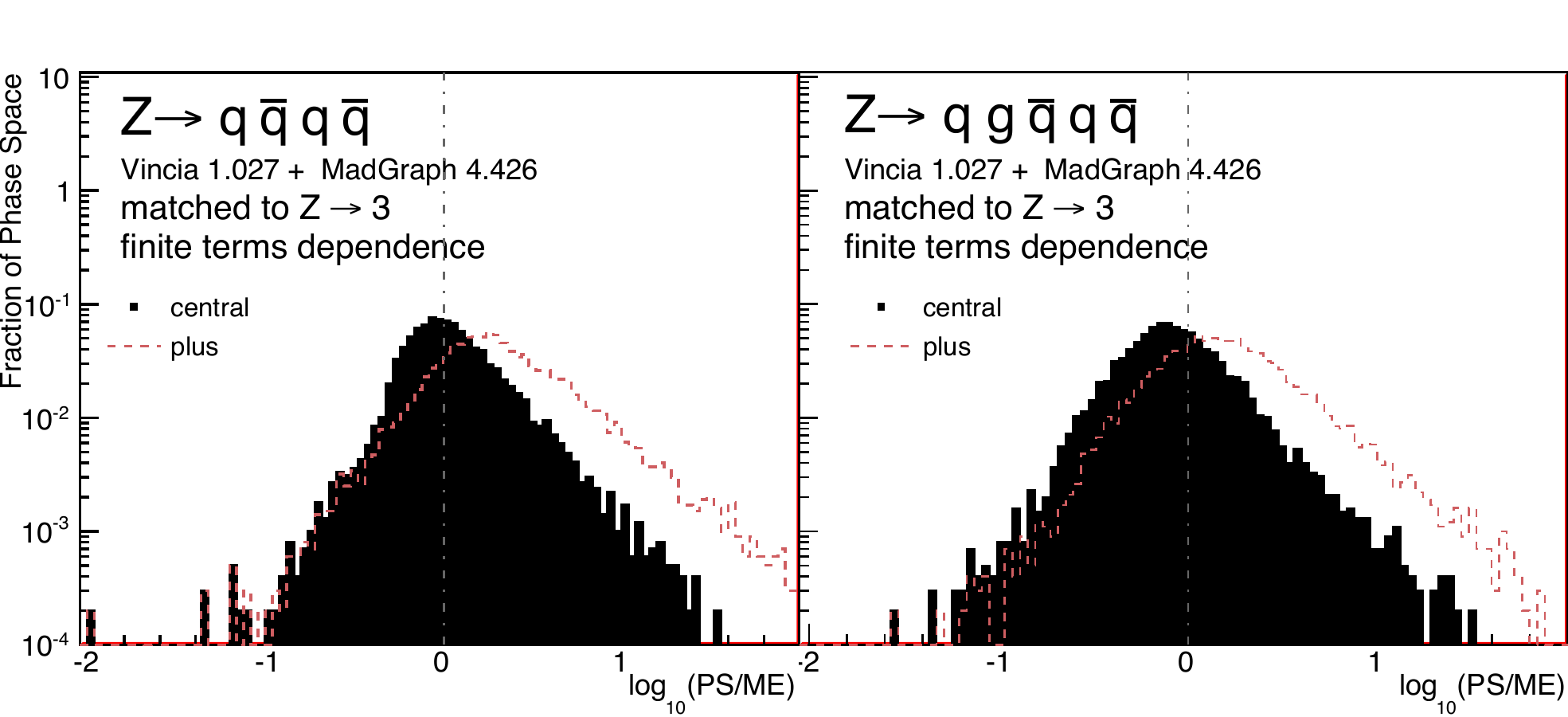}
\caption{Illustration of the impact of finite-term variations, for
  processes involving a $g\to q\bar{q}$ splitting. The
  default sector antennae are shown in the solid filled histogram,
  with ``minus'' (solid lines) and ``plus'' (dashed lines) 
  variations defined in the text. 
}
\label{finiteterms_quarks_pureshower}
\end{figure}

Note that this plus/minus variation is not intended to
represent any conservative max/min range, but merely to illustrate
what the consequence of moderate finite-term variations is for the
matrix-element comparisons that were considered in the previous subsection. 
This is done in \figsRef{finiteterms_gluons_pureshower} and
\ref{finiteterms_quarks_pureshower}, for gluon emission and gluon
splitting, respectively.

These distributions do not include matching beyond $Z\to 3$
partons, hence the variation grows with multiplicity. For gluon
emission, we see that the central choice stays relatively well centred
on $\log(R)\sim 0$, while for gluon splittings, a choice intermediate
between the central and the minus variation would appear to generate
the best agreement, for this particular process. 
We emphasize that matching to matrix
elements removes these ambiguties, up to the matched order. Also note
that we are showing flat phase-space scans, which do not represent the
actual weighing induced by the shower, where soft and collinear
regions (in which the agreement is generally better) are strongly
privileged.  

\subsection{Choice of Sector Decomposition \label{sec:decomposition}}

The default variable we use to partition phase-space into
sectors\footnote{I.e., to decide which $ijk \to IK$ clustering to
  perform, or,
equivalently,  whether to accept a given $IK \to ijk$ 
trial branching during the
shower.} 
was defined in \eqRef{eq:QSvariable}. It basically amounts to finding
the sector with the smallest value of $p_\perp$
for gluon emissions, which is modified to a $p_\perp$-weighted virtuality for
gluon splittings\footnote{Since the virtuality, $s_{ij}$, only
involves two partons, the virtuality alone cannot be used to 
distinguish between two neigboring 3-parton clusterings that share the
same small invariant. The choice represented by \eqRef{eq:QSvariable}
is therefore essentially the geometric mean of \pT and
$s_{ij}$.}. That choice is not unique. The basic 
criterion is that if any of the partons of the configuration is
approaching the soft limit, or a pair of them approaches the collinear
limit, we must select an antenna that contains the appropriate
divergent terms. This ensures that the shower will achieve at least LL 
precision in every phase-space point. 
Beyond that, different choices will lead to
different subleading behavior.

For simplicity, we first focus on gluon emission only, i.e., without
the additional complication of interleaved gluon splittings. Since the
sector-decomposition variable must isolate the leading singular
regions, we have explored three possible variations that can be 
constructed from the soft eikonal factor. Thus, the prescription is to
select the sector which minizes either of the three following measures:
\begin{enumerate}
\item{Transverse momentum, $\pT^{2}=y_{ij}y_{jk}s$,}
\item{Scaled transverse momentum, $y_E = \pT^{2}/s = y_{ij}y_{jk}$
  (dimensionless),} 
\item{Inverse eikonal, $p_{Eik}^2 \equiv \frac{\pT^{2}}{y_{ik}}=\frac{y_{ij}y_{jk}}{y_{ik}}s$.} 
\end{enumerate}
The difference between these choices can be characterized as follows. 
For a branching that occurs inside a small-mass dipole-antenna, 
the dimensionful \pT will always associate
a small scale, even if the branching is relatively hard compared with
the parent mass, while the scaled variant
only considers the hardness of the branching relative to its 
parent. The inverse eikonal represents a variation of \pT
which has the same singular limit but which goes to infinity along the
(non-singular) boundary $y_{ik}\to 0$, while \pT remains
bounded by $\sqrt{s}/2$.

\begin{figure}[t]
\centering
\includegraphics*[scale=0.75]{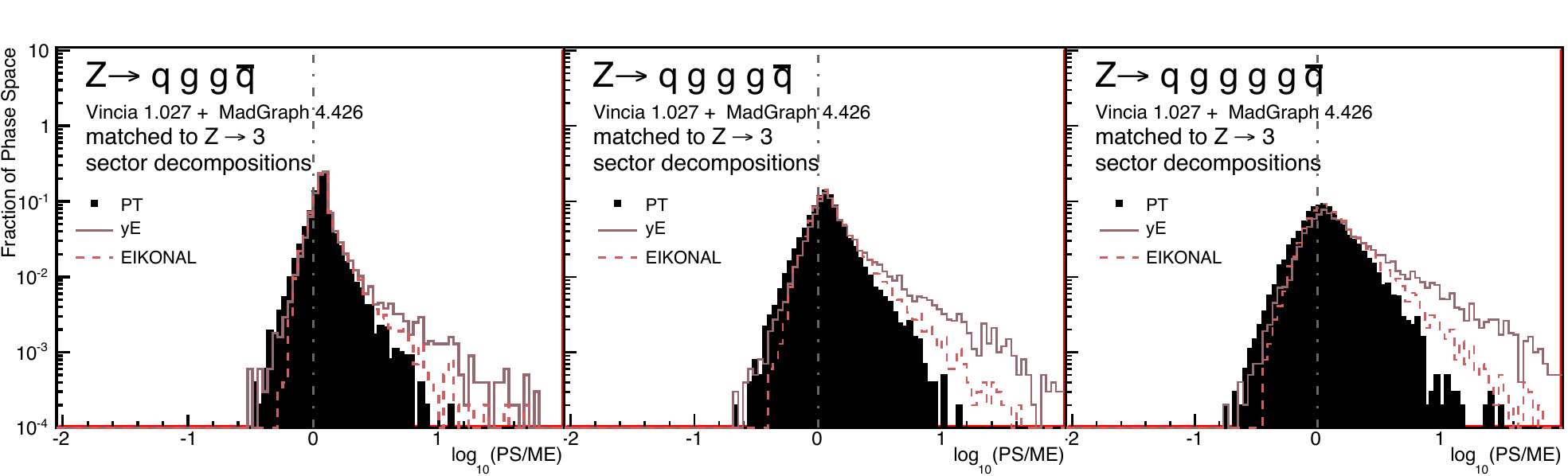}
\caption{Illustration of various sector decomposition variables, for $Z\to q\bar{q}\,+\,$gluons: dimensionful \pT, dimensionless $\pT/s$ and $p_{Eik}$. } 
\label{sectorclustering_gluons_pureshower}
\end{figure}
The comparison of the sector-shower expansion to matrix elements, 
using each of these choices, is illustrated in
\figRef{sectorclustering_gluons_pureshower}. We see that the
dimensionless choice, $y_E$ (thin solid 
lines), produces the worst description, with large tails towards
overcounting of the matrix elements. 
We ascribe this to the scaled $y_E$ only
including information about the unresolved limit within the current
antenna (it always prioritizes the most singular one, regardless of
size) 
while the presence of $s$ in the dimensionful \pT (solid
filled histogram), introduces an additional information, the size of 
the $ijk$ dipole-antenna itself, which is implicitly related to the
singularity structure of the previous branching. Changing between \pT
and the full eikonal (dashed histograms) has a smaller effect, with
\pT coming out slightly better, at least for this process. This
is the motivation for using \pT as the sector-decomposition variable
for gluon emissions.

\begin{figure}[t]
\centering
\includegraphics*[scale=0.5]{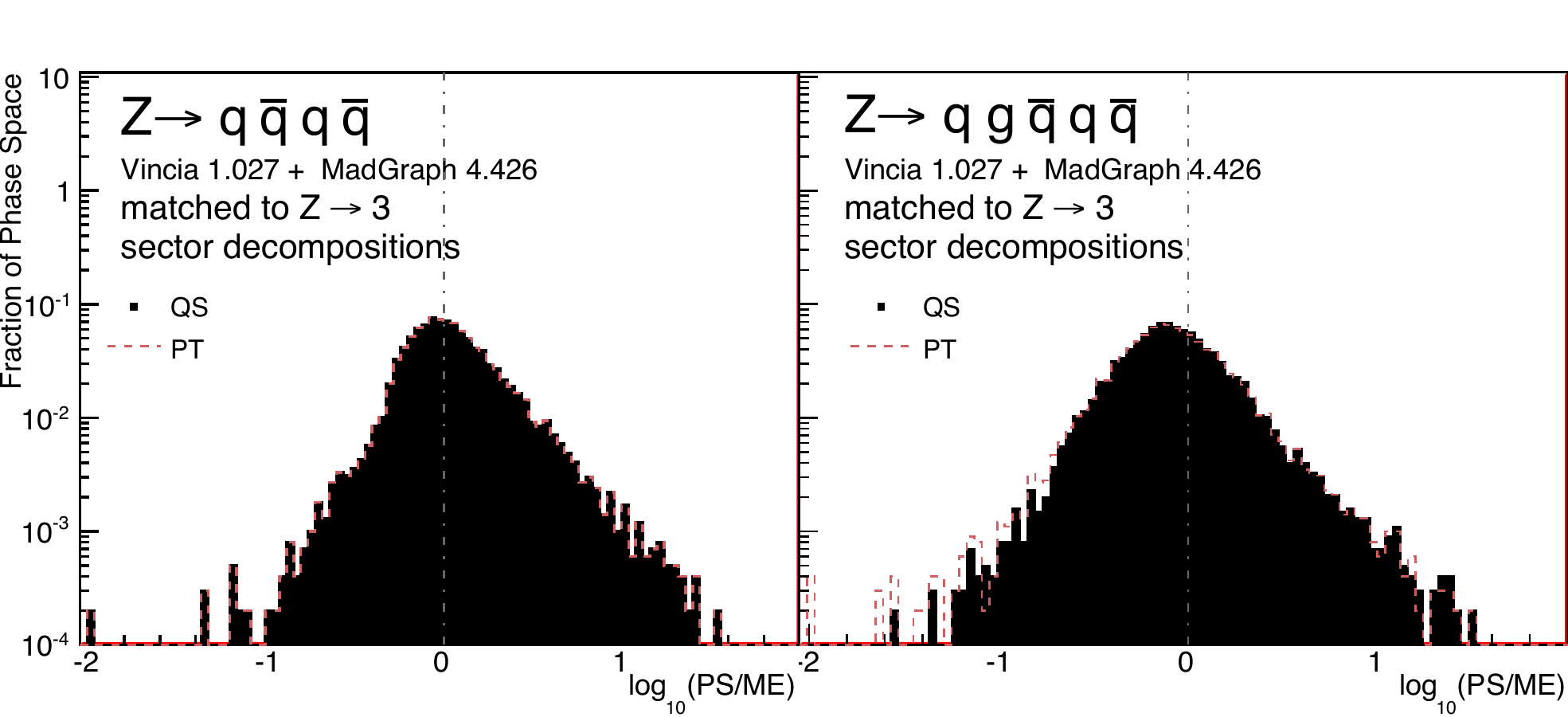}
\caption{illustration of two sector decomposition variables: \pT and
  $Q_{S}$, for processes involving a $g\to q\bar{q}$ splitting,
  including the Ariadne factor in the shower approximation.
}\label{sectorclustering_quarks_pureshower}
\end{figure}

To include gluon splittings, the simplest would be to just use \pT for 
all partons. Alternatively, the \Vc~default choice defined in
\eqRef{eq:QSvariable}, attempts to reflect the 
different structure of gluon splittings in the choice of 
measure computed for clusterings involving such a splitting. These two
choices are compared in a flat phase-space scan in
\figRef{sectorclustering_quarks_pureshower}. One basically sees no
difference between them. Note, however, that there is really no
competition going on between different sectors until the $Z\to 5$
level. For $Z\to 4$ (in the left-hand pane), 
the evolution sequence is fixed to a gluon
emission followed by a gluon splitting. \pT and $Q_S$ then produce the
same sectors, as is also evident from the plot. 
At $Z\to 5$, the $g\to q\bar{q}$ splitting can happen either in the
second or third evolution step, with $Q_S$ and \pT now classifying the
sectors differently. Nonetheless, only very small differences are
visible also on the right-hand pane of
\figRef{sectorclustering_quarks_pureshower}. 

However, for the parametrization of the gluon-splitting antennae we
have chosen, there \emph{is} actually an important subtlety connected
with this choice, which can be illustrated by considering 
the color-ordered structure $X-g-\qbar-q$, with $X$ an arbitrary
colored parton. The $g\qbar$-collinear limit, $s_{g\bar{q}}\to 0$, 
is singular in the $X\qbar\to Xg\qbar$ antenna, but not in the 
$gg\to g\qbar q$ one. Since the parametrization chosen for our
gluon-splitting antennae does not allow any ``spillover terms'' from
neigboring gluon-emission sectors, the entire $g\qbar$-collinear limit
should therefore be classified as belonging to the $X\qbar\to Xg\qbar$
sector, in order to correctly reproduce the full collinear gluon-emission 
singularity. Since this is only a single pole, as compared to
the leading double pole for gluon emission, it does not show up 
clearly in \figRef{sectorclustering_quarks_pureshower}.  

\begin{figure}[t]
\centering
\includegraphics*[scale=0.28]{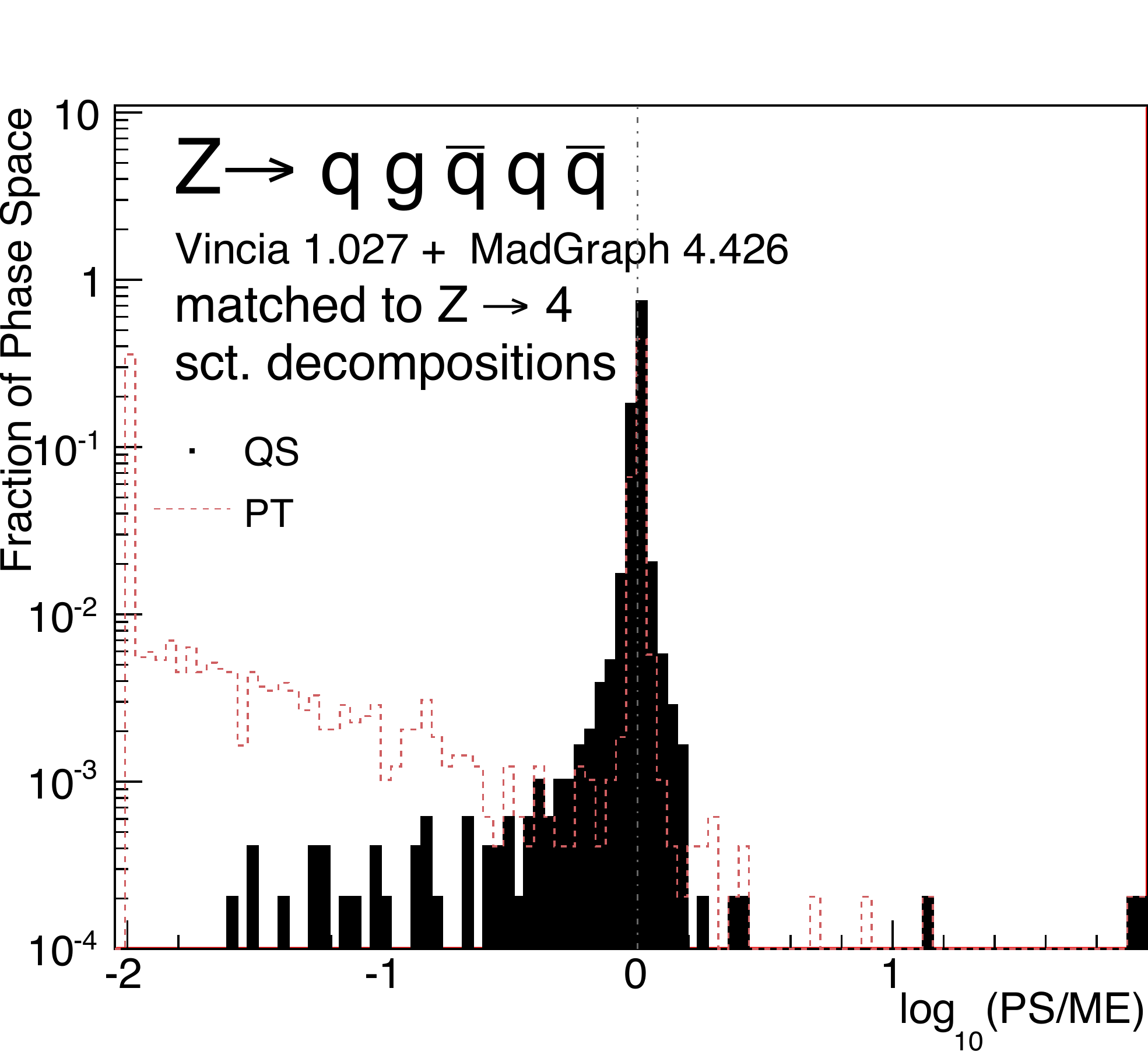}
\caption{Illustration of two sector decomposition variables,  \pT and $Q_{S}$, in a particular region of phase-space: the collinear region of the pair $g \qbar$ in the configuration $q g \qbar q \qbar$. The actual values for the figure are $m_{g\qbar}^2<0.02\,$GeV$^2$ and $E_{\mrm{cm}}=70\,$GeV. 
  The spike on the far left represent the underflow bin. 
  }
\label{sectorclustering_specialregion_qgqqq_matched}
\end{figure}
We may isolate 
the potentially problematic region in phase-space, by plotting only
phase-space points for which $m_{g\bar{q}}^2 < 0.02\,$GeV$^2$. This is
done in \figRef{sectorclustering_specialregion_qgqqq_matched}. Once we
``zoom in'' on the problematic region in this way, it is immediately
apparent that using \pT only produces the ``correct''
answer for half of the accepted phase-space points (the dotted
histogram does still have a peak at $\log(R)\sim 0$, but only half of
the phase-space points populate it), while the other
half (those corresponding to the ``wrong'' clustering, which does not
have a $g\bar{q}$ singularity) is significantly undercounted. This is
the fundamental reason we choose $Q_S$ as the partitioning variable
for the sector-shower implementation in
\Vc\footnote{We thank D.~Kosower for pointing out this subtlety and
  for suggesting the modification  necessary to cure it.}.

\section{The Shower Algorithm \label{sec:shower}}

The implementation of the sector shower in \Vc~is based on the
global shower setup. The latter is extensively discussed in
\cite{Giele:2007di,Giele:2011cb} and will not be repeated here. For a
general introduction to shower Monte Carlos including use of the veto
algorithm and related topics, see \cite{Buckley:2011ms}. 
Here, we focus exclusively on the modifications to the showering
algorithm that occur when going from the global to the sector case. In
\secRef{sec:2to3}, we consider the basic sector shower, built from
sequences of $2\to 3$ branchings. In \secRef{2ton_matching}, we
describe the small modification that is 
required to adapt GKS matching to the sector case. 

\subsection{$(2\to3)$: The Basic Trial Generator \label{sec:2to3}}

Our fundamental building block for showering purposes is the evolution
integral:
\begin{equation}
\IntA(s,Q^2_{E1},Q^2_{E2})  = \int_{Q^2_{E2}}^{Q^2_{E1}} 
\frac{\d{s_{ij}} \d{s_{jk}}}{16\pi^2 s} \ a(s,s_{ij},s_{jk})~~~;~~~Q^2_{E2}<Q^2_{E1}~, \label{eq:ev1}
\end{equation}
which represents the integrated tree-level splitting probability
between the scales $Q_{E1}$ and $Q_{E2}$, for an arbitrary ``infrared
sensible'' \cite{Skands:2009tb} definition of the evolution
variable $Q_E$. 
As in \cite{Giele:2011cb}, we perform a change of variables to recast the
integral in such a way that the evolution variable appears explicitly
as an integration variable, 
\begin{equation} 
\IntA(s,Q^2_{E1},Q^2_{E2}) =\frac{1}{16\pi^2 s} \int_{Q^2_{E2}}^{Q^2_{E1}} 
\! \! \d{Q^2_E} \d{\zeta} \ |J| \ 
 a(s,s_{ij},s_{jk})~, \label{eq:ev2b}
\end{equation}
where  $|J|$ 
is the Jacobian associated with the transformation from
$(s_{ij},s_{jk})$ to $(Q^2_E,\zeta)$. 
The default choice in \Vc\ is to use $Q_E=2\pT$ for gluon emission
and $Q_E=m_{q\qbar}$ for gluon splitting, with phase-space contours as
illustrated in \secRef{sec:conventions}. In the global case, several 
alternative options have been implemented for gluon emission, while
the choice of $Q_E$ for gluon 
splitting is fixed, see \cite{massive}. In the sector implementation,
we have so far only considered the default choices for both antenna
types.  We return to the choice of $\zeta$
below, for which we shall require some extensions relative to  the global case. 

As in all shower implementations, we make use of the veto algorithm to replace
the integrand, $a$, by a simpler function,  $\atrial$, called
the ``trial function". Provided our trial function is larger than the
actual integrand, the veto algorithm will allow us to recover the
exact integral post facto. So far, we also rely on the veto
algorithm to implement the restriction to phase-space sectors; that
is, for each  antenna we start by 
generating trial branchings over all of phase-space (as in the
global shower), and then veto those 
which do not have the smallest value of $Q_S$ in their respective would-be 
post-branching parton configurations. 

The simplest case to describe is actually that of gluon splitting, for which
the only difference with respect to the global case (apart from the
sector veto) is the overall
factor of 2 on both trial and ``physical'' antenna functions,
cf.\ \tabRef{tab:antenna_coefficients}. Since applying a
multiplicative factor to the branching generator is trivial, we refer the reader
to \cite{massive}, where the formalism for generating gluon splittings
is described in detail for the global shower. 

For gluon emission, the additional gluon-collinear terms that appear
in the sector case, see \secRef{sec:poles}, necessitate a further
manipulation of the shower 
algorithm. Essentially, we shall treat the additional
terms as separate sub-antenna functions, assigning them  their own 
 trial functions and $\zeta$ definitions.
The remaining terms, which include the eikonal, 
correspond exactly to the global case and are carried over directly
from there. 

The $q\bar{q}\to qg\bar{q}$ antenna does not change, since none of the
parents are gluons. The trial function for this antenna is therefore identical to
the one used for all gluon emission-antennae in the global case,
\begin{eqn}
\frac{\atrialemit^\mrm{sct}}{16\pi^2} = \frac{\atrialemit^\mrm{gl}}{16\pi^2} = \frac{\hat{\alpha}_s}{4 \pi} C_A
\ \frac{2s}{s_{ij}s_{jk}}~. 
\end{eqn} 

For $qg\to qgg$, we split the physical sector antenna function into two
sub-antennae, consisting of the global part and an additional
gluon-collinear piece, 
\begin{eqn}
\frac{\hat{\alpha}_{s}}{4\pi}C_{A}\ \left(\bar{a}^\mrm{gl}_{g/qg}
 + 
\frac{1}{s}\left[\frac{2}{y_{jk}(1-y_{ij})} + \frac{-2}{y_{jk}}+\frac{y_{ij}}{y_{jk}}+\frac{-y_{ij}^{2}}{y_{jk}}\right]\right) ~,
\label{eq:physK}
\end{eqn} 
with the trial function for the global part the same as in the
global case (i.e., identical to the one
for $q\bar{q}\to qg\bar{q}$ above), and the one for the additional
piece being 
\begin{eqn}
\frac{a_{\mrm{trial-coll}-K}^{\mrm{sct}}}{16\pi^2} = 
\frac{\hat{\alpha}_{s}}{4\pi}C_{A}\ \frac{2s}{s_{jk}(s-s_{ij})}
\label{eq:trialK}
\end{eqn} 

Finally, we split the $gg \to ggg$ antenna into three sub-antennae, again
consisting of a global part but now with two additional collinear
pieces, corresponding to each of the parent gluons, $I$ and $K$,
respectively. The physical and 
trial terms are defined analogously to those in \eqsRef{eq:physK} and
\eqref{eq:trialK}, respectively, with the $I$-collinear ones obtained
by the replacement $i \leftrightarrow k$.

One can check that the sum of the coefficients of the same powers of 
$y_{ij}$ and $y_{jk}$ among the sub-antennae makes up the total
coefficients of the sector antennae 
displayed in tab.~\ref{tab:antenna_coefficients}. The fact that the
first sub-antenna of each process corresponds to the global case makes
it possible to rely on the properties that have already been put to
test in the global shower implementation, simplifying the sector
shower case to the addition of the extra $I$- and $K$-collinear
sub-antennae. In fact, the ultimate reason for this splitting is that
the integrals for these sub-antennae are separately treatable in an
analytic way.   

The overall normalization of the trial function can be
adjusted, should the finite terms associated with, e.g.,
matrix-element matching, render the physical function 
bigger than the trial one in some corner of phase-space. 
We emphasize that there is no trace of the
overestimator present in the final results, and the only sensitivity
to its shape and normalization 
is in the \emph{speed} of the calculation. 

As mentioned above, we restrict our attention to $Q_E = 2\pT$ for
gluon emissions. For the $\zeta$ variable appearing in \eqRef{eq:ev2b}, we make
a separate choice for each type of trial function, 
\begin{eqnar}
\zeta \ = \ \left\{\begin{array}{lll}\zeta_-=y_{ij} &
\mbox{$K$-collinear trial function}\\[1mm] \zeta_0 = y_{ij}/(y_{ij}+y_{jk})&
\mbox{Eikonal (global) trial function}\\[1mm]\zeta_+ =
y_{jk} & \mbox{$I$-collinear trial function}\end{array}\right.~\label{eq:zetadefinitions}. 
\end{eqnar}
The associated Jacobians for these different cases are, correspondingly, 
\begin{eqnar}
|J| = \left\{\begin{array}{lll}|J_-|=\frac{s}{4y_{ij}}\\[2mm] |J_0| =\frac{ s(y_{ij}+y_{jk})^2}{8\ y_{ij}y_{jk}}\\[2mm] |J_+| = \frac{s}{4y_{jk}}\end{array}\right.~\label{eq:zetadefinitionsJ}.
\end{eqnar}
These $\zeta$ definitions share the same limits of the
$\zeta$-integrals in expression (\ref{eq:ev2b}), since the relevant
boundary of phase-space is defined by the condition $y_{ij}+y_{jk}=1$,
as can be inferred, e.g., from the illustration of \pT contours that was given in 
\figRef{fig:variables}. Specifically, we
have 
\begin{eqn}
\zeta_{\mrm{min}}(Q^2_E) = \frac{1-\sqrt{1-Q^2_E/s}}{2}~~~,~~~
\zeta_{\mrm{max}}(Q^2_E) = \frac{1+\sqrt{1-Q^2_E/s}}{2}~
\label{eq:zetaminmaxfunction}
\end{eqn}

To derive an analytical expression for the $\zeta$ integral in
\eqRef{eq:ev2b}, we make two simplifications. 
First, we neglect any possible dependence of $\alpha_{s}$ on
$\zeta$ (i.e., we shall take the trial $\hat{\alpha}_s$ either to be a
constant or to depend only on $Q_E$). 
Second, we shall generate trial branchings 
in a larger phase-space region than the physically allowed 
one, again using the veto algorithm to reject trials that are
generated in the unphysical region. 
\begin{table}[t]
\begin{center}
\scalebox{\figscale}{
\begin{tabular}{rlcrc}
\multicolumn{5}{c}{\textsc{Vincia Evolution Windows}}\\\toprule
$i$ & $[Q_{E\mrm{min}}$&,&$Q_{E\mrm{max}}]$ & $n_f$\\\midrule
0   & $[0$&,&$m_c]$ & 3 \\
1   & $[m_c$&,&$m_b]$ & 4 \\
2   & $[m_b$&,&$\sqrt{m_bm_t}]$ & 5 \\
3   & $[\sqrt{m_bm_t}$&,&$m_t]$ & 5 \\
4   & $[m_t$&,&$\infty]$ & 6  \\\bottomrule
\end{tabular}}
\capt{The evolution windows used in \textsc{Vincia}, with the $Q_{E}$ boundaries and active number of flavors corresponding to each. The number of active 
flavors is the same for windows 2 and 3, but the $\zeta$ boundaries for trials
 are different, due to the different $Q_{E\mrm{min}}$
 values. This improves the efficiency of the generator.
The first window will not actually extend down to 
zero in practice, but will instead be cut off by the hadronization scale. 
\label{tab:windows}}  
\end{center}
\end{table}
The overestimate of phase-space is
divided into several distinct windows in $Q_E$, given in table \ref{tab:windows};
 in each such window, we replace the $Q_E$-dependent $\zeta$ limits in
 the $\zeta$-integral of (\ref{eq:ev2b}) by constant ones,  
\begin{eqn}
\zeta_{\mrm{min}}(Q^2_E) = \zeta_{\mrm{min}}(Q^2_{E\mrm{min}})~~~,~~~
\zeta_{\mrm{max}}(Q^2_E) = \zeta_{\mrm{max}}(Q^2_{E\mrm{min}})~,
\label{eq:zetaminmax}
\end{eqn}
where $Q_{E\mrm{min}}$ is the value of $Q_E$ at the end of the current
window (e.g., the next flavor threshold or, 
ultimately, the hadronization scale). 
\begin{figure}[t]
\centering
\includegraphics*[scale=0.5]{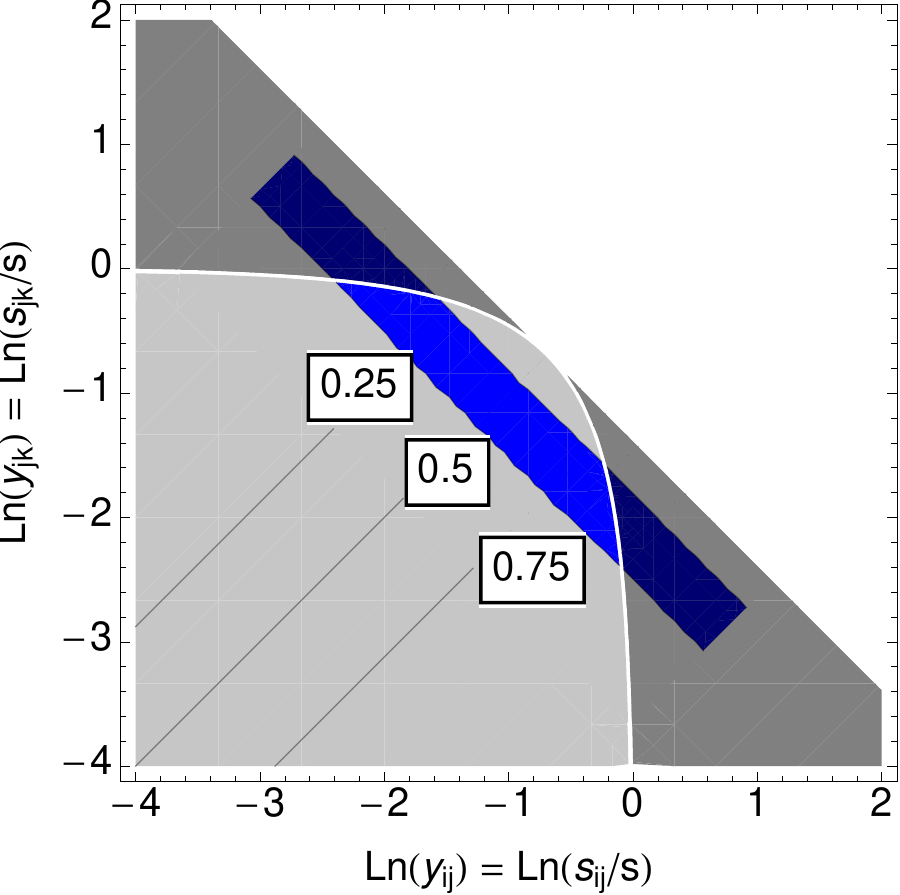}\hspace*{8mm}
\includegraphics*[scale=0.5]{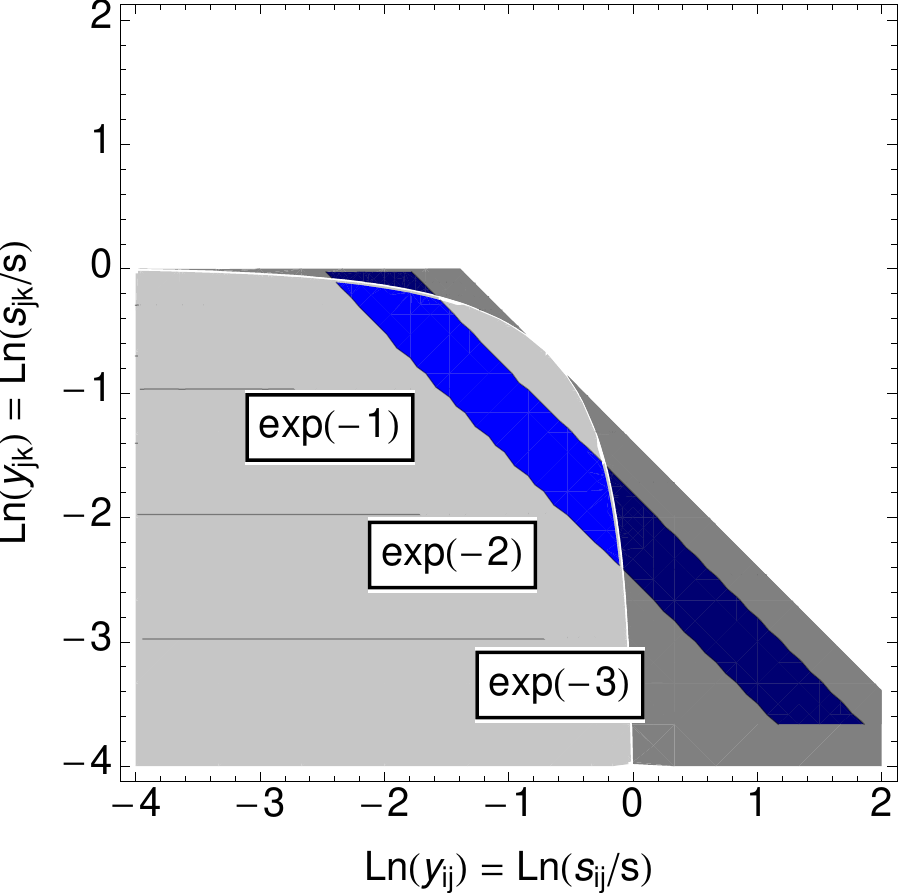}
\caption{Illustrations of the $\zeta$ choice for (left) the global antenna
  part, $\zeta_0$, and (right)  the additional $I$-collinear sector
  trial function, $\zeta_+$ (the $K$-collinear one, $\zeta_-$, is
  obtained by swapping the
  invariants). Axes are logarithmic in the $y_{ij}$
  and $y_{jk}$ phase-space variables. The physical phase-space is
  shown with lighter shading, while the overestimate of phase-space
  used for trial branchings is shown with darker shading.}
\label{ordering-variables_zeta}
\end{figure}
This is illustrated in \figRef{ordering-variables_zeta}, for the
$\zeta_0$ (left) and $\zeta_+$ (right) definitions, with the physical
region of phase-space shown with lighter shading and the unphysical
one with darker shading. Note: the axes are logarithmic in the
scaled invariants $y_{ij}$ and $y_{jk}$, hence the boundary of the
physical phase-space does not look like a triangle here. The dark
diagonal strips correspond to a window of trial generations, again
with the lighter part corresponding to trials inside the physical
phase-space and the darker part to ones outside it. In the right-hand
pane, the tail of trial generations extending towards large $y_{ij}$
and small $y_{jk}$ is not a problem for efficiency, since it is only
used in combination with the $I$-collinear trial function, which is
strongly peaked in the opposite region of $y_{ij}$. 

During the evolution, the progression between different evolution windows
happens as follows; if none of the generated trials fall within the
current evolution window, the evolution is 
restarted at $Q_E=Q_{E\mrm{min}}$, upon which the $Q_{E\mrm{min}}$ and $\zeta$
boundaries is updated to correspond to those of the next
evolution window. 

With these simplifications, the $\zeta$ integrals are
\begin{eqnarray}
I_{\zeta,0} & \equiv &  \int_{\zeta_{min}}^{\zeta_{max}} d\zeta_{0}
\frac{1}{\zeta_{0}(1-\zeta_{0})} =
\ln\left(\frac{\zeta_{max}(1-\zeta_{min})}{\zeta_{min}(1-\zeta_{max})}\right) \label{eq:zeta0int}\\[2mm] 
I_{\zeta,-} & \equiv & \int_{\zeta_{min}}^{\zeta_{max}} d\zeta_{-} \frac{1}{(1-\zeta_{-})} = \ln\left(\frac{1-\zeta_{min}}{1-\zeta_{max}}\right) \label{eq:zeta-int}\\[2mm]
I_{\zeta,+} & \equiv & \int_{\zeta_{min}}^{\zeta_{max}} d\zeta_{+}
\frac{1}{(1-\zeta_{+})} = 
\ln\left(\frac{1-\zeta_{min}}{1-\zeta_{max}}\right) \label{eq:zeta+int}
\end{eqnarray}

Defining a one-loop running $\hat{\alpha}_s$ for trial branchings by
\begin{equation}
\hat{\alpha}_s(k_\mu Q_E) = \frac{1}{b_0\ln\left(x_E^2\right)}~,\label{eq:aS}
\end{equation}
with  
\begin{equation}
 b_0 = \frac{33 - 2n_f}{12\pi}~, \label{eq:b0}
\end{equation}
\begin{equation}
x_E = \frac{k_\mu Q_E}{\Lambda_\mrm{QCD}}~,  \label{eq:xE}
\end{equation}
and $k_\mu$ an arbitrary scale
factor that can be used to adjust the 
effective renormalization scale up or down, 
the integrals over $Q_E$, defined in
\eqRef{eq:ev1}, can now finally be expressed as 
\begin{itemize}
\item{
for $\frac{\hat{\alpha}_{s}}{4\pi}C_{A}\frac{2s}{s_{ij}s_{jk}}$ ($\zeta_0$ used):
\begin{equation}
\IntA(s,Q^2_{E1},Q^2_{E2})  =
\frac{C_{A}}{4\pi
b_0}\left[\ln\left(\frac{\ln(x_{E1}^{2})}{\ln(x_{E2}^{2})}\right)
  I_{\zeta,0}(\zeta_{min},\zeta_{max})\right], \label{eq:splitprob0}
\end{equation}
}
\item{
for $\frac{\hat{\alpha}_{s}}{4\pi}C_{A}\frac{2s}{s_{jk}(s-s_{ij})}$ ($\zeta_-$ used):
\begin{equation}
\IntA(s,Q^2_{E1},Q^2_{E2})  = \frac{C_A}{4\pi b_0}\left[2\ln\left(\frac{\ln(x_{E1}^{2})}{\ln(x_{E2}^{2})}\right) I_{\zeta,-}(\zeta_{min},\zeta_{max})\right], \label{eq:splitprob-}
\end{equation}
}
\item{
for $\frac{\hat{\alpha}_{s}}{4\pi}C_{A}\frac{2s}{s_{ij}(s-s_{jk})}$ ($\zeta_+$ used):
\begin{equation}
\IntA(s,Q^2_{E1},Q^2_{E2})  = \frac{C_A}{4\pi b_0}\left[2\ln\left(\frac{\ln(x_{E1}^{2})}{\ln(x_{E2}^{2})}\right) I_{\zeta,+}(\zeta_{min},\zeta_{max})\right]~, \label{eq:splitprob+}
\end{equation}
}
\end{itemize}
in which the $\ln(\ln(x))$ structure comes from folding the trial-function
singularities with the Landau pole in $\hat{\alpha}_s$.
Note: to use a constant trial $\hat{\alpha}_s$ instead in these expressions, make the
replacements $1/b_0 \to \hat{\alpha}_s$ and $\ln(x_E) \to Q_E$. To
include running beyond one loop in the trial function, see \cite{Giele:2011cb}.

For gluon splitting, we again emphasize that the only change is a factor 2
relative to the global case, and refer to \cite{massive} for details. 

The actual generating function for the shower is constructed from
these integrals via the Sudakov form factor: 
\begin{eqn}
\Delta(Q_{E1}^2,Q^2_{E2}) = \exp\left(
- \IntA(Q_{E1}^2,Q^2_{E2})
\right)~,\label{eq:Sudakov}
\end{eqn}
where we may substitute for $\IntA$ either of the expressions
eqs.~(\ref{eq:splitprob0}, \ref{eq:splitprob-}, \ref{eq:splitprob+}). 
Trial branchings are generated according to this Sudakov by solving 
the equation
\begin{eqn}
R = \Delta(Q_{E1}^2,Q^2_{E2}) \label{eq:QofR}
\end{eqn}
for $Q_{E2}$, where $R\in [0,1]$ is a uniform random number and
$Q_{E1}$ is the ``(re)starting'' scale for the evolution. 
If the evolution is being started from scratch, the (re)start scale is
$\sqrt{s}$, the invariant mass of the dipole-antenna. If the evolution
is being continued after an accepted branching, the restart scale is
likewise set to $\sqrt{s}$. This is equivalent to the ``unordered'' global
case, discussed in \secRef{sec:tests}, but here with the sector veto
protecting us from overcounting, as was illustrated in 
\figRef{secvsglo_gluons_pureshower_glUNORDvssector}.  
In practice, since the sector veto will reject any trial generated
above the smallest $Q_S$ scale that remains unchanged by the branching, 
the restart scale after a preceding accepted trial is actually reduced to
$Q_{S\mrm{min}}^{\mrm{unc}}\le\sqrt{s}$, defined as the smallest $Q_S$
scale among all possible clusterings not involving
any of the parent partons of the dipole-antenna under
consideration. This speeds up the algorithm by eliminating the time
spent generating trials in the region above
$Q_{S\mrm{min}}^{\mrm{unc}}$, none of which would be
accepted anyway. Lastly, if the preceding trial was rejected,
the restarting scale is the scale of that failed branching. 

Due to the simple structure of the trial Sudakov, \eqRef{eq:Sudakov}, solving 
\eqRef{eq:QofR} is straightforward, yielding solutions of the type
\cite{Giele:2011cb} 
\begin{equation}
x_{E2}^2 = (x_{E1}^2)^{R^{B'}}
\end{equation}
for a one-loop running trial $\hat{\alpha}_s$, with $x_E$ defined by
\eqRef{eq:xE}, and the exponents 
\begin{eqn}
B'_0 = \frac{4\pi b_0}{C_A I_{\zeta,0}(\zeta_{\mrm{min}}(Q^2_{E\mrm{min}}),\zeta_{\mrm{max}}(Q^2_{E\mrm{min}}))} ~,
\end{eqn}
\begin{eqn}
B'_{\pm} = \frac{4\pi b_0}{2 C_A I_{\zeta,\pm}(\zeta_{\mrm{min}}(Q^2_{E\mrm{min}}),\zeta_{\mrm{max}}(Q^2_{E\mrm{min}}))} ~,
\end{eqn}
for each of the trial-function types, respectively, while 
for a constant $\hat{\alpha}_s$, the solution is even simpler,
\begin{equation}
Q_{E2}^2 = Q_{E1}^2 R^{B}~, 
\end{equation}
with the exponents
\begin{eqn}
B_0 = \frac{4\pi}{\hat{\alpha}_s C_A I_{\zeta,0}(\zeta_{\mrm{min}}(Q^2_{E\mrm{min}}),\zeta_{\mrm{max}}(Q^2_{E\mrm{min}}))} ~,
\end{eqn}
\begin{eqn}
B_{\pm} = \frac{4\pi}{2\hat{\alpha}_s C_A I_{\zeta,\pm}(\zeta_{\mrm{min}}(Q^2_{E\mrm{min}}),\zeta_{\mrm{max}}(Q^2_{E\mrm{min}}))} ~.
\end{eqn}
Note that the coefficients $B_0$ and $B_0'$ for the global trial
function are identical to those 
denoted $b$ and $b'$ in \cite{Giele:2011cb}. We used capital letters 
here in order  not to confuse the exponents with the $b_0$ coefficient
used in the running of $\alpha_s$, \eqRef{eq:aS}. 

Given any set of branching variables
$(Q^2_E,\zeta)$ we may obtain the invariants $(s_{ij},s_{jk})$ without
ambiguity. Thus, the next step is to generate a random
$\zeta$ value distributed according to the integrand of the $I_\zeta$
integrals,
eqs.~(\ref{eq:zeta0int},\ref{eq:zeta-int},\ref{eq:zeta+int}). This is
done by solving the 
\begin{eqn}
R_\zeta =
\frac{I_\zeta(\zeta_{\mrm{min}},\zeta)}{I_\zeta(\zeta_{\mrm{min}},\zeta_{\mrm{max}})} 
\label{eq:zGeneration}
\end{eqn}
for $\zeta$, where $R_\zeta\in [0,1]$ is another uniform random number
and $\zeta_{\mrm{min}}(Q_{E\mrm{min}})$ is given by the 
evolution windows, \tabRef{tab:windows}, and by the $\zeta$ limits, 
\eqRef{eq:zetaminmaxfunction}.  

Following \cite{Giele:2011cb}, 
we solve eq.~(\ref{eq:zGeneration})  by first translating
to an auxiliary variable $r$, extending the treatment to cover also the
new $\zeta_\pm$ variables,
\begin{eqn}
  r_{0,\mrm{max}} = \frac{1}{1-\zeta_{0,\mrm{max}}}~~~,~~~
  r_{0,\mrm{min}} = \frac{1}{1-\zeta_{0,\mrm{min}}}~~~,
\end{eqn}
\begin{eqn}
  r_{\pm,\mrm{max}} = \frac{1}{1-\zeta_{\pm,\mrm{max}}}~~~,~~~
  r_{\pm,\mrm{min}} = \frac{1}{1-\zeta_{\pm,\mrm{min}}}~~~;
\end{eqn}
we then generate a random value for $r$
\begin{eqn}
  r = r_{\mrm{min}} \left(\frac{r_{\mrm{max}}}{r_{\mrm{min}}}\right)^{R_\zeta}~,
\end{eqn}
and finally solve for $\zeta$,
\begin{eqn}
\zeta_0 = \frac{r_0}{1+r_0} ~,
\end{eqn}
\begin{eqn}
\zeta_{\pm} = 1-\frac{1}{r_{\pm}} ~.
\end{eqn}

If the $\zeta$ generated in this way falls outside the physical phase
space, 
\begin{eqn}
\zeta < \zeta_{\mrm{min}}(Q^2_E) ~~~\vee~~~
\zeta > \zeta_{\mrm{max}}(Q^2_E)
\end{eqn}
the branching is vetoed and a new one generated, with $Q_E$ as restart
scale.

If the branching is inside the physical phase-space, the next step is
to obtain values for the pair of phase-space
invariants $(s_{ij},s_{jk})$ in terms of which we cast the original
evolution equation, eq.~(\ref{eq:ev1}). We quote here the relevant inversions:
\begin{itemize} 
\item{
for $\zeta_0$
\begin{eqn}
s_{ij} \ = \ \frac{Q_E \sqrt{s} \sqrt{\zeta_0}}{2 \sqrt{1-\zeta_0}}~~~~~~;~~~~~~s_{jk}\ = \ \frac{Q_E \sqrt{s} \sqrt{1-\zeta_0}}{2 \sqrt{\zeta_0}} 
\end{eqn}
}
\item{
for $\zeta_-$
\begin{eqn}
s_{ij} \ = \  s \ \zeta_- ~~~~~~;~~~~~~s_{jk}\ = \ \frac{Q_E^2}{4 \zeta_-} 
\end{eqn}
}
\item{
for $\zeta_+$
\begin{eqn}
s_{ij} \ = \ \frac{Q_E^2}{4 \zeta_+} ~~~~~~;~~~~~~s_{jk}\ = \ s \ \zeta_+ 
\end{eqn}
}
\end{itemize}
Finally, the full kinematics (4-momenta) for the trial branching can
be constructed, from the explicit formulae given in
\cite{Giele:2007di,massive}. The last step is to check the sector
veto, i.e., whether the sector represented by partons $ijk$ has the
smallest value of $Q_S$ in the tentative $(n+1)$-parton momentum
configuration that would arise if the branching is accepted. 
If not, the trial is rejected and a new one generated
starting from $Q_E$. 

To obtain an LL shower from the trial branchings generated according
to the expressions above, it suffices to
accept each trial branching with a probability 
\begin{eqn}
P^{LL} = \frac{\alpha_s}{\hat{\alpha}_s}  
                \frac{{\cal C}_{ijk}}{\hat{\cal C}_{ijk}}
\frac{\abar^\mrm{sct}_\mrm{LL}(s,s_{ij},s_{jk})}{\abartrial(s,s_{ij},s_{jk})}~,
\label{eq:PLL}
\end{eqn}
where the $\alpha_s/\hat{\alpha}_s$ ratio takes into account the
possibility that the trial generator could be using a nominally larger
$\alpha_s$ than the physically desired one, 
the ${\cal C}/\hat{\cal C}$ factor represents the same for color factors, 
and the antenna function ratio matches the trial function onto
the desired physical splitting antenna for the relevant $2\to 3$
branching.  
We must also require $\abar_\mrm{LL}$ to be non-negative in order that the ratio
here be interpretable as probability. If the branching is accepted,
partons $I$ and $K$ are replaced by partons $i$, $j$, and $k$ and the
evolution is restarted as discussed previously. 

\subsection{$(2\to n)$: Unitary Matrix-Element Corrections} \label{2ton_matching}

Briefly summarized, the GKS strategy \cite{Giele:2011cb} for matching
to leading-order 
matrix elements  is as follows. Similarly to the
\Py~\cite{Bengtsson:1986hr} and \textsc{Geneva}~\cite{Bauer:2008qj}
approaches, the \Vc matching formalism 
relies on the antenna shower itself to provide an all-orders phase-space
generator that  captures the leading
behavior of full QCD by construction. At each trial branching in the shower, the
accept/reject probability can then be augmented by a multiplicative factor
that goes to unity in the collinear and soft limit, but which 
modifies the branching probability outside those limits. The
modification factor for global showers is given in
\cite{Giele:2011cb}. Since only a single path contributes to each
phase-space point in the sector case, the corresponding matching
factor is simpler, and is given by
\begin{equation}
P^{\mrm{sct}}_{\mrm{ME}}(\{p\}_n) = \frac{|M_{n}
  (\{p\}_n)|^2}
 {g_s^2 {\cal C}_{j/IK} \ \bar{a}^\mrm{sct}_{\mrm{LL}}(p_i,p_j,p_k) 
   |M_{n-1}(\{\hat{p}\}_{n-1})|^2}~,
\end{equation}
with post- and pre-branching parton configurations denoted by
\begin{equation}
\{p\}_n = (p_1,\ldots,p_i,p_j,p_k,\ldots) ~~~\mbox{and}~~~
\{\hat{p}\}_{n-1} = (p_1,\ldots,p_I,p_K,\ldots)~,
\end{equation}
respectively. The $P_{\mrm{ME}}$ factor is thus 
constructed precisely such that the shower
approximation is matched (up or down) to the LO matrix-element squared
at each order. The prescription to include full-color matrix
elements, by scaling the expression above by the ratio of
color-summed full- to leading-color  matrix elements squared, is not 
modified from the global case as given in \cite{Giele:2011cb}. 

Note also that since $P_{\mrm{ME}}$ multiplies the trial-accept
probability, \eqRef{eq:PLL}, the factor $\bar{a}^\mrm{sct}_{\mrm{LL}}$ actually
cancels in the product, leaving no trace of the LL antenna function in
the final answer. The color-ordered 
matrix elements themselves instead act as the $2\to n$ sector-antenna
functions, up to the matched orders.

The approach relies heavily on unitarity and is qualitatively
different from other multi-leg approaches in the literature,   
such as the MLM~(see \cite{Alwall:2007fs} for a description) and
CKKW~\cite{Catani:2001cc} ones. An important technical
difference is that \Vc only requires a Born-level phase-space
generator, with all higher multiplicities being generated by
the shower. There is therefore no need for separate phase-space
generators for the higher-multiplicity matrix elements, which 
can result in significant speed gains, both in terms of initialization
time (virtually zero in \Vc), and in terms of running speed. We return
to this issue in \secRef{sec:results} below.
We refer the reader to \cite{Giele:2011cb} for further details on the
GKS formalism. 

\section{Results \label{sec:results}}

\begin{figure}[t]
\centering
\includegraphics*[scale=0.4]{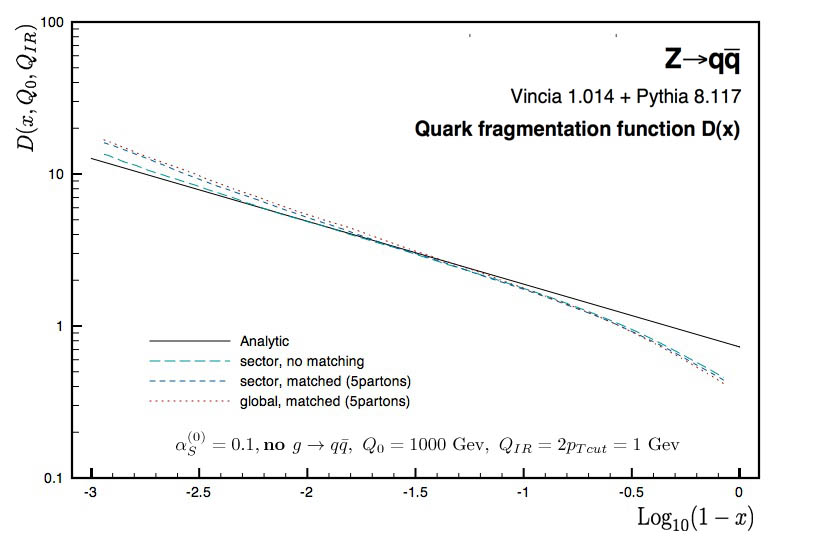}
\caption{The quark fragmentation function, $D(x)$, in hadronic $Z$
  decays. Comparison of 
  an analytic LL resummation \cite{Skands:2009tb} (solid line) to 
  \Vc with sector showers, without (long dashes) and with (short
  dashes) matching through $Z\to 5$ partons, and to the default
  (matched) global shower in \Vc (dots).\label{fig:plotFF_filled}}
\end{figure}
In addition to the LO matrix-element comparisons given in \secRef{sec:tests}, 
we have performed two basic tests of the all-orders sector shower
implementation in \Vc\ interfaced to \Py~8. 
First, we compare results obtained with just the perturbative \Vc
shower (i.e., without switching on \Py's hadronization model) 
to a leading-logarithmic analytical resummation of the quark
fragmentation function, similarly 
to what was done in \cite{Skands:2009tb}. 
We recall that the energy fraction is defined as
\begin{equation}
x=\frac{2E_q}{\sqrt{s}}~.
\end{equation}
We use a constant value of $\alpha_s = 0.1$, a starting scale of
$\sqrt{s}=1000\,$GeV, and an ending scale of
$Q_{IR}=1\,$GeV, for a perturbative evolution spanning three orders of
magnitude in $x$. 
This comparison is shown in
\FigRef{fig:plotFF_filled}, with and without matching, and also
compared to the default (matched) global result, as a function of
$\log_{10}(1-x)$ on the $x$ axis. The region on the right-hand side of the plot,
$x \to 0$, is dominated by hard emissions and is not expected to be
well reproduced by the analytical soft resummation. Likewise, 
energy-momentum conservation effects are important, included in \Vc
but neglected in the analytical
resummation. It is therefore not
surprising the analytical
calculation differs from all  of the \Vc ones in that region. On the
left half of the plot, soft emissions dominate. One observes 
that the unmatched sector shower 
is quite close to the analytical result. The matching correction
actually increases the difference slightly, which we interpret as due
to our matching corrections being applied also in the soft
region. The difference is consistent with similar variations
observed by varying the LL finite terms in \cite{Skands:2009tb}. One
also notes that the two matched calculations (global and sector) are
consistent with each other.

\begin{figure}[t]
\centering
\includegraphics*[scale=0.37]{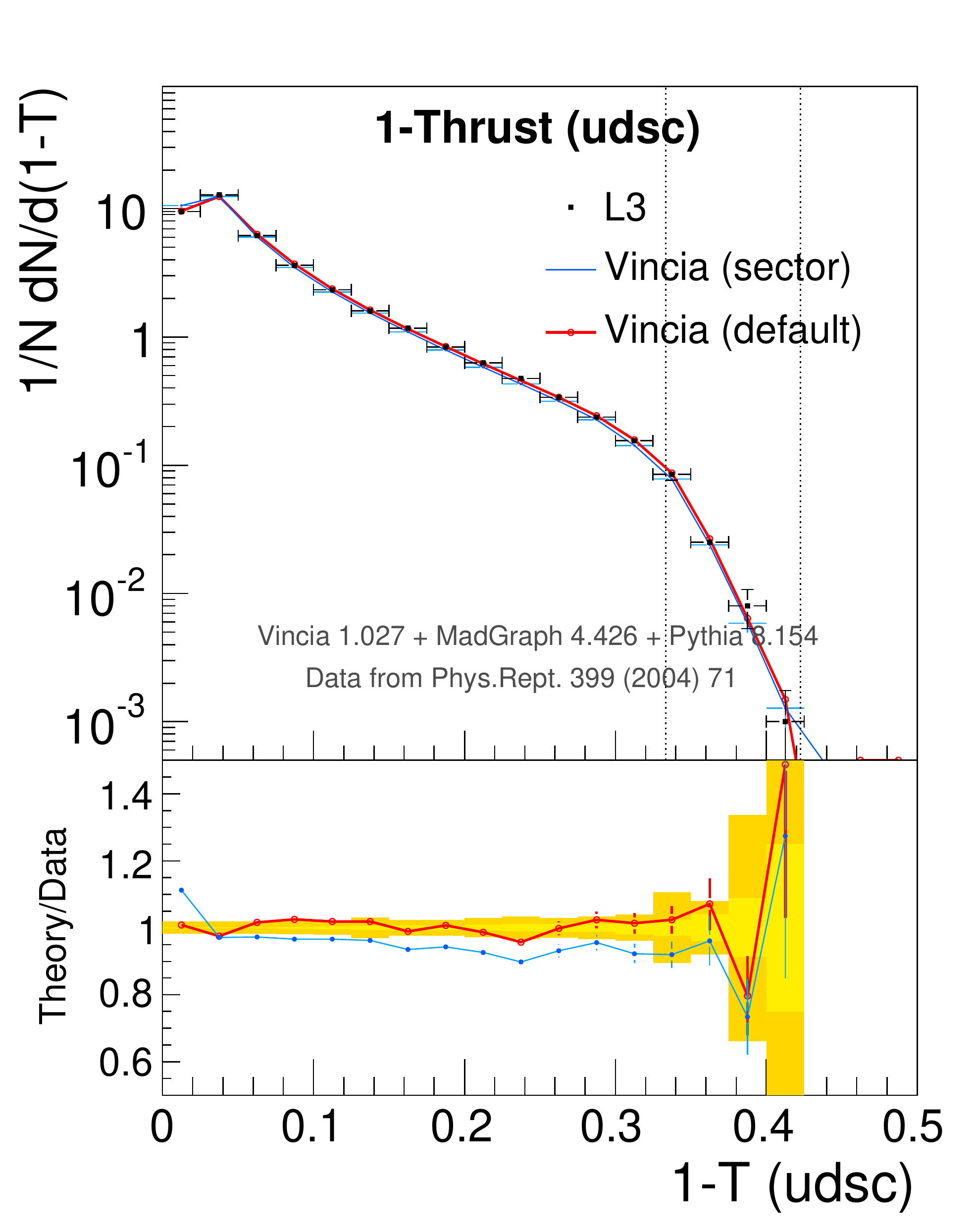}
\includegraphics*[scale=0.37]{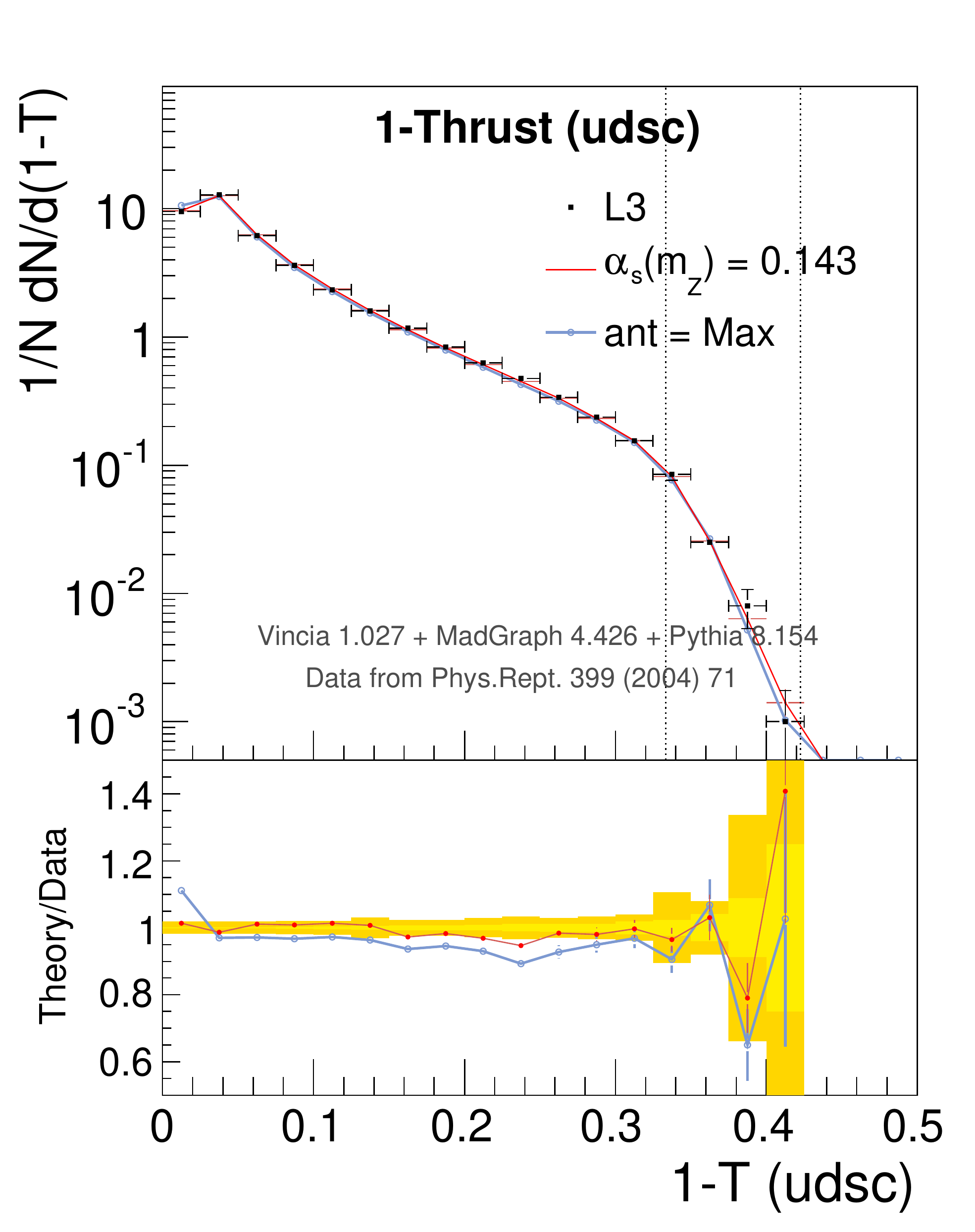}
\caption{Normalized Thrust ($1-T$) distribution. \Vc compared to L3 data for
  light-flavor $Z$ decays \cite{Achard:2004sv}. {\sl Left:} sector (thin)
  vs.\ global (thick) showers, using default (global) \Vc tune. {\sl
    Right:} sector shower using $\alpha_s(m_Z)=0.143$ (thin)
  vs.\ ``Max'' antenna functions \cite{Ridder:2011dm} (thick). 
\label{fig:t}}
\end{figure}

\begin{figure}[t]
\centering
\includegraphics*[scale=0.37]{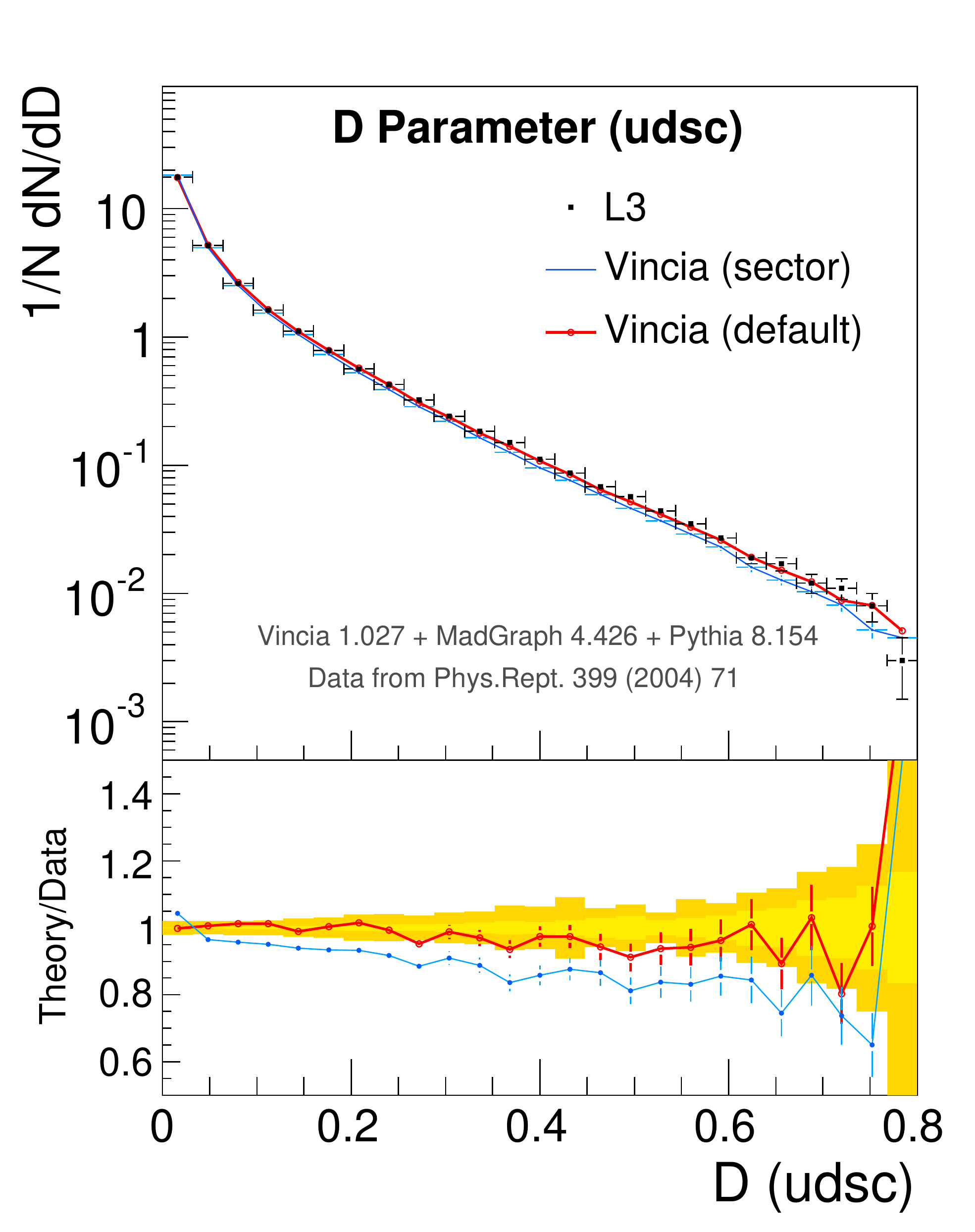}
\includegraphics*[scale=0.37]{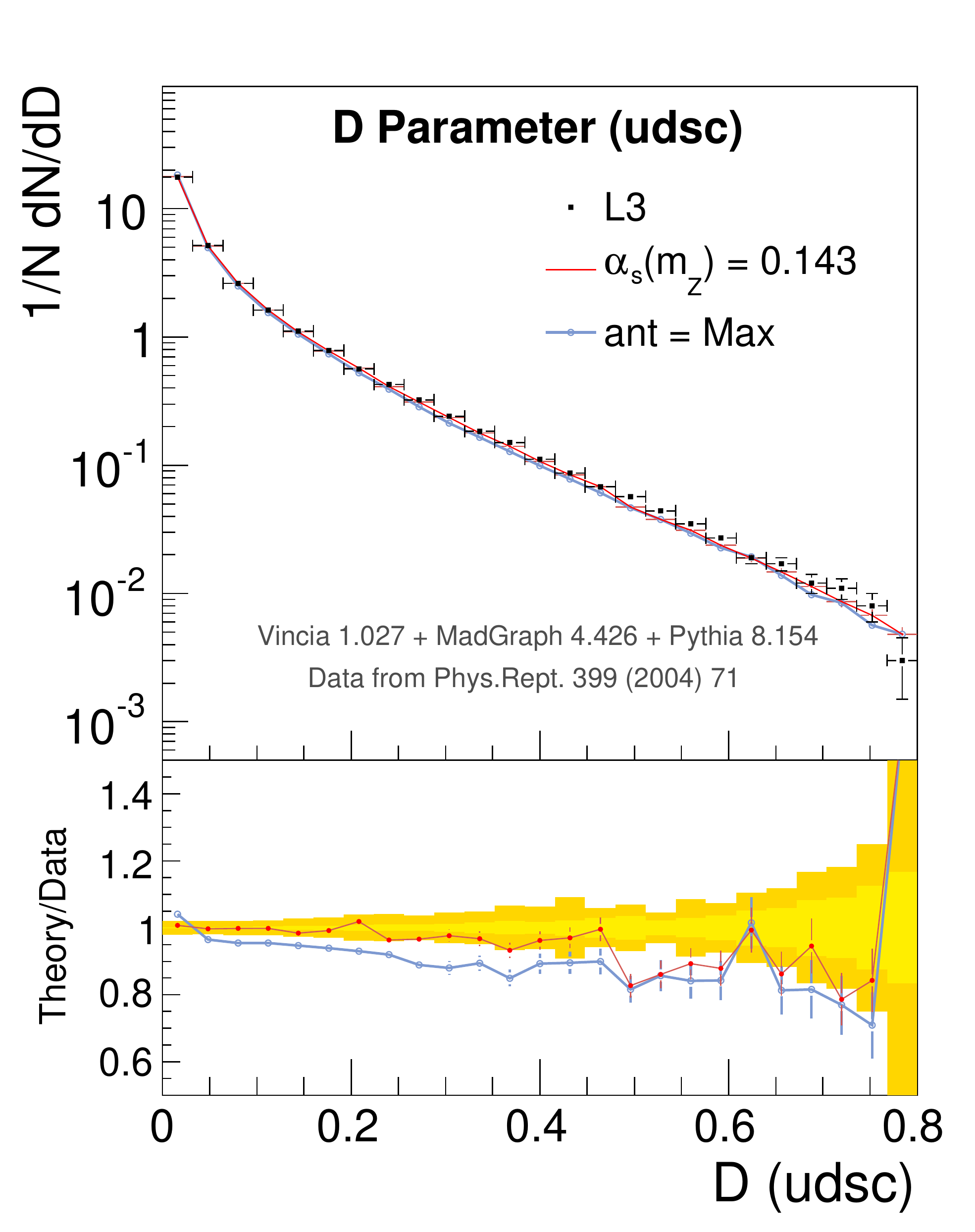}
\caption{Normalized $D$-parameter distribution. \Vc compared to L3 data for
  light-flavor $Z$ decays \cite{Achard:2004sv}. {\sl Left:} sector (thin)
  vs.\ global (thick) showers, using default (global) \Vc tune. {\sl
    Right:} sector shower using $\alpha_s(m_Z)=0.143$ (thin)
  vs.\ ``Max'' antenna functions \cite{Ridder:2011dm} (thick). \label{fig:d}}
\end{figure}

As a second cross check, we include some comparisons to LEP
event-shape data at $\sqrt{s} = m_Z$, using light-flavor ($udsc$) data
taken by the L3 collaboration \cite{Achard:2004sv}. 
In all cases, we include tree-level matching through $Z\to 5$ partons,
the default in \Vc. Between 1 and 2 million unweighted 
events were generated for each generator setting.
These comparisons necessarily
include the effects of hadronization. We have not attempted to do a
full-fledged tuning of \Py's non-perturbative hadronization
parameters for use with the sector shower. Instead, the default \Vc
tune (summarized in Appendix \ref{app:tune}) is used, with the same
value of the infrared cutoff (1 GeV) as in the global case.  

With this setup, replacing the default global shower by 
the sector one with antenna functions as defined in this paper, 
we find that the sector shower 
produces slightly softer event shapes than the global one. A first
illustration of this 
is given in the left-hand panes of \figsRef{fig:t} and \ref{fig:d}, 
in which we compare the sector and global
shower implementations in \Vc to measurements of the Thrust and
$D$-parameter event-shape variables, which arise at ${\cal O}(\alpha_s)$
and ${\cal O}(\alpha_s^2)$ respectively (see \cite{Achard:2004sv} for
a definition). For reference, 
the $C$ parameter,
qualitatively similar to Thrust, is included
in Appendix \ref{app:plots}, as are the Wide and Total Jet Broadening
parameters.   
The result obtained with the sector shower is shown with
thin (blue) lines, the global one with thick (red) lines. The
upper pane of each plot shows the normalized event-shape distribution
and the lower pane the ratio of the calculations to data.  

In all the event shapes, the sector shower peaks 
at lower values than the corresponding global distribution.  
Since both showers include matching through $Z\to 5$ partons, 
their tree-level expansions are equal up to the
first three orders in $\alpha_s$. We therefore do not believe
finite-term contributions alone could be responsible for the apparent 
 ``softness'' of the sector shower relative to the global one. 
This conclusion is corroborated by
the line labeled ``ant=Max'' in the right-hand pane of the figures 
(thick blue line), 
for which we replaced the sector antenna functions defined in this
paper by the ``Max'' ones defined in \cite{Ridder:2011dm}, which have
large finite terms;  the result can be seen not to vary substantially
from the sector curve shown in the left-hand panes, indicating that it
is stable under finite-term variations.

Tentatively, our conclusion is that the difference between the
distributions produced by the two shower models owes to a difference
between the perturbative corrections generated beyond tree level, such
as their $\alpha_s$ choices and Sudakov form factors. To 
illustrate this, the thin (red) curves in the right-hand panes of 
\figsRef{fig:t} and \ref{fig:d}, labeled $\alpha_s(m_Z)=0.143$, 
show what happens if the value used to define the 1-loop
running coupling in \Vc is changed from 0.139 to 0.143. (Note that
these values should be interpreted in an LO scheme defined by \Vc,
hence they are not immediately interpretable as, e.g.,
$\overline{\mbox{MS}}$ values.) This corresponds to
a change in the 5-flavor value of $\Lambda_{\mrm{QCD}}$ 
from $\sim250\,$MeV to $\sim295\,$MeV and is sufficient to bring the sector shower
into agreement with the result obtained with the global one. 
We note that this change may be connected with the question of 
whether it is still correct to use 
$p_\perp$ as the argument for $\alpha_s$ in the sector shower, an
issue we plan to return to in the context of a separate study \cite{lisa}. 

\begin{figure}[tp]
\centering
\subfloat{\includegraphics*[scale=0.37]{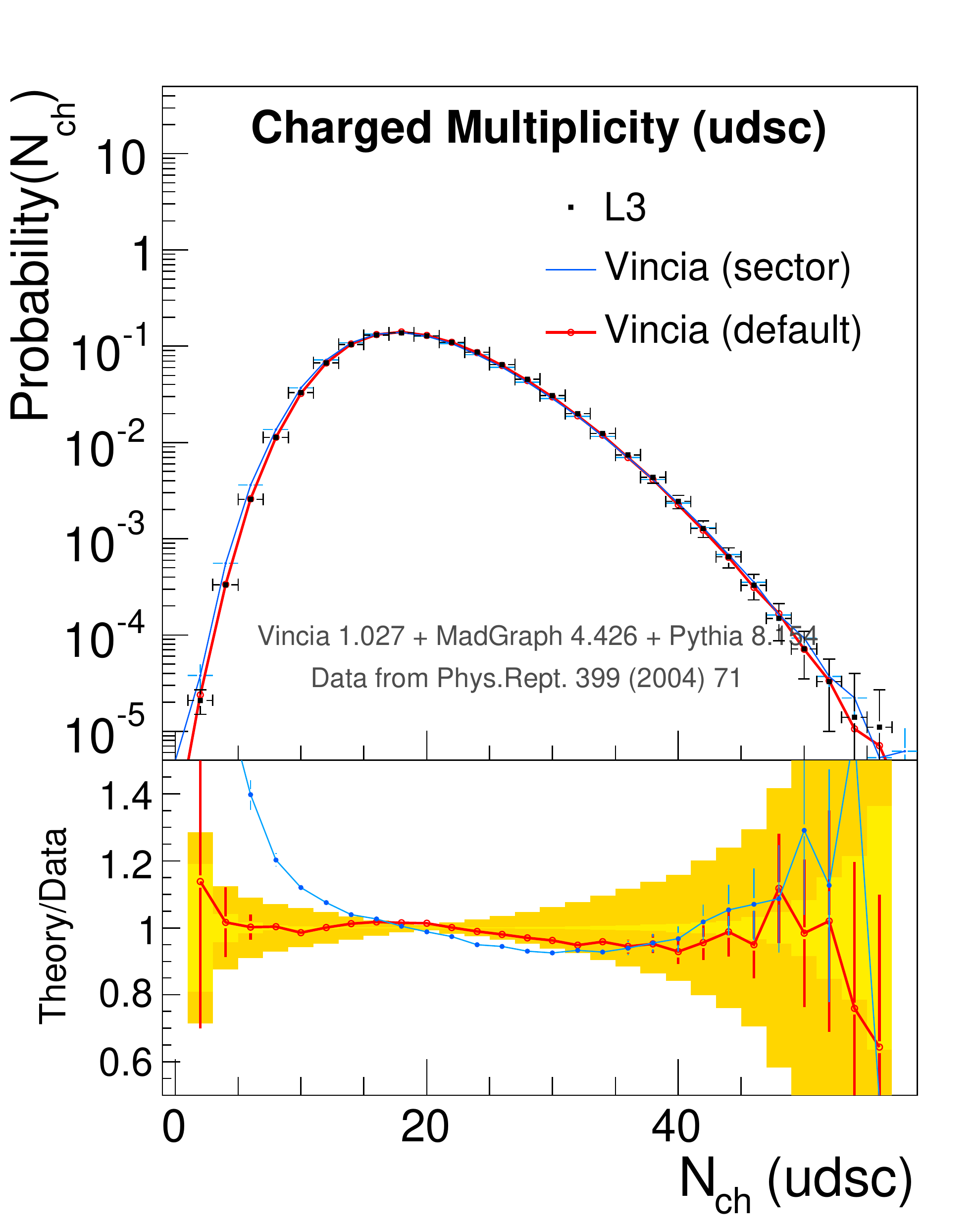}}
\subfloat{\includegraphics*[scale=0.37]{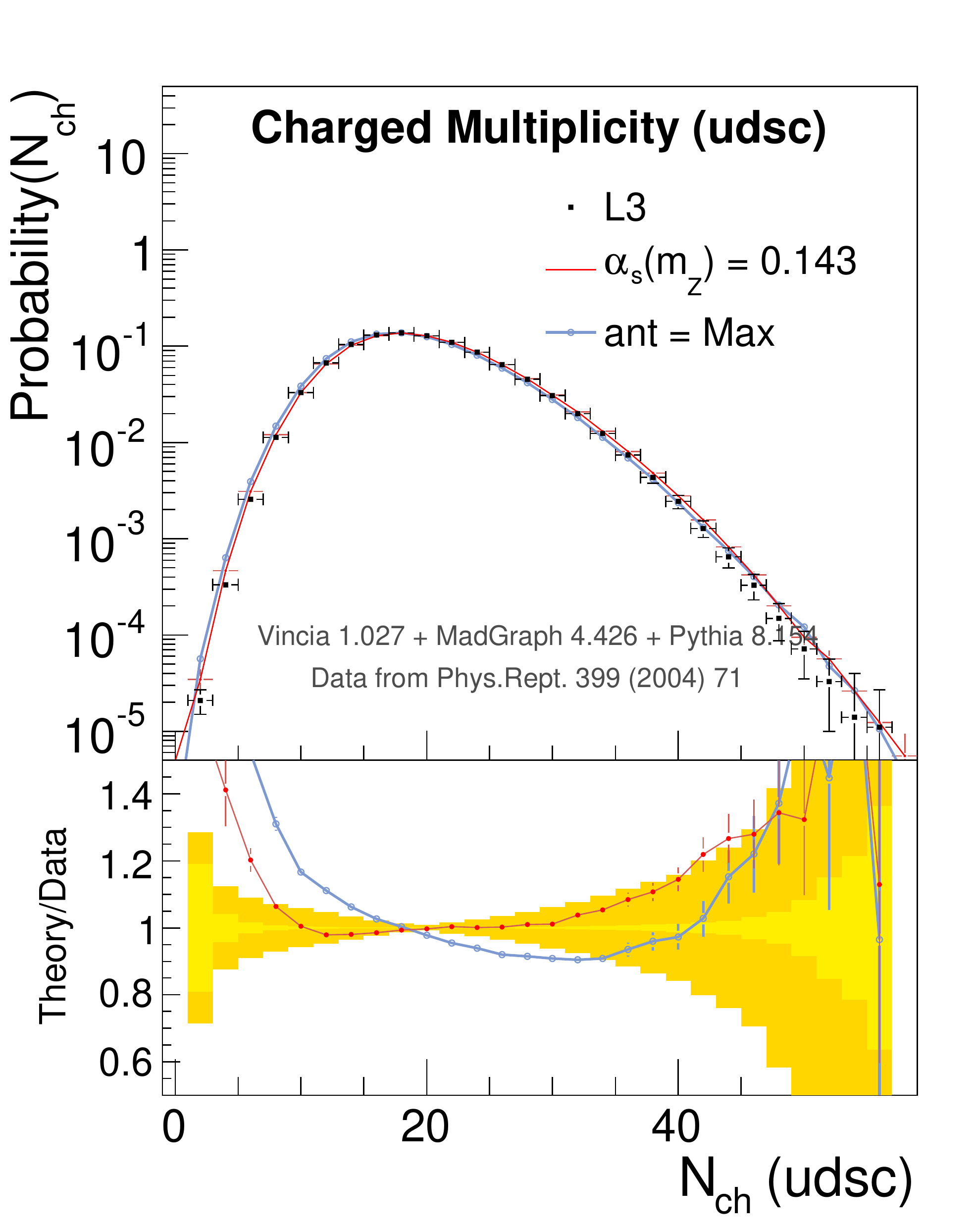}}\\
\subfloat{\includegraphics*[scale=0.37]{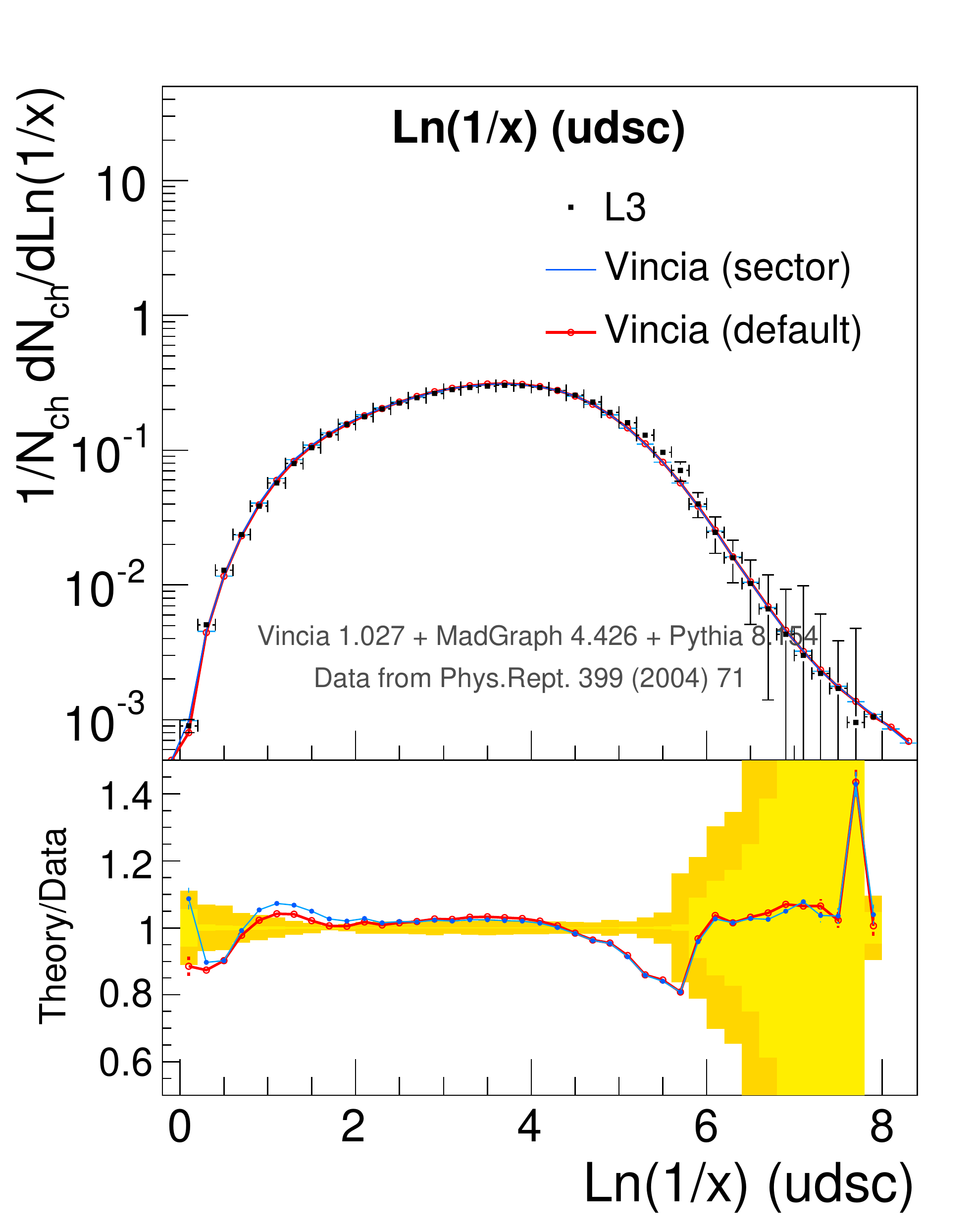}}
\subfloat{\includegraphics*[scale=0.37]{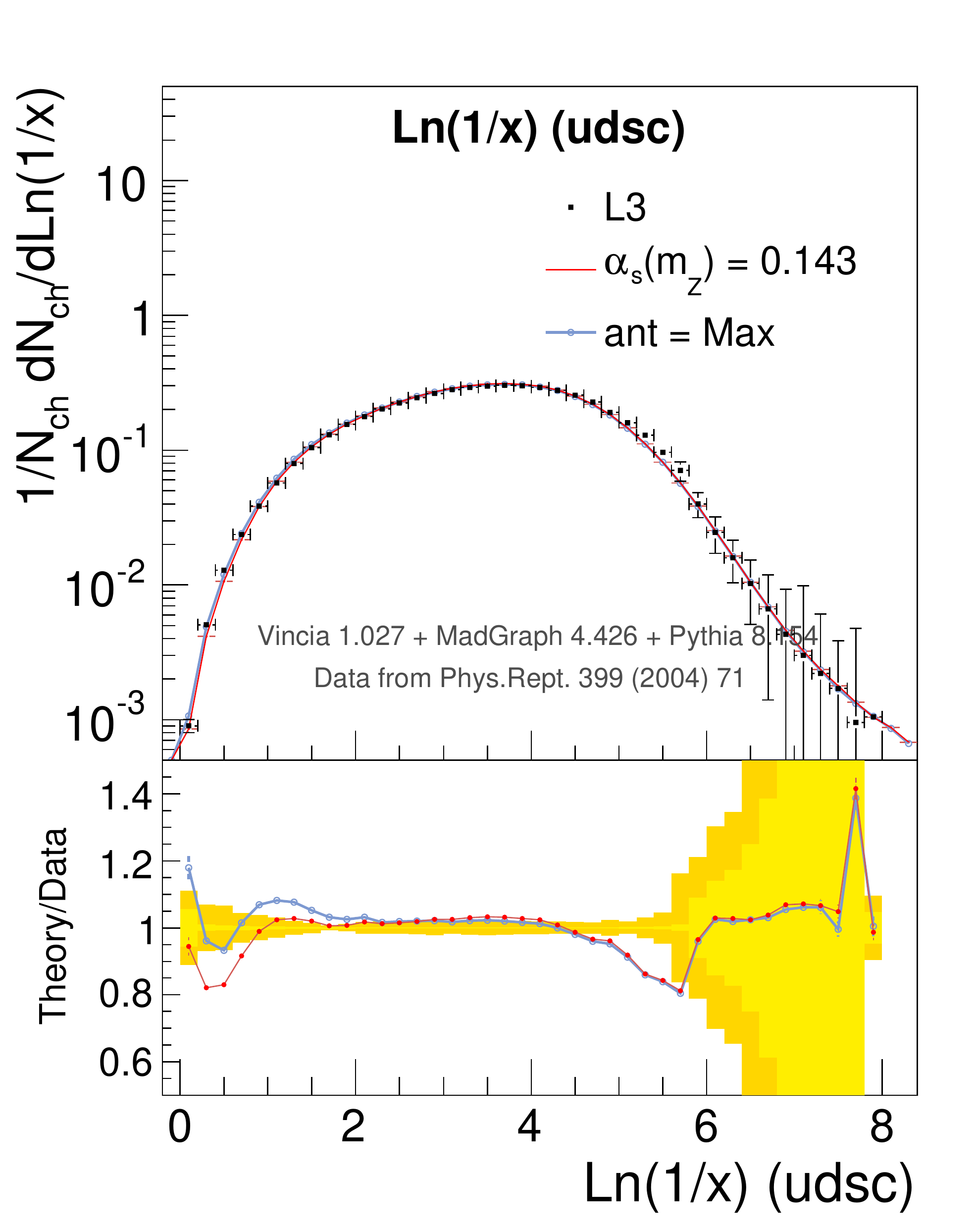}}
\caption{Normalized charged-particle-mutiplicity (top)
  and momentum-fraction (bottom) distributions, with the latter using 
  $x=x_p=2|p|/\sqrt{s}$. \Vc
  compared to L3 data for 
  light-flavor $Z$ decays \cite{Achard:2004sv}. {\sl Left:} sector (thin)
  vs.\ global (thick) showers, using default (global) \Vc tune. {\sl
    Right:} sector shower using $\alpha_s(m_Z)=0.143$ (thin)
  vs.\ ``Max'' antenna functions \cite{Ridder:2011dm} (thick).
\label{fig:np}}
\end{figure}
At the non-perturbative level, the intrinsic softness of the sector
shower also has consequences, as illustrated in the left-hand panes of
\figRef{fig:np}, where we compare to the distributions of the number
 (top row) and  momentum fraction (bottom row) of charged particles,
with $x=2|p|/\sqrt{s}$. The sector shower defined in this paper (thin
blue lines) produces a wider multiplicity distribution than the global
one, with slightly more particles having $x\sim 1$. Again, the
right-hand panes illustrate what happens when choosing larger finite
terms (thick curve) and when choosing a larger $\alpha_s(m_Z)$ value
(thin lines). Similarly to above, the variation of antenna-function finite terms
does not lead to substantial differences, while changing the value of
$\alpha_s$ does. It is evident that the sector shower, even with
$\alpha_s(m_Z)=0.143$, could benefit from a slight retune of its
non-perturbative parameters, e.g., to suppress the slightly overpopulated 
tails of low- and high-multiplicity events.

Further event-shape comparisons, and the production ratios of certain
meson and baryon species, normalized to the average charged
multiplicity, are given in Appendix \ref{app:plots}.

\begin{table}[t]
\centering
\includegraphics*[scale=0.4]{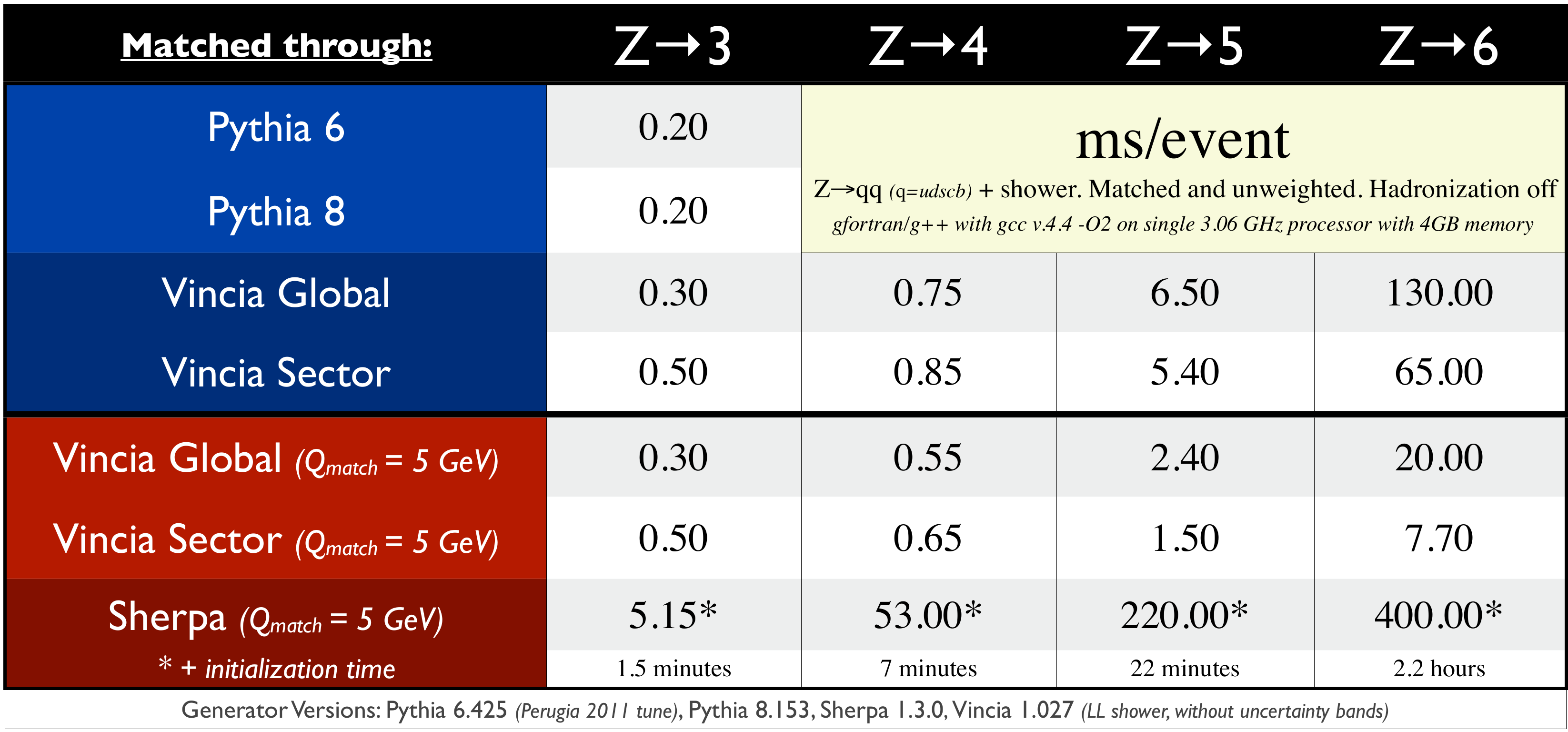}
\caption{Comparison of the dependence of generator speed (measured in
  milliseconds per shower under standardized circumstances, see text) 
  on the number of matched orders for \Py~6, \Py~8, \Vc~(global and
  sector, with and without a ``matching scale''), and 
  \Sh~\cite{Gleisberg:2008ta}.
\label{fig:speed}}
\end{table}
As emphasized in the preceding sections, one of the chief advantages
of the sector shower approach is the fact that it only generates a
single contributing term per phase-space point, for each order in
perturbation theory. This makes it ideally suited for the GKS matching
strategy \cite{Giele:2011cb}, which requires at least one
matrix-element-evaluation per contributing shower path. In
\tabRef{fig:speed}, we compare the number of milliseconds it takes to
generate one event, between various programs and matching algorithms, using a
standardized set-up. Since we are not interested in the speed of
hadronization or hadron decay algorithms, we leave hadronization
switched off and define an ``event'' as a 
perturbative cascade  starting from a $q\bar{q}$ dipole at
$\sqrt{s}=m_Z$ (with $q=udscb$ and using massive matrix elements for $b$
quarks), evolved down to the default perturbative cutoff scale in the
respective code, which is of order 1 GeV in all cases. 

In \Py~6~\cite{Sjostrand:2006za} and 8~\cite{Sjostrand:2007gs}, 
only matching to the $Z\to3$ decay matrix element is
available, hence speeds for higher multiplicity-matching are not
shown in the table. We note that, at least in this test, \Py~8 generates events as
fast as \Py~6. 

The default (global, smoothly ordered) shower implementation in \Vc\
(third row) is slightly slower than the one in \Py, which we regard as
a consequence of the greater flexibility built into \Vc (smooth
ordering, with generic evolution variables, antenna functions, and
kinematics maps) combined with
its more elaborate 
matching setup (taking explicit ratios of leading- and full-color
\Mg\ matrix elements \cite{Alwall:2007st} and evaluating them using
\textsc{Helas}~\cite{Murayama:1992gi} routines). Using the GKS 
matching, however, \Vc's matching may be extended to higher
multiplicities, shown in the third to sixth columns. (All comparisons
use matching to full-color matrix elements.) Since the number
of matrix-element evaluations grows linearly with the multiplicity in 
the global approach, we see that the speed decreases quite
rapidly with multiplicity. 

In the sector approach, on the other hand
(fourth row), the perturbative evolution is slightly slower at low
matched multiplicities (basically since each gluon now requires two separate trial
functions, and since no optimization of the trial phase-space has yet
been implemented, relative to the global case), 
but the increase with multiplicity is less severe,
resulting in the sector approach being faster than the global one
starting from matching through $Z\to 5$. 

In order to compare directly with other multileg approaches, such as
the CKKW one \cite{Catani:2001cc} implemented in
\Sh~\cite{Gleisberg:2008ta}, we have included the optional possibility
to stop applying matching corrections below a specific value of the
evolution scale in \Vc, thus emulating the ``matching scale'' that is
present in the CKKW strategy. This obviously speeds up the calculation
somewhat, since lots of soft emissions no longer need to have matching
coefficients evaluated. For both \Vc\ and \Sh, we set the matching
scale equal to 5 GeV, ignoring that the 
phase-space contours defined by that value do not exactly match
between the two codes. In the comparison to \Sh, it is furthermore necessary
to divide the total event-generation time up on a non-negligible 
initialization stage and a subsequent per-event time. In both \Py\ and
\Vc, the initialization time is essentially zero, while it grows
 with final-state complexity in \Sh, due to the necessity
of computing cross sections and initializing phase-space generators
for each matrix-element configuration separately. 
The total event-generation time for
\Sh~(bottom row in \tabRef{fig:speed}) is therefore divided up on two
numbers, with the initialization duration reported separately below
the main per-event time. Even if one  neglects the initialization
component, however, it is clear that there is a significant speed
difference between the two methods. Including the initialization time,
the differences become even more pronounced. For example, 
during the 22 minutes it takes to initialize the CKKW generator for
matching through $Z\to 5$ partons (the default matching level in \Vc), 
the GKS implementation has time to generate almost 1 million
matched showers. We stress that \Sh\ is still obviously a much more
versatile tool than \Vc, and hence this comparison is not intended as
an advertisement for one code over another, rather its purpose 
is to test the dependence of the algorithmic speed on multiplicity, 
of the two matching prescriptions implemented in the respective
generators. Note also that, while \Vc\ currently relies on \Mg\ and
\textsc{Helas} for its matrix elements, \Sh\ uses the 
\textsc{Comix} generator \cite{Gleisberg:2008fv}. 
We did not attempt to calibrate for this difference in this
comparison, hence it is possible that the 
\Vc\ numbers, in particular for high matched multiplicities, 
could be reduced somewhat by implementing a faster
matrix-element method.

Note also that the \emph{relative} increase in per-event time for
\Sh\ actually becomes smaller with multiplicity. For instance, the 
per-event time only increases by a factor 2 when going from 5 to 6
partons, compared to factors of 4 and 10 at each of the preceding
orders, respectively. We interpret this as being due to the
fact that the corresponding cross sections, for $n$
exclusively\footnote{For the highest matched multiplicity, replace 
exclusively by inclusively.} resolved partons
above the matching scale, 
are becoming increasingly small. Thus, once e.g.\ the 6-parton 
cross section has been computed (during initialization, the time for which
still increases by an order of magnitude from 5 to 6 partons), the
time to actually \emph{generate} additional events does not increase
substantially. This is very different from \Vc, in which the
initialization time (zero) does not increase substantially, 
but the per-event time does. 
  
\clearpage
\section{Conclusions \label{sec:conclusion}}

We have presented a formalism for parton
showers based on sector antennae, accompanied by an implementation in the \Vc
plug-in \cite{Giele:2007di} to the \Py~8 event generator \cite{Sjostrand:2007gs}. 
The main distinguishing feature of
such showers is that only a single radiation antenna contributes to each
phase-space point, as compared to a sum over all radiators in
traditional ``global'' showers
\cite{Gustafson:1987rq,Giele:2007di,Winter:2007ye,Buckley:2011ms}.  
The coefficients of the single poles of gluon antennae
are modified to reflect this reorganization. A similar formalism
including mass and polarization corrections has been developed in 
\cite{Larkoski:2009ah,Larkoski:2011fd}, but has not yet been
implemented in a publicly available event generator. 

At the analytical level, we have tested the formalism 
by comparing tree-level expansions of it to fixed-order matrix
elements for $Z\to 4$, $5$, and $6$ partons. We find that 
the global shower, with its many terms, is able
to deliver a somewhat better average agreement at the multileg level, in
particular for processes involving $g\to q\bar{q}$ splittings. 
To our minds, the advantage of the sector approach is therefore
at present mainly a computational one, to be sought in the
consequences of its simpler structure. 
Since the sector shower only produces a single term per
phase-space point, it gives a speed advantage over the global approach
when combined with the ``GKS'' matching formalism developed in
\cite{Giele:2011cb}, which requires at least one matrix-element-evaluation per
contributing path. We demonstrate this speed gain by comparing global
and sector showers matched to LO matrix elements through up to four
branchings in \Vc. 
For reference, we also compare to an implementation of the CKKW method
for multileg matching \cite{Catani:2001cc}, using the \Sh~generator
\cite{Gleisberg:2008ta}. 

As a final cross-check, we have also compared the sector shower
implementation in \Vc, with and without matching, to an analytic
resummation of the quark fragmentation function and to experimental
measurements of event shapes and related quantities at LEP. We find
that the present sector shower implementation appears to be consistent
with these distributions, within the expected precision, and hence
consider it validated and ready to be used for other phenomenology
studies. 

Nevertheless, since the sector shower a priori produces slightly 
more particles with $x\to 1$ and somewhat softer event-shape
distributions, we recommend to increase the value of $\alpha_s(m_Z)$ from
0.139 in the default tune to $\sim 0.143$ for use with the sector
shower. This results in good agreement with event shapes but still
generates a slightly too broad charged-particle multiplicity
distribution. Depending on the application, 
a further iteration of the non-perturbative tuning, focusing
specifically on sector showers, could therefore also 
be interesting to explore.  

\subsection*{Acknowledgments}
We thank A.~Gehrmann-de-Ridder, W.~Giele, D.~Kosower, A.~Larkoski, 
M.~Peskin, M.~Ritzmann, and J.~Winter, for many
enjoyable discussions on the singularity structure of antenna
showers. We thank D.~Kosower in particular for 
pointing out the subtlety connected with the choice of sector
decomposition variable for gluon splittings and
  for suggesting the modification  necessary to cure it.

This work was supported in part by the Marie Curie FP6 research training
network ``MCnet'' (contract number
MRTN-CT-2006-035606), as well as by the European Commission (HPRN-CT- 200-00148), FPA2009-09017 (DGI del MCyT, Spain) and S2009ESP-1473 (CA Madrid). J.J. L-V is supported by a MEC grant, AP2007-00385, and wants to thank the CERN Theory Division for its hospitality.

\appendix
\section{Additional LEP Comparisons \label{app:plots}}
\begin{figure}[t]
\centering
\includegraphics*[scale=0.37]{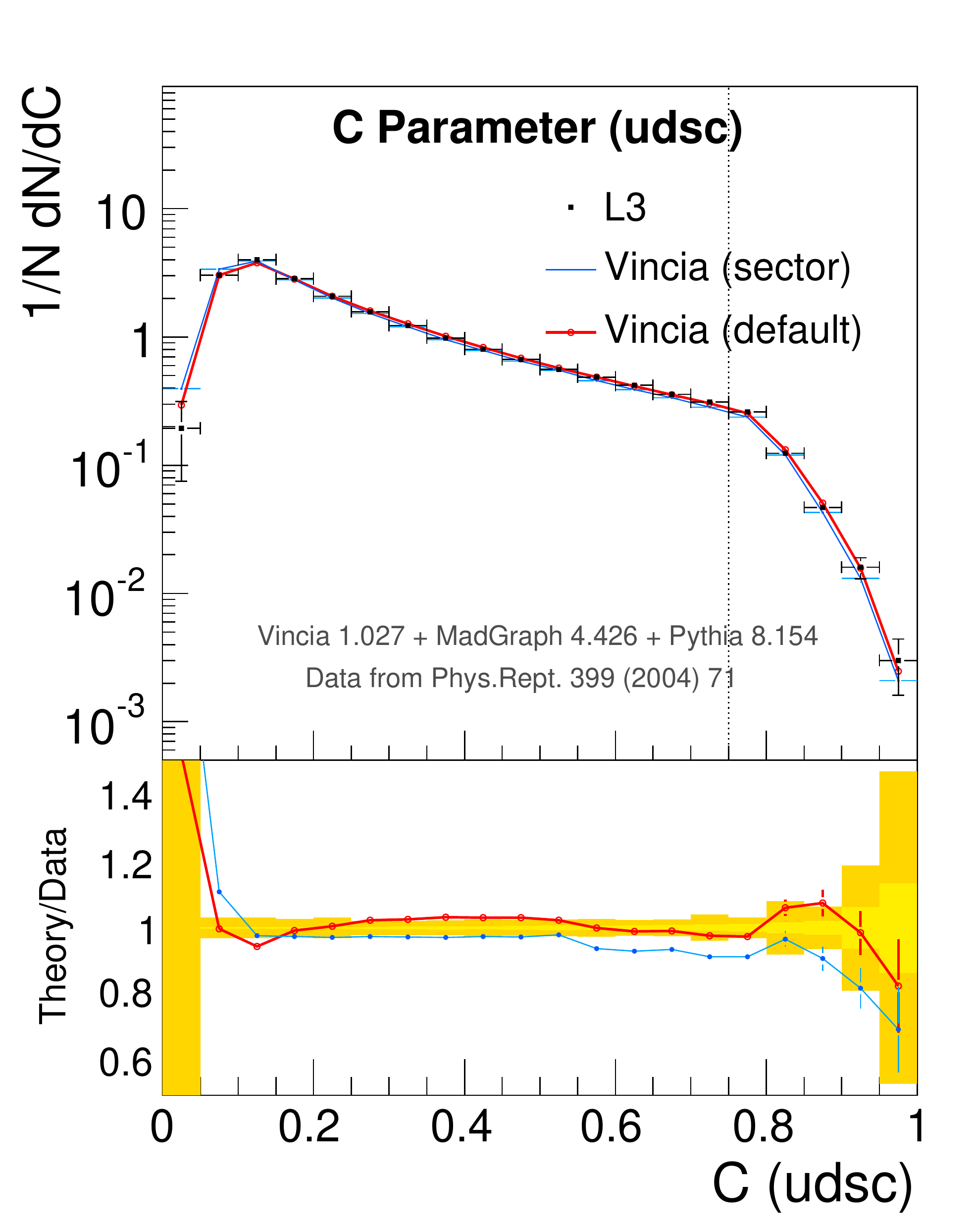}
\includegraphics*[scale=0.37]{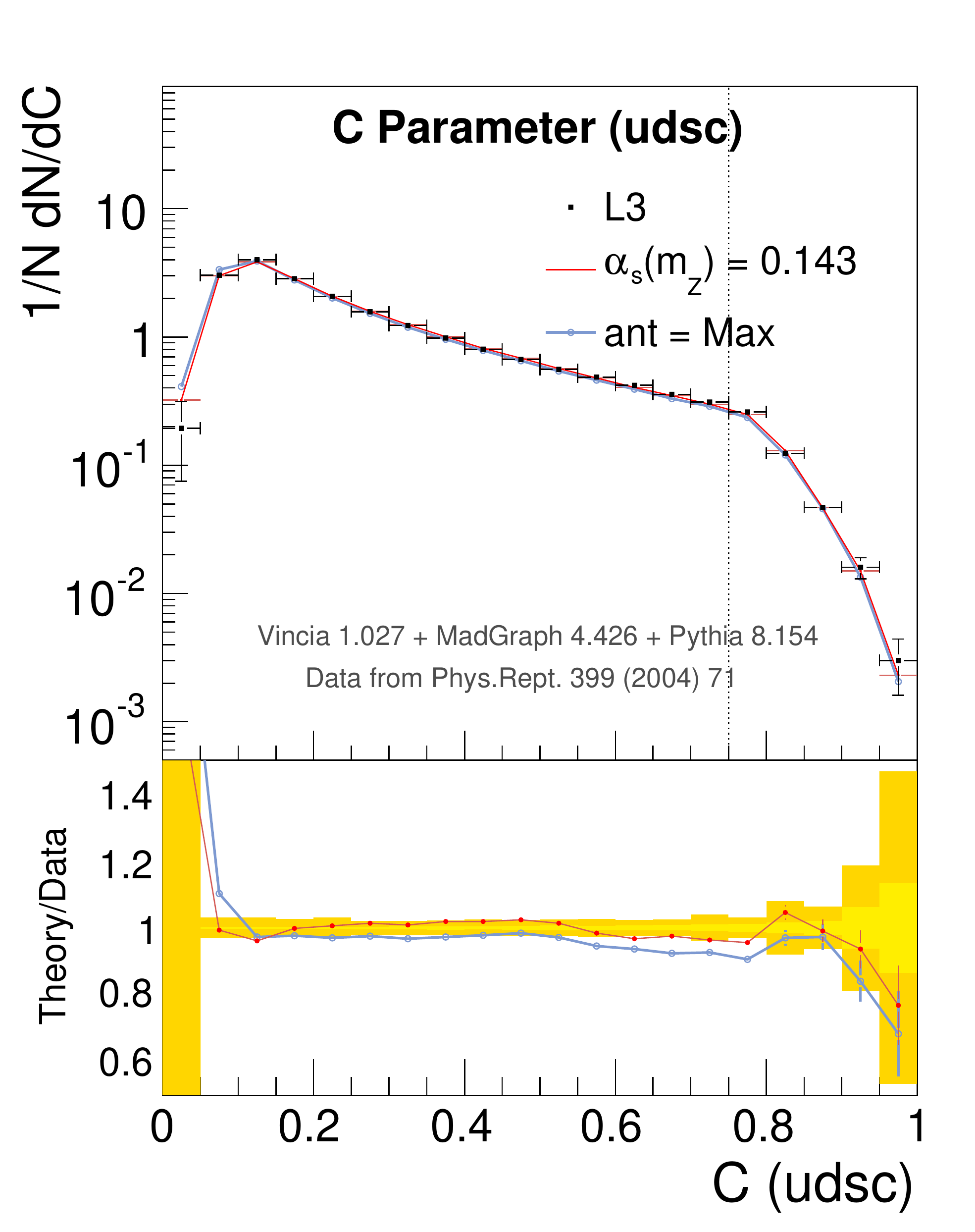}
\caption{Normalized $C$-parameter distribution. \Vc compared to L3 data for
  light-flavor $Z$ decays \cite{Achard:2004sv}. {\sl Left:} sector (thin)
  vs.\ global (thick) showers, using default (global) \Vc tune. {\sl
    Right:} sector shower using $\alpha_s(m_Z)=0.143$ (thin)
  vs.\ ``Max'' antenna functions \cite{Ridder:2011dm} (thick).
 \label{fig:C}}
\end{figure}

\begin{figure}[t]
\centering
\includegraphics*[scale=0.37]{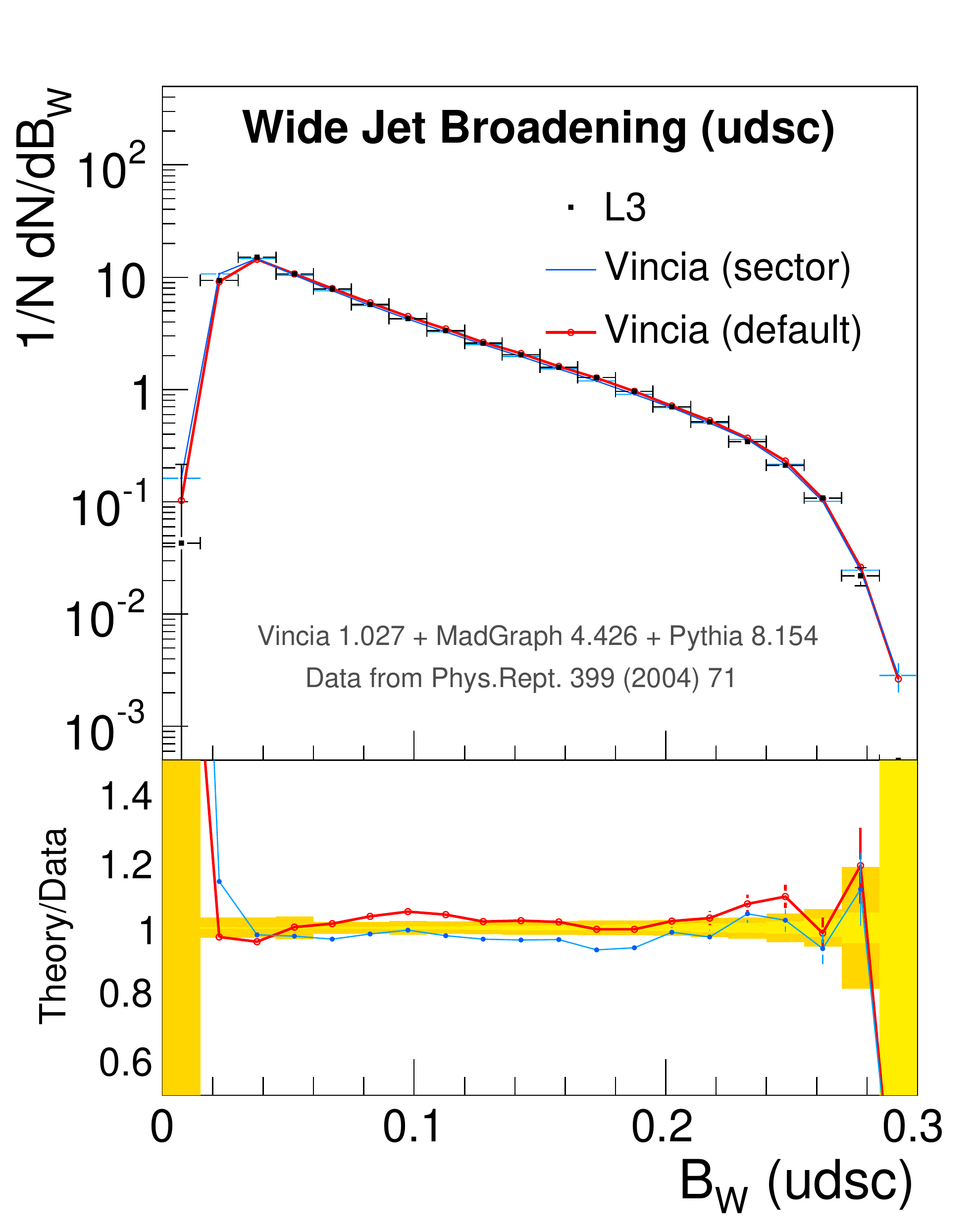}
\includegraphics*[scale=0.37]{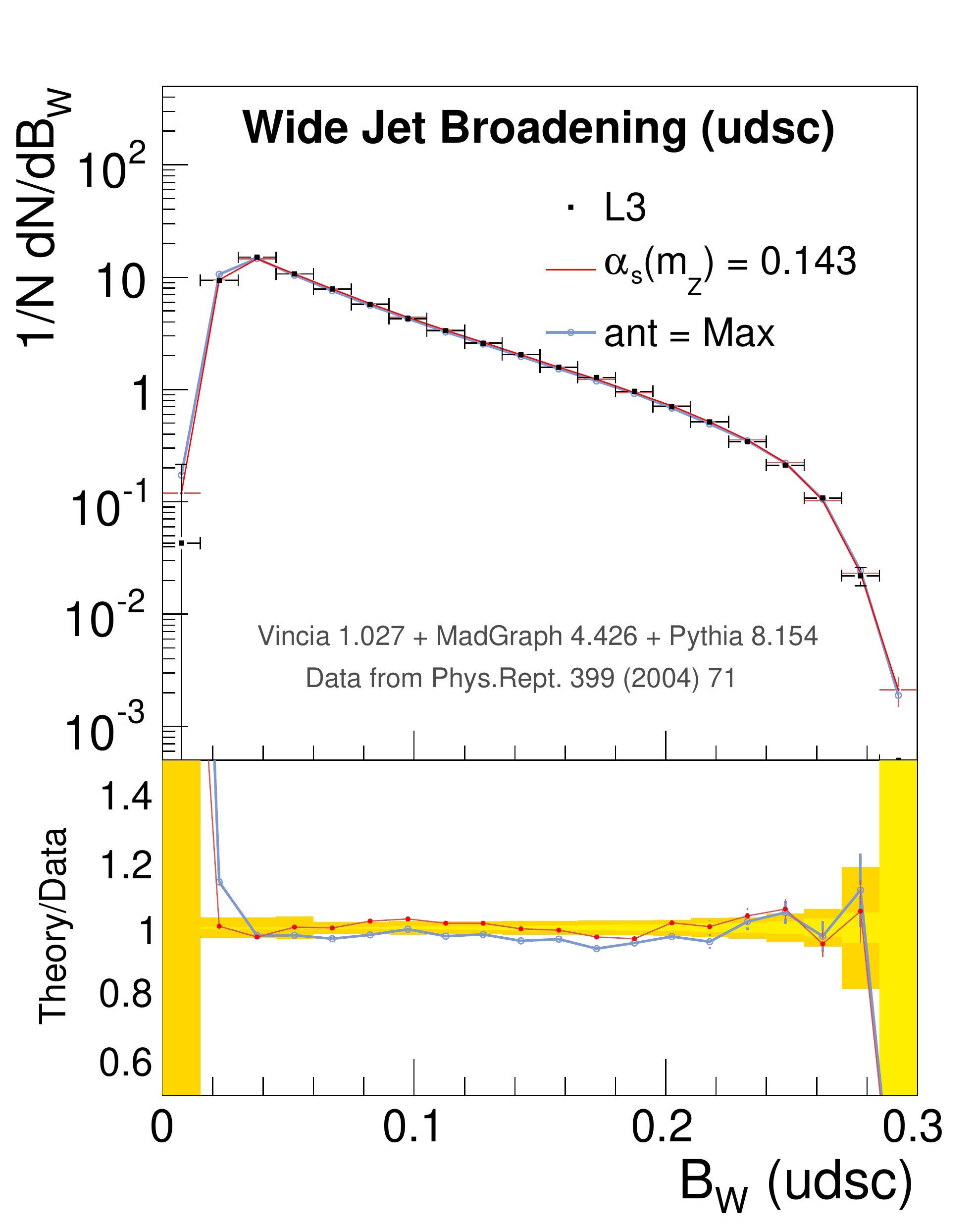}
\caption{Normalized Wide-Jet-Broadening ($B_W$) 
  distribution. \Vc compared to L3 data for
  light-flavor $Z$ decays \cite{Achard:2004sv}. {\sl Left:} sector (thin)
  vs.\ global (thick) showers, using default (global) \Vc tune. {\sl
    Right:} sector shower using $\alpha_s(m_Z)=0.143$ (thin)
  vs.\ ``Max'' antenna functions \cite{Ridder:2011dm} (thick).
 \label{fig:bw}}
\end{figure}

\begin{figure}[t]
\centering
\includegraphics*[scale=0.37]{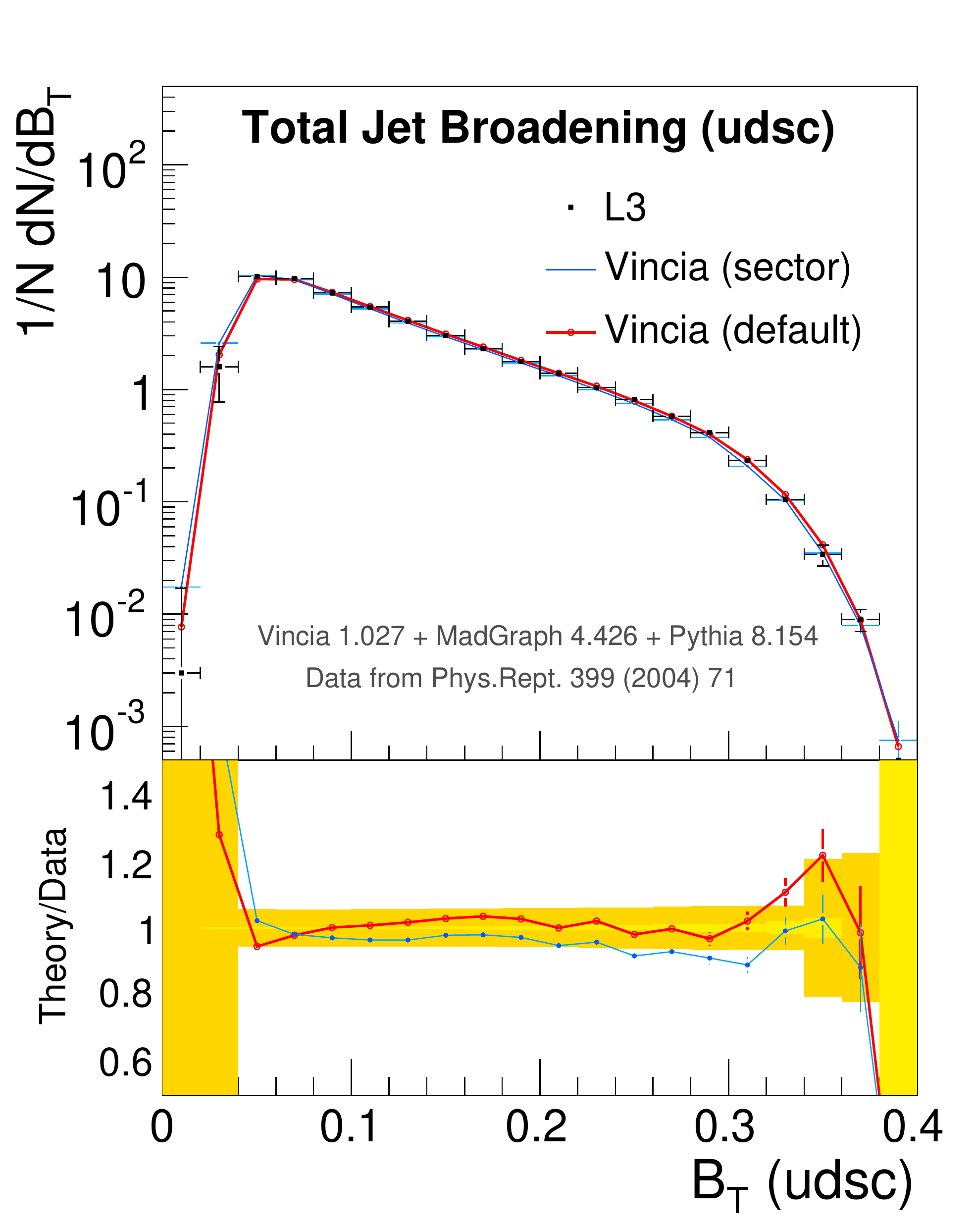}
\includegraphics*[scale=0.37]{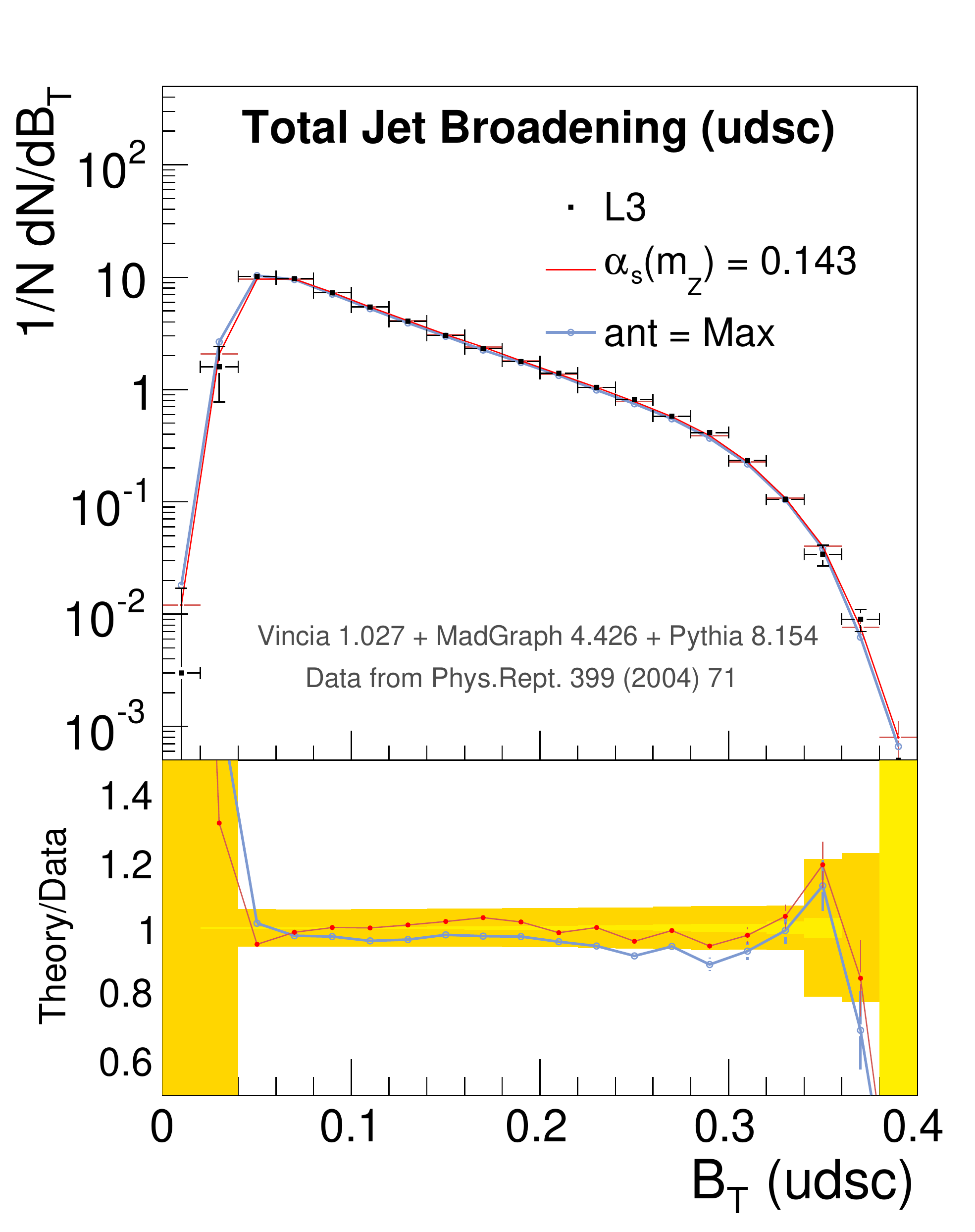}
\caption{Normalized Total-Jet-Broadening ($B_T$) 
  distribution. \Vc compared to L3 data for
  light-flavor $Z$ decays \cite{Achard:2004sv}. {\sl Left:} sector (thin)
  vs.\ global (thick) showers, using default (global) \Vc tune. {\sl
    Right:} sector shower using $\alpha_s(m_Z)=0.143$ (thin)
  vs.\ ``Max'' antenna functions \cite{Ridder:2011dm} (thick).
\label{fig:bt}}
\end{figure}

\begin{figure}[t]
\centering
\includegraphics*[scale=0.37]{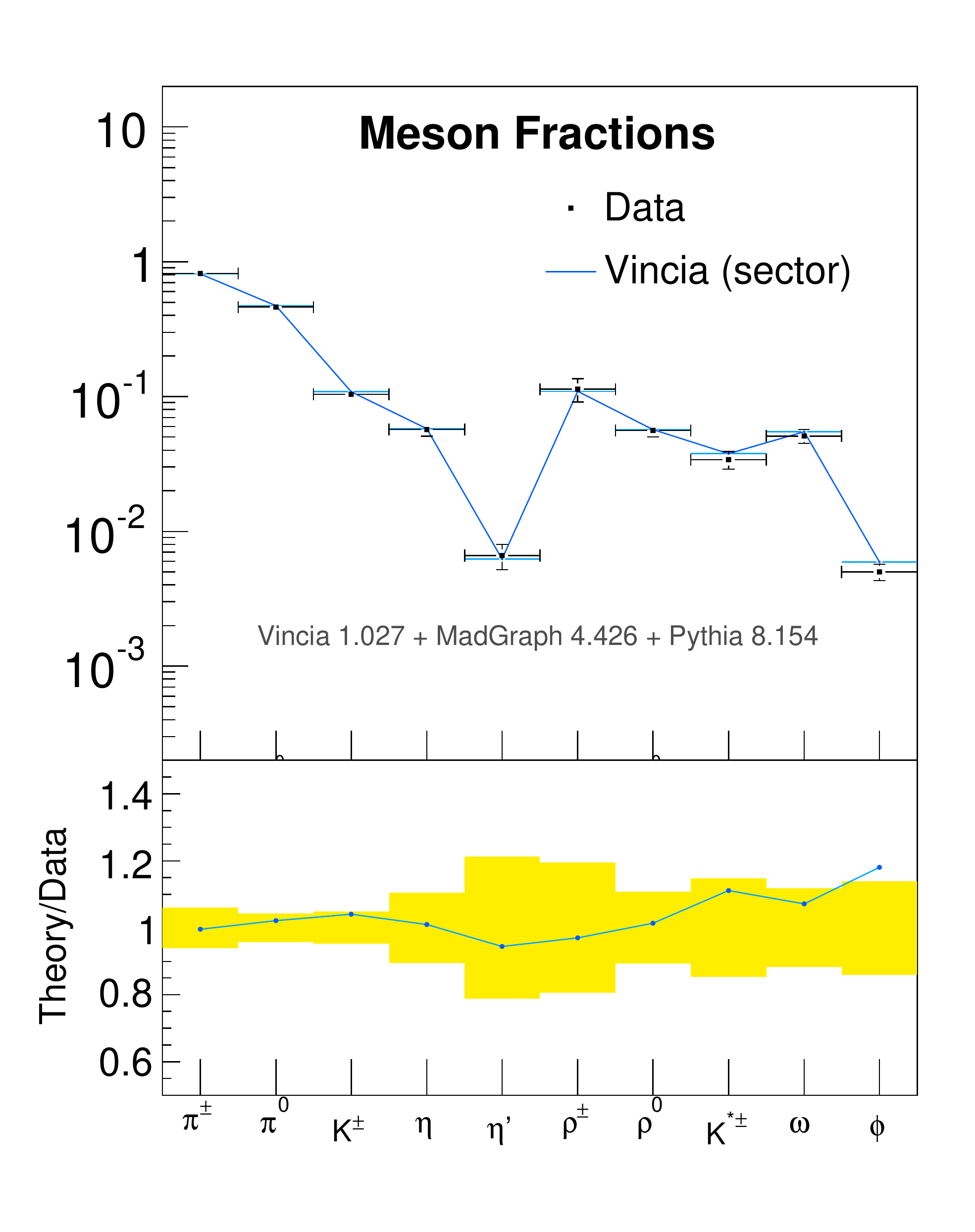}
\includegraphics*[scale=0.37]{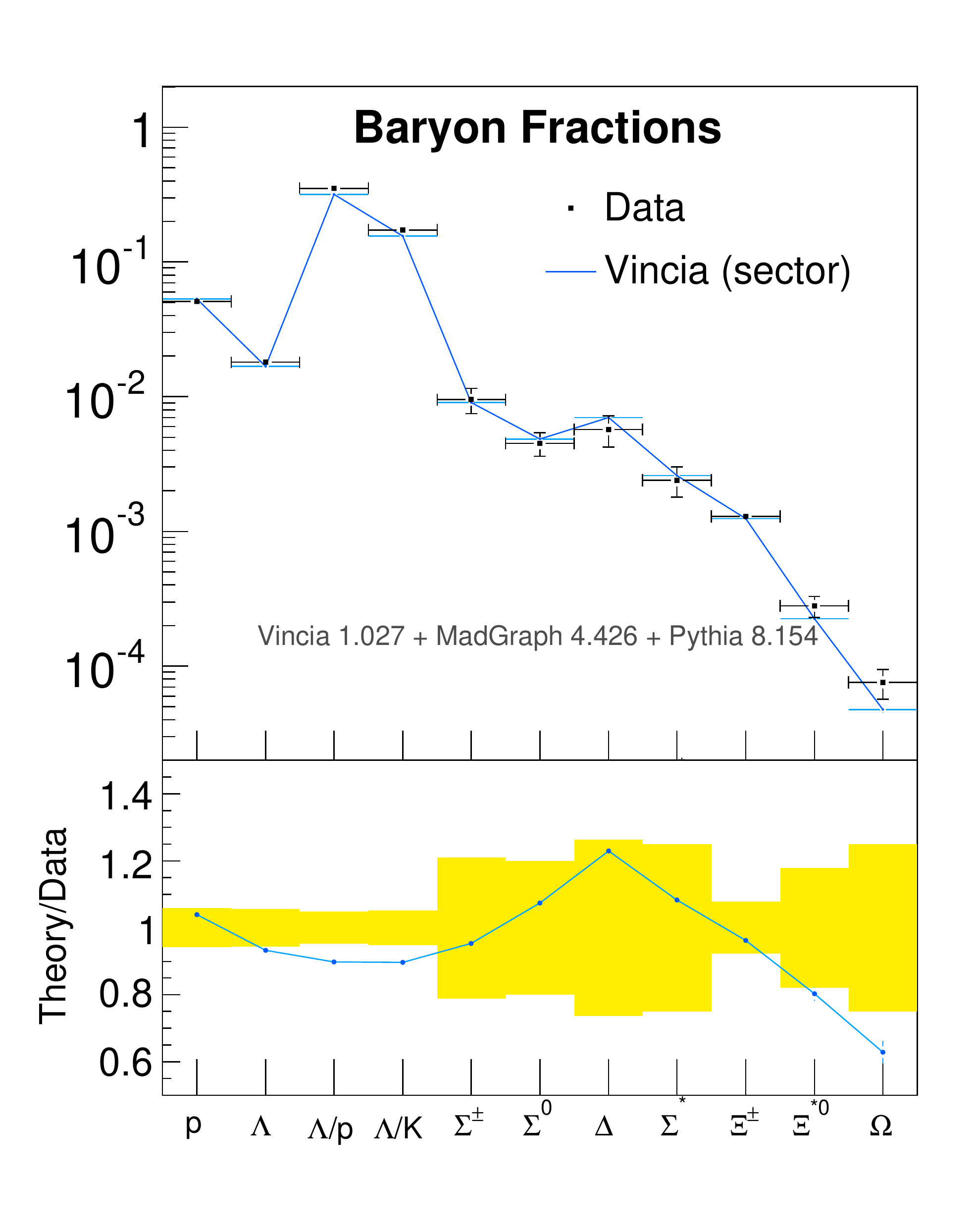}
\caption{Selected meson and baryon rates, compiled from the numbers in 
  \cite{Lafferty:1995jt,Nakamura:2010zzi}, expressed as fractions of the average
  charged multiplicity. \label{fig:rates}} 
\end{figure}

This appendix contains some further comparisons of the sector shower
with LEP distributions, as follows: the $C$ event-shape variable 
(\figRef{fig:C}), and the Wide and Total Jet broadening parameters
(\figsRef{fig:bw} and \ref{fig:bt}, respectively), defined as in 
\cite{Achard:2004sv}, to which we compare. We also compare the 
production rates of selected identified baryon and meson species,
normalized to the average charged-particle multiplicity, 
 to our own average over the various identified-particle measurements
 performed at LEP~\cite{Lafferty:1995jt,Nakamura:2010zzi}~(\figRef{fig:rates}). 
\section{Tune Parameters \label{app:tune}}
The default tune of \Vc 1.0.27 is ``Jeppsson4'', an update of the
original ``Jeppsson'' 
  tune presented in \cite{Giele:2011cb}. The parameters are optimized
  for use with the global shower (the default in \Vc) but are here
  used for the sector shower as well, with
comments as given in \secRef{sec:results}. 
The Jeppsson4 tune is characterized by the following parameters:
{\small
\begin{verbatim}
! * VINCIA alphaS
Vincia:alphaSValue        = 0.139
Vincia:alphaSscaleFactor  = 0.5
Vincia:alphaSorder        = 1
Vincia:alphaSmode         = 3  
! * VINCIA Shower cutoff scale
Vincia:cutoffType         = 1
Vincia:cutoffScale        = 1.0     
! * PYTHIA String fragmentation parameters
StringZ:aLund             = 0.55   
StringZ:bLund             = 0.95   
StringZ:aExtraDiquark     = 1.0    
StringPT:sigma            = 0.275  
StringPT:enhancedFraction = 0.01   
StringPT:enhancedWidth    = 2.0    
! * PYTHIA String breakup flavor parameters
StringFlav:probStoUD     = 0.20    
StringFlav:mesonUDvector = 0.45    
StringFlav:mesonSvector  = 0.7     
StringFlav:probQQtoQ     = 0.085   
StringFlav:probSQtoQQ    = 1.00    
StringFlav:probQQ1toQQ0  = 0.035   
StringFlav:decupletSup   = 1.0     
StringFlav:etaSup        = 0.68    
StringFlav:etaPrimeSup   = 0.085   
\end{verbatim}
}
\bibliography{sector}
\end{document}